\documentclass[%
 reprint,
 amsmath,amssymb,
 aps,
pra,
floatfix,
]{revtex4-2}
\pdfoutput=1
\usepackage{graphicx}
\usepackage{dcolumn}
\usepackage{bm}
\usepackage{float}
\usepackage{subfloat}
\usepackage{amsfonts}
\usepackage{graphicx}
\usepackage{mathtools}
\usepackage{braket}
\usepackage{subfig}
\usepackage{hyperref}
\hypersetup{colorlinks,allcolors=black}
\usepackage{csquotes}
\usepackage{xcolor, soul}
\usepackage{amsthm}
\newtheorem*{theorem}{Theorem}
\sethlcolor{yellow}
\DeclareMathOperator{\tr}{Tr}

\begin{document}

\preprint{APS/123-QED}

\title{Information capacity analysis of fully correlated multi-level amplitude damping channels}

\author{Rajiuddin Sk}
\email{skrajiuddin@gmail.com}
\author{Prasanta K. Panigrahi}%
 \email{pprasanta@iiserkol.ac.in}
\affiliation{%
Department of Physical Sciences, Indian Institute of Science Education and Research Kolkata, Mohanpur 741246, West Bengal, India}%

\date{\today}

\begin{abstract}
The primary objective of quantum Shannon theory is to evaluate the capacity of quantum channels. In spite of the existence of rigorous coding theorems that quantify {the} transmission of information through quantum channels, superadditivity effects limit our understanding of the channel capacities. In this paper, we mainly focus on a family of {channels} known as multi-level amplitude damping {channels}. We investigate {some of} the information capacities of the simplest member of multi-level Amplitude Damping Channel, a qutrit channel, in {the} presence of correlations between successive applications of the channel.
{We find the upper bounds of the single-shot classical capacities and calculate the quantum capacities associated with a specific class of maps after investigating the degradability property of the channels.}
Additionally, the quantum and classical capacities of the channels have been computed in entanglement-assisted scenarios.
\end{abstract}

\maketitle


\section{\label{sec:level1}Introduction}
{In quantum information theory, information is encoded in a quantum system and transmitted from one party to another via a quantum communication channel. However, the communication process is subject to imperfections because of the presence of noise within the channel.}
Hence, the {optimal rate of} information that can be efficiently transmitted via a quantum channel is a topic of importance for {the} practical implementation of information processing tasks. The term channel capacity is used to quantify this information, and it is the main focus of our paper \cite{nielsen2002quantum,holevo2019quantum}.
From the pioneering work of Shannon, one can compute the capacity of classical channels using the principle and framework of classical information theory \cite{shannon2001mathematical}. 
{A quantum channel can have multiple capacity definitions based on whether classical, private classical, or quantum information is being sent, as well as whether there are any additional resources shared between the sender and receiver. The shared entanglement between the communicating parties can have the potential to enhance the communication capability, and the corresponding capacity is commonly referred to as the entanglement-assisted capacity} \cite{watrous2018theory,wilde2013quantum,gyongyosi2018survey,bennett1999entanglement}. 
 
In quantum operational theory, the noisy quantum channel is defined by a completely positive and trace-preserving linear map (CPTP). {Over the last three decades, the various capacities of qubit channels have been investigated} 
\cite{poshtvan2022capacities,buscemi2010quantum,fanizza2021estimating}. {However, the different capacities for qubit channels are not completely computable. For instance, while the classical capacity of a depolarizing channel is known, the quantum capacity is still unknown over a certain parameter range, specifically when the decoherence parameter is in the interval $0<p<1/4$} \cite{king2003capacity}. 
The detection scheme of quantum capacity for different quantum channels \cite{singh2022detecting,macchiavello2016detecting,macchiavello2016witnessing} has been experimentally verified by Cuevas {\it et al.}, \cite{cuevas2017experimental}, as reported in their study. 
In the simplest case, a quantum channel is memoryless, {\it i.e.}, the channel acts independently on successive applications. However, in practical scenarios, memory effects or correlations may be present between successive applications of the channel.
When the information transmission rate is high, such memory effects become unavoidable and have been experimentally investigated in optical fibres \cite{banaszek2004experimental} and solid-state devices \cite{paladino2002decoherence,makhlin2001quantum}. 
{The perfect memory channel or fully correlated channel can be realized physically when the distance between two information carriers is negligible such that the time interval $\tau$ between the interactions of the local environment with successive carriers is much shorter than the dissipation time $\tau_E$ of the environment, {\it i.e.}, $\tau<<\tau_E$ } \cite{giovannetti2005dynamical}.
{Recently, an experimental method to detect the lower bound of the quantum capacity of a correlated dephasing channel was reported in Ref.} \cite{cimini2020experimental}, {where the correlated channels were realized using liquid crystals that affect the polarization of photons.} Over the last two decades, quantum memory channels have gained significant attention from researchers \cite{caruso2014quantum,macchiavello2002entanglement,bowen2004quantum,kretschmann2005quantum,plenio2007spin,caruso2010teleportation,lupo2010capacities,sk2022protecting,hu2020protecting,xu2019quantum}. Moreover, it has been shown that the effect of memory between successive applications of the channel can enhance the quantum capacity of a channel \cite{d2007quantum,plenio2007spin}.
{The primary challenge in computing the capacities stems from the fact that the associated Holevo quantity and coherent information obey the superadditivity property, and the calculation of capacities involves a regularization process.}
Because of this difficulty, there are only a few memory channels whose capacities have been analyzed completely \cite{d2007quantum,plenio2007spin,lupo2010capacities}. 

{The amplitude damping channel (ADC) is a well-known example of a non-unital channel, and the information capacity of qubit ADC has been investigated under different scenarios}
\cite{giovannetti2005information,jahangir2015quantum,khatri2020information}. 
{The quantum capacity of a two-level ADC is well comprehended. However, the classical capacity remains unclear, with only knowledge of the single-shot classical capacity} \cite{giovannetti2005information}.
The {single-shot classical capacity} and quantum capacity of a two-level ADC have been studied by D’Arrigo {\it et al.} when the channel is fully correlated \cite{d2013classical}. 
{A comprehensive analysis of the capacity of multi-level channels is essential, given that these physical noises are unavoidable when qudit states are used for information processing and computational purposes. However, the information capacities of the multi-level amplitude damping (MAD) channel have not been explored well. Recently, Chessa} {\it et al.} {have investigated the quantum capacity of a three-level amplitude damping channel} \cite{chessa2021quantum}. {Subsequently, there are few other works that focus on the higher dimensional amplitude damping noise model, particularly in terms of their capacity analysis} \cite{chessa2021partially,chessa2023resonant}.

The focus of this paper is on the fully correlated multi-level amplitude damping (MAD) channel, in which two qutrits simultaneously {relax from high-energy states to the lower energy states}. Making use of {the} Lindblad master equation and finding out the Kraus operators, we have characterised the fully correlated MAD channel.
We specifically analyze the fully correlated MAD channel on the qutrit space and {systematically examine the quantum capacity of different associated maps obtained by imposing some constraints on the decay parameters}. Before calculating the capacities, we find the conditions under which these quantities can be determined. We also compute an {upper bound of} the single-shot classical capacities in some regions of the damping parameters space. Finally, the quantum and classical capacities have been analyzed in the entanglement-assisted scenario.

{The paper is organized in the following manner}. In Sec. \ref{Sec_model}, we have discussed the MAD channel and the model corresponding to a fully correlated MAD channel. Sec. \ref{Sec_channel_property} deals with an overview of complementary channels along with the degradability property of the channel. This section also addresses the covariance property of the channel. In Sec. \ref{Sec_CC}, we have derived {the upper bound of} single-shot classical capacity in special cases of fully correlated MAD channels, while Sec. \ref{Sec_QC} contains the analysis of quantum capacity in different scenarios. Sec. \ref{Sec_EAC} is dedicated to the analysis of the capacities in entanglement-assisted scenarios. Finally, concluding remarks are given in Sec. \ref{Sec_conclusion}.
\begin{figure}[ht]
    \centering
    \includegraphics{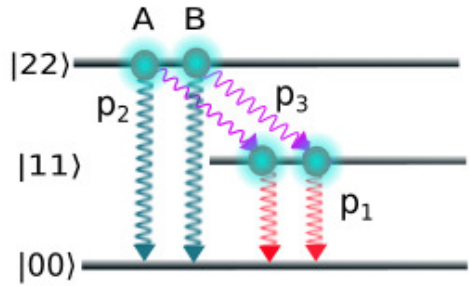}
    \caption{Schematic representation of a fully correlated MAD channel for a three-level system. A and B are two qutrits that undergo these relaxations when they are fully correlated.}
    \label{Fig_1}
\end{figure}
\section{The Model}\label{Sec_model}
In this section, we begin with a brief review of the multi-level amplitude damping noise model, including memoryless and correlated scenarios \cite{nielsen2002quantum,holevo2019quantum}. We also analyze the fully correlated MAD channel for a three-level system from the Lindblad master equation approach.
A MAD channel is a linear mapping which is completely positive and trace preserving (CPTP) map \cite{watrous2018theory,wilde2013quantum}. A $d$-dimensional MAD channel is described by the following set of Kraus operators \cite{chessa2021quantum}:
\begin{equation}
\begin{gathered}
E_0 \equiv|0\rangle\langle 0|+\sum_{1 \leq l \leq d-1} \sqrt{1-\zeta_l}\ket{k}\bra{k}\\
E_{k l} \equiv \sqrt{p_{l k}}\ket{k}\bra{l},\\
\end{gathered}
\label{Eq_MADkraus}
\end{equation}
where $\{\ket{l} \}$ are the set of orthonormal basis of the Hilbert space $\mathcal{H}_S$ with
$0 \leq k,l \leq d-1$, $p_{l k}$ are the decay parameter and $\zeta_l=\sum_{0\leq k < l} p_{lk}\leq 1$.
In this work, we will limit our examination to the particular category of MAD channels illustrated in Eq. (\ref{Eq_MADkraus}) that are linked to a three-level system or qutrit system. 
For a single qutrit system, the evolution of the density matrix is calculated by the relation $\rho_{t}= \sum E_n \rho E_n^\dagger$, where $E_n$ is defined in Eq. (\ref{Eq_MADkraus}) and $n$ can take values from 0 to 3.

For two consecutive uses of the memoryless MAD channel, the evolution is 
\begin{equation}
\rho_{t}=\Phi(\rho)=\sum_{ij} E_i^m\otimes E_j^m \rho {E_i^m}^\dagger\otimes {E_j^m}^\dagger,
\end{equation}
where $E^m$ refers to the Kraus operator corresponding to an uncorrelated channel or memoryless channel.
 If the subsequent action of channels has some correlations, it is not possible to write the Kraus operators simply as the tensor product of individual single qutrit Kraus operators \cite{yeo2003time}. The transformation of the density matrix $\rho$ for two consecutive applications of the channel with arbitrary degrees of memory can be written as
 \begin{equation}
     \rho_t=(1-\mu)\sum_{ij} E_i^m\otimes E_j^m \rho {E_i^m}^\dagger\otimes {E_j^m}^\dagger+\mu\sum_k E_k^c \rho {E_k^c}^\dagger,
 \end{equation}
 where $\mu$ is the memory parameter, and $E^c$ corresponds to the Kraus operator for a fully correlated channel. For $\mu=0$, the channel is said to be a memoryless or uncorrelated channel, whereas for $\mu=1$, the channel is a fully correlated or perfect memory channel. We find the information capacity here for this fully correlated channel.
The explicit form of the Kraus operators describing the fully correlated MAD channel can be obtained from the solution of the Lindblad master equation, which we will derive below methodically.
The evolution of a three-level system over time is described by the following Lindblad master equation:
\begin{equation}
\dot{\rho}=\mathcal{L}\rho=\mathcal{L}_c(\rho)+\mathcal{D}(\rho)=-i[H,\rho]+\mathcal{D}(\rho),
\label{LM}
\end{equation}
where $\rho$ is the density matrix for a three-level system and $-i[H,\rho]$ represents the coherent evolution which is unitary in nature, and the $\mathcal{D}(\rho)$ is the damping part which indicates the non-unitary evolution.\\
For a Markov quantum channel, a stochastic map can be expressed as follows:
\begin{equation}
    \rho\rightarrow \rho_t=\Phi(\rho)=e^{\mathcal{L}t}\rho.
\end{equation}
The above equation gives the dynamics of the system coupled with the reservoir.
The non-unitary part, which gives the dissipation of the density matrix, is
\begin{equation}
\begin{split}
 \mathcal{D}(\rho)= \frac{\Gamma_3}{2} \left(2 \sigma_{12} \rho \sigma_{21}-\sigma_{22} \rho-\rho \sigma_{22}\right)+\frac{\Gamma_2}{2}(2\sigma_{02} \rho \sigma_{20}-\\
 \sigma_{22}\rho-\rho \sigma_{22})+\frac{\Gamma_1}{2}\left(2\sigma_{01} \rho \sigma_{10}-\sigma_{11}\rho-\rho \sigma_{11}\right).\\
 \end{split}
\end{equation}
\newline
In the above equation, $\Gamma_3$, $\Gamma_2$ and $\Gamma_1$ are the spontaneous decay rate corresponding to the transition of atom from $\ket{2}\rightarrow \ket{1}$, $\ket{2}\rightarrow \ket{0}$ and $\ket{1}\rightarrow \ket{0}$ respectively. The transition operator $\sigma_{kl}$ indicates the transition of atom from $\ket{k}\rightarrow \ket{l}$, {\it i.e.}, $\sigma_{kl}=\ket{k}\bra{l}$. These transitions are governed by the interaction between the system (S) and environment ($\mathcal{E}$).

We now extend the Lindblad equation for the case of two three-level atoms, where the action of the amplitude damping is fully correlated:
\begin{equation}\label{Eq_disspator}
\begin{split}
 \mathcal{D}^c (\rho)= \frac{\Gamma_3}{2} \left(2 \mathcal{S}_{12} \rho \mathcal{S}_{21}-\mathcal{S}_{22} \rho-\rho \mathcal{S}_{22}\right)+\frac{\Gamma_2}{2}(2\mathcal{S}_{02} \rho \mathcal{S}_{20}-\\
 \mathcal{S}_{22}\rho-\rho \mathcal{S}_{22})+\frac{\Gamma_1}{2}\left(2\mathcal{S}_{01} \rho \mathcal{S}_{10}-\mathcal{S}_{11}\rho-\rho \mathcal{S}_{11}\right),\\
 \end{split}
\end{equation}
where $\mathcal{S}_{kl}=\sigma_{kl}\otimes\sigma_{kl}$. The decay process associated with two qutrits, A and B, is displayed in Fig. (\ref{Fig_1}). 
There are several methods available to solve the master equation of the form shown in Eq. (\ref{Eq_disspator}). We adopt the method proposed by Briegel {\it et al.} \cite{briegel1993quantum}, wherein they have used left, $\{\mathbb{L}_i \}$ and right, $\{\mathbb{R}_i \}$, damping basis with damping eigenvalue $\lambda_i$ for a Lindblad super-operator that yields the image of a trace-preserving, completely positive map:
\begin{equation}\label{Eq_7}
    \rho_t=\Phi(\rho)=\sum_i \tr(\mathbb{L}_i \rho) e^{\lambda_i t} \mathbb{R}_i.
\end{equation}
Here, the left and right eigenoperators, $\{\mathbb{L}_i \}$ and $\{\mathbb{R}_i \}$ have the same eigenvalue $\lambda_i$ and they satisfy the eigenvalue equations, $\mathbb{L}\mathcal{D}=\lambda \mathbb{L}$ and  $\mathcal{D}\mathbb{R}=\lambda \mathbb{R}$ respectively. They also obey the duality relation $\tr\{ \mathbb{L}_i \mathbb{R}_j \}=\delta_{ij}$.

Let us consider the initial density matrix of a two qutrit system in the absence of any interaction with the environment as
\begin{equation}\label{Input}
   \rho=\left(\begin{array}{ccccc}\rho_{00} & \cdots & \rho_{04} & \cdots & \rho_{08} \\ \vdots & & \vdots & & \vdots \\ \rho_{40} & \cdots & \rho_{44} & \cdots & \rho_{48} \\ \vdots & & \vdots & & \vdots \\ \rho_{80} & \cdots & \rho_{84} & \cdots & \rho_{88}\end{array}\right).
\end{equation}
We solve Eq. (\ref{Eq_disspator}) by converting the initial density matrix in Hilbert space into the state in Fock-Liouville space and finding the corresponding Lindblad super-operator. After that, we find the left and right eigenbasis and corresponding eigenvalues, which will give the output density matrix according to Eq. (\ref{Eq_7}). In our two qutrit system the Lindblad super-operator is $81\times81$ matrix with the eigen values ($\lambda_i$) are $(0)_{49}$, $-(\Gamma_2+\Gamma_3)_1$, $-((\Gamma_2+\Gamma_3)/2)_{14}$, $(\Gamma)_1$, $(-\Gamma_1/2)_{14}$ and $((\Gamma_1+\Gamma_2+\Gamma_3)/2)_{2}$  respectively. Note that the subscripts attached to eigenvalues correspond to the frequency of occurrence of each specific eigenvalue.

\begin{figure}[ht]
    \centering
    \includegraphics[width=8cm, height=7cm]{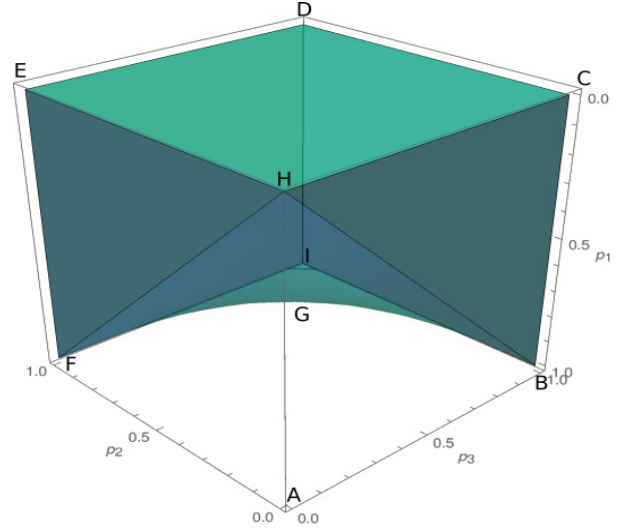}
    \caption{The accessible domain of the damping parameters. The coloured region gives the values of the damping parameters for which the CPTP property of the maps is satisfied.}
    \label{Region}
\end{figure}

Evidently, the dynamical evolution of the input density matrix has the following form:
\begin{widetext}
\begin{equation}
\resizebox{\textwidth}{!}{$
\rho_t=
\begin{pmatrix}
 \Tilde{\rho}_{00} & \rho_{01} & \rho_{02} & \rho_{03} & e^{-\Gamma_{1}t/2}\rho_{04} & \rho_{05} & \rho_{06} & \rho_{07} & e^{-\frac{\Gamma_{2}+\Gamma_{3}}{2}t}\rho_{08}\\
\rho_{10} & \rho_{11} & \rho_{12} & \rho_{13} & e^{-\Gamma_{1}t/2}\rho_{14} & \rho_{15} & \rho_{16} & \rho_{17} & e^{-\frac{(\Gamma_{2}+\Gamma_{3})}{2}t}\rho_{18}\\
\rho_{20} & \rho_{21} & \rho_{22} & \rho_{23} & e^{-\Gamma_{1}t/2}\rho_{24} & \rho_{25} & \rho_{26} & \rho_{27} & e^{-\frac{(\Gamma_{2}+\Gamma_{3})}{2}t}\rho_{28}\\
\rho_{30} & \rho_{31} & \rho_{32} & \rho_{33} & e^{-\Gamma_{1}t/2}\rho_{34} & \rho_{35} & \rho_{36} & \rho_{37} & e^{-\frac{\Gamma_{2}+\Gamma_{3}}{2}t}\rho_{38}\\
e^{-\Gamma_{1}t/2}\rho_{40} & e^{-\Gamma_{1}t/2}\rho_{41} & e^{-\Gamma_{1}t/2}\rho_{42} & e^{-\Gamma_{1}t/2}\rho_{43} & \Tilde{\rho}_{44} & e^{-\Gamma_{1}t/2}\rho_{45} & e^{-\Gamma_{1}t/2}\rho_{46} & e^{-\Gamma_{1}t/2}\rho_{47} & e^{-\frac{\Gamma_{1}+\Gamma_{2}+\Gamma_{3}}{2}t}\rho_{48}\\
\rho_{50} & \rho_{51} & \rho_{52} & \rho_{53} & e^{-\Gamma_{1}t/2}\rho_{54} & \rho_{55} & \rho_{56} & \rho_{57} & e^{-\frac{\Gamma_{2}+\Gamma_{3}}{2}t}\rho_{58}\\
\rho_{60} & \rho_{61} & \rho_{62} & \rho_{63} & e^{-\Gamma_{1}t/2}\rho_{64} & \rho_{65} & \rho_{66} & \rho_{67} & e^{-\frac{\Gamma_{2}+\Gamma_{3}}{2}t}\rho_{68}\\
 \rho_{70} & \rho_{71} & \rho_{72} & \rho_{73} & e^{-\Gamma_{1}t/2}\rho_{74} & \rho_{75} & \rho_{76} & \rho_{77} & e^{-\frac{\Gamma_{2}+\Gamma_{3}}{2}t}\rho_{78}\\
e^{-\frac{\Gamma_{2}+\Gamma_{3}}{2}t}\rho_{80} & e^{-\frac{\Gamma_{2}+\Gamma_{3}}{2}t}\rho_{81} & e^{-\frac{\Gamma_{2}+\Gamma_{3}}{2}t}\rho_{82} & e^{-\frac{\Gamma_{2}+\Gamma_{3}}{2}t}\rho_{83} & \ e^{-\frac{\Gamma_{1}+\Gamma_{2}+\Gamma_{3}}{2}t}\rho_{84} & e^{-\frac{\Gamma_{2}+\Gamma_{3}}{2}t}\rho_{85} & e^{-\frac{\Gamma_{2}+\Gamma_{3}}{2}t}\rho_{86} & e^{-\frac{\Gamma_{2}+\Gamma_{3}}{2}t}\rho_{87} & e^{-(\Gamma_{2}+\Gamma_{3})t}\rho_{88}  
\end{pmatrix}.
$}
\end{equation}
\end{widetext}
In the above density matrix, $$
\begin{aligned}
\Tilde{\rho}_{00}=&\rho_{00}+\rho_{44}(1-e^{-\Gamma_1 t})+\rho_{88}(1-e^{-(\Gamma_{2}+\Gamma_{3})t}\\
&- \Theta(\Gamma)e^{-\Gamma_1 t}+ \Theta(\Gamma)e^{-(\Gamma_2+\Gamma_3)t})
\end{aligned}
$$ 
and $$\Tilde{\rho}_{44}=e^{-\Gamma_1 t}\rho_{44}+\Theta(\Gamma)(e^{-\Gamma_1 t}-e^{-(\Gamma_2+\Gamma_3)t}),$$
where $\Theta(\Gamma)=\Gamma_3/(\Gamma_3+\Gamma_2-\Gamma_1)$.
As we have discussed earlier, the dynamics of a two-qutrit state $\rho$ subject to a MAD channel with full Markovian memory can be written in terms of Kraus representation as  $\rho_t=\sum_n E_{n}\rho E^{\dagger}_n$.
It is evident that the Kraus operators $E_{n}$, can be computed from the relation $\sum_i \tr(\mathbb{L}_i \rho) e^{\lambda_i t} \mathbb{R}_i=\sum_n E_n\rho E^{\dagger}_n$. The explicit expression of the Kraus operators is obtained by solving the correlated Lindblad equation, which is presented as follows:
\begin{align}\label{kraus}
E_{00}=&\ket{00}\bra{00}+\sqrt{1-p_1}\ket{11}\bra{11}\nonumber\\
&+\sqrt{(1-p_2)(1-p_3)}\ket{22}\bra{22}+\sum_{\substack{i,j=0\\i\neq j}}^{2}\ket{ij}\bra{ij}\nonumber\\
E_{11}=&\sqrt{p_1}\ket{00}\bra{11}\nonumber\\
E_{22}=&\sqrt{p_1+(1-\Theta(\Gamma))p_{123}}\ket{00}\bra{22}\nonumber\\ E_{33}=&\sqrt{\Theta(\Gamma)p_{123}}\ket{11}\bra{22},
\end{align} 
where $\Theta(\Gamma)=\Gamma_{3}/(\Gamma_3+\Gamma_2-\Gamma_1)$, $p_{123}=e^{-\Gamma_1 t}-e^{-(\Gamma_2+\Gamma_3)t}=(1-p_1)-(1-p_2)(1-p_3)$ and $\Gamma_2+\Gamma_3>\Gamma_1$. We have written the Kraus operators in matrix form in the Appendix \ref{AP_deg} to make its structure easier to understand. The CPTP condition of the transformation is satisfied when $(1-p_1)\geq(1-p_2)(1-p_3)$, which gives accessible values of $p_1$, $p_2$ and $p_3$, shown in colored region of Fig. (\ref{Region}). The Kraus operators mentioned above fulfil the completeness relation $\sum E_n^\dagger E_n = I$. Conversely, the non-equivalence, $\sum E_n E_n^\dagger \neq I$, implies that the channel is non-unital.
\section{Channel properties}\label{Sec_channel_property}
{In the calculation of the quantum capacity, it is essential to optimize the coherent information, which is determined from the entropy of the output states of the quantum channel and its complementary map.}
In this section, we provide a brief description of the complementary channels, including the degradability and anti-degradability characteristics of quantum channels. Additionally, we demonstrate the covariance property of the channel.
\subsection{Complementary channel and degradability}\label{Sec_IIIA}
Let $\mathcal{O}(\mathcal{H})$ denote the space of positive linear operators on a Hilbert space $\mathcal{H}$. A quantum channel $\Phi$ maps the input state of the system $S$ into the output state of the system $S'$: $\mathcal{O}(\mathcal{H}_S)\rightarrow\mathcal{O}(\mathcal{H}_{S'})$. If $\mathcal{E}$ and $\mathcal{E}'$ represent the corresponding environment of the input system and output system, then from the Stinespring representation \cite{choi1975completely,stinespring1955positive}, one can define a quantum channel as
 \begin{equation}
   \Phi(\rho_S)=\tr_{\mathcal{E}'}(V\rho_S V^\dagger),
 \end{equation}
where $V$ represents an isometry: $\mathcal{H}_S\rightarrow\mathcal{H}_{S'}\otimes\mathcal{H}_{\mathcal{E}'}$. In this configuration, the complementary map $\Tilde{\Phi}$, which maps the input system to the output environment: $\mathcal{O}(\mathcal{H}_S)\rightarrow\mathcal{O}(\mathcal{H_{\mathcal{E}'}})$, is defined as
\begin{equation}\label{Eq_cmap}
  \Tilde{\Phi}(\rho_S)=\tr_{S'}(V\rho_S V^\dagger).
\end{equation}
If $E_k$ are the Kraus operators which characterize the map $\Phi$, and {the basis states of the environment are $\ket{k}_{\mathcal{E}}$}, then  the operator $V$ can be expressed as
\begin{equation}
    V=\sum_k E_k \otimes \ket{k}_{\mathcal{E}}.
\end{equation}
We can express Eq. (\ref{Eq_cmap}), equivalently as follows:
\begin{equation}\label{Eq_commap}
\Tilde{\Phi}(\rho_S)=\sum_{k,l} \tr_{S'}[E_k\rho_S E_k^\dagger]\ket{k}\bra{l}_{\mathcal{E}}.
\end{equation}
Let us revisit the definitions of a degradable channel and an anti-degradable channel \cite{devetak2005capacity}. A quantum channel $\Phi$ is degradable, when there exists another channel $\Phi_D$: $\mathcal{O}(\mathcal{H_{S'}})\rightarrow\mathcal{O}(\mathcal{H_{\mathcal{E}'}})$ such that 
\begin{equation}\label{Eq_degrad}
\Tilde{\Phi}=\Phi_D\circ\Phi.
\end{equation}
In the above equation, the symbol \enquote{$\circ$} represents the channel concatenation. On the contrary, the channel is anti-degradable when there exists another map $\Phi_{AD}$: $\mathcal{O}(\mathcal{H_{\mathcal{E}'}})\rightarrow\mathcal{O}(\mathcal{H_{S'}})$ such that 
\begin{equation}\label{Eq_antidegrad}
\Phi=\Phi_{AD}\circ\Tilde{\Phi}.
\end{equation}

If the mapping $\Phi$ is invertible, then simply we can make the inversion of it to construct the super-operators $\Tilde{\Phi}\circ\Phi^{-1}$ or $\Phi\circ\Tilde{\Phi}^{-1}$ and checking the CPTP of these super-operators we can conclude whether the channel is degradable or anti degradable. The complete positivity of the super-operators can be determined by examining the positivity of its Choi matrices \cite{smith2007degenerate}.
One can represent quantum channels as a matrix in the vector space since it connects the vector space of linear operators. This can be done by vectorization of density matrices: 
\begin{equation}
\begin{aligned}
\rho_{S}= & \sum_{k l} \rho_{k l}|k\rangle_{S}\langle l|\longrightarrow| \rho\rangle\rangle=\sum_{k l} \rho_{k l}|k\rangle_{S} \otimes|l\rangle_{S}\\
& \Phi\left(\rho_S\right) \longrightarrow \mathcal{M}_{\Phi}|\rho\rangle\rangle.
\end{aligned}
\end{equation}
In the above equation, $\mathcal{M}_{\Phi}$ is a $d^2_{S'}\times d^2_{S}$ dimension matrix, connects $\mathcal{H}_{S'}^{\otimes 2}$ and $\mathcal{H}_{S}^{\otimes 2}$.
Hence, starting from Eq. (\ref{Eq_degrad}), one can write the following identity:
\begin{equation}
  \mathcal{M}_{\Tilde{\Phi}}=\mathcal{M}_{\Phi_D}\mathcal{M}_{\Phi}. 
\end{equation}
Using this equality one can represent the super-operator $\Tilde{\Phi}\circ\Phi^{-1}$ as $\mathcal{M}_{\Tilde{\Phi}} \mathcal{M}^{-1}_{\Phi}$ provided that $\mathcal{M}_{\Phi}$ is invertible.

\subsection{Covariance property}
{Here, we inspect the covariance properties of the MAD channel with respect to certain unitary transformations. To begin with, we assume three unitary matrices}
$$
 \mathtt{V}_1= \begin{pmatrix}
     1 & 0 & 0\\
     0 & 1 & 0\\
     0 & 0 & -1\\
 \end{pmatrix}, 
 \mathtt{V}_2= \begin{pmatrix}
     1 & 0 & 0\\
     0 & -1 & 0\\
     0 & 0 & 1\\
 \end{pmatrix},
 \mathtt{V}_3= \begin{pmatrix}
     -1 & 0 & 0\\
     0 & 1 & 0\\
     0 & 0 & 1\\
 \end{pmatrix}.
$$
Now, we define sixteen unitary operations ($\mathbf{U}_0,\dots,\mathbf{U}_{15}$) using above three matrices as follows:
$$\mathbf{U}_i=\mathtt{V}_m\otimes\mathtt{V}_n \quad\forall m,n\; s.t\; 0\leq m\;\leq 3\;\text{and}\;0\leq n\;\leq 3,$$
where m and n take values 0,1,2,3 and note that $\mathtt{V}_0=\mathbb{I}_{3\times3}$. In the upcoming section, we will demonstrate how these unitaries can effectively eliminate the off-diagonal elements of the density matrix.

All the unitary operations $\mathbf{U}_i$ either commute or anti-commute with the Kraus operators given in Eq. (\ref{kraus}).
For a particular $\mathbf{U}_i$, for instance $\mathbf{U}_1=\mathtt{V}_0\otimes\mathtt{V}_1$, the Kraus operators $E_{00}$ and $E_{11}$ commute with the $\mathbf{U}_1$, whereas,  $E_{22}$ and $E_{33}$ anti-commute with $\mathbf{U}_1$:
$E_{00}\mathbf{U}_1= \mathbf{U}_1E_{00}$, $E_{11}\mathbf{U}_1= \mathbf{U}_1E_{11}$, $E_{22}\mathbf{U}_1= -\mathbf{U}_1E_{22}$, and $E_{33}\mathbf{U}_1= -\mathbf{U}_1E_{33}$.

Using these commutation and anti-commutation relations, it is straightforward to prove that
\begin{equation}
\begin{aligned}
&\Phi(\mathbf{U}_1\rho\mathbf{U}_1)=\\
&E_{00}\mathbf{U}_1\rho\mathbf{U}_1E_{00}^\dagger+E_{11}\mathbf{U}_1\rho\mathbf{U}_1E_{11}^\dagger\\
&+E_{22}\mathbf{U}_1\rho\mathbf{U}_1E_{22}^\dagger+E_{33}\mathbf{U}_1\rho\mathbf{U}_1E_{33}^\dagger\\
&=\mathbf{U}_1 E_{00}\rho E_{00}^\dagger\mathbf{U}_1+\mathbf{U}_1 E_{11}\rho E_{11}^\dagger\mathbf{U}_1\\
&+(-\mathbf{U}_1 E_{22})\rho (-E_{22}^\dagger\mathbf{U}_1)+(-\mathbf{U}_1E_{33}) \rho (-E_{33}^\dagger\mathbf{U}_1)\\
&=\mathbf{U}_1\Phi(\rho)\mathbf{U}_1.
\end{aligned}
\end{equation}
In the same way, we can prove the covariance under others $\mathbf{U}_i$.
Now, we find some unitary matrices that will swap some of the diagonal entries of the density matrix with each other. These unitary matrices have the following form:
\begin{equation}\label{swap}
\begin{aligned}
\mathbf{V}_1=&\ket{00}\bra{00}+\ket{01}\bra{02}+\ket{10}\bra{12}+\ket{02}\bra{01}+\ket{20}\bra{21}\\
&+\ket{11}\bra{11}+\ket{12}\bra{10}+\ket{21}\bra{20}+\ket{22}\bra{22}\\
\mathbf{V}_2=&\ket{00}\bra{00}+\ket{01}\bra{21}+\ket{10}\bra{02}+\ket{02}\bra{10}+\ket{20}\bra{12}\\
&+\ket{11}\bra{11}+\ket{12}\bra{20}+\ket{21}\bra{01}+\ket{22}\bra{22}\\
\mathbf{V}_3=&\ket{00}\bra{00}+\ket{01}\bra{10}+\ket{10}\bra{01}+\ket{02}\bra{20}+\ket{20}\bra{02}\\
&+\ket{11}\bra{11}+\ket{12}\bra{21}+\ket{21}\bra{12}+\ket{22}\bra{22}\\
\mathbf{V}_4=&\ket{00}\bra{00}+\ket{01}\bra{12}+\ket{10}\bra{20}+\ket{02}\bra{21}+\ket{20}\bra{10}\\
&+\ket{11}\bra{11}+\ket{12}\bra{01}+\ket{21}\bra{02}+\ket{22}\bra{22}\\
\mathbf{V}_5=&\ket{00}\bra{00}+\ket{01}\bra{20}+\ket{10}\bra{21}+\ket{02}\bra{12}+\ket{20}\bra{01}\\
&+\ket{11}\bra{11}+\ket{12}\bra{02}+\ket{21}\bra{10}+\ket{22}\bra{22}.
\end{aligned}
\end{equation}
The action of the unitaries defined above is to swap the position of the diagonal without affecting the states $\ket{00}$, $\ket{11}$ and $\ket{22}$. It is straightforward to confirm that $\mathbf{V}_i$ commutes with the Kraus operators. Therefore, $\Phi$ is a covariant channel with respect to the unitaries $\mathbf{V}_i$:
$$
\Phi(\mathbf{V}_i\rho\mathbf{V}_i)=\mathbf{V}_i\Phi(\rho)\mathbf{V}_i.
$$
We are operating these swap unitaries to make the optimization procedure easier, which will be clear in the next section.
\section{Classical capacity} \label{Sec_CC}
{The classical capacity $\mathcal{C}$ is determined by the maximum amount of classical information that can be transmitted reliably through the quantum channel per single use of the channel.}
{The calculation of classical capacity involves optimization of the Holevo quantity over multiple uses of the channel:} 
\begin{equation}\label{Eq_CC}
\mathcal{C}(\Phi)=\lim _{n \rightarrow \infty} \frac{1}{n} \bar{\chi}\left(\Phi^{\otimes n}\right),
\end{equation}
where $\bar{\chi}(\Phi)=\max_{\xi_j,\rho_j} \chi\left\{\Phi,\left(\xi_j,\rho_j\right)\right\}$ and $\chi(\Phi, \left\{\xi_j, \rho_j\right\})$ is the Holevo quantity over single uses of the channel. Generally, the Holevo quantity obeys the super-additivity {property} \cite{hastings2009superadditivity}. {The regularization process in Eq.} (\ref{Eq_CC}) {is an essential step to find out the classical capacity.}

In this section, we mainly focus on single-shot classical capacity $\mathcal{C}_1$ of the fully correlated three-level ADC. The single-shot classical capacity $\mathcal{C}_1$ is determined by optimizing the Holevo quantity $\chi$ over single uses of the channel $\Phi$ and over possible ensembles $\{\xi_j,\rho_j\}$, which is
\begin{equation}
\begin{aligned}
\mathcal{C}_1=&\max _{\xi_j, \rho_j \in \mathcal{H}} \chi (\Phi,\{\xi_j,\rho_j\})\\
=&\max _{\xi_j, \rho_j \in \mathcal{H}}\{S\left(\Phi(\rho)\right)-\sum_j \xi_j S\left(\Phi\left(\rho_j\right)\right)\}
\end{aligned}
\end{equation}
with $\{\xi_j\}$ probability distribution and the average transmitted message $\rho=\sum_j \xi_j\rho_j$. $S(\rho)=-\tr[\rho\log_2\rho]$ is the von Neumann entropy of the state $\rho$. The first term of $\chi$ corresponds to the entropy of the channel output for the input quantum state $\rho$, while the second term indicates the average entropy of the channel output. One can find an ensemble of pure input states for any ensemble of mixed input states, such that the resulting output states have a value of $\chi$ that is equal to or greater than the original ensemble \cite{schumacher1997sending}. Hence, we define an ensemble of pure state $\{\xi_j,\ket{\psi_j}\}$ with the single-shot classical capacity,
\begin{equation}
\begin{aligned}
\mathcal{C}_1=\max _{\xi_j, \rho_j \in \mathcal{H}}\{S\left(\Phi(\rho)\right)-\sum_j \xi_j S\left(\Phi\left(\ket{\psi_j}\bra{\psi_j}\right)\right)\},
\end{aligned}
\label{eq_max_CC}
\end{equation}
where $\rho=\sum_j\xi_j\ket{\psi_j}\bra{\psi_j}$.
Now, our primary aim is to search for the ensemble which will maximize $\chi$. In the following section, we give a comprehensive description of the method we used.

The channel covariance properties discussed in the earlier section are employed in this section to find the form of the ensembles $\{\xi_j,\ket{\psi_j} \}$ that solve the maximization problems (\ref{eq_max_CC}).
In order to make the optimization simpler, we first find another ensemble $\{\xi'_j,\ket{\psi'_j} \}$ by replacing each state $\ket{\psi_j}$ of the ensemble $\{\xi_j,\ket{\psi_j} \}$ with the set $\{\mathbf{U}_0\ket{\psi_j},\dots,\mathbf{U}_{15}\ket{\psi_j}\}$, where each state occurs with probability $\xi'_j=\xi_j/16$ \cite{dorlas2008calculating}. The corresponding density operator of the new ensemble, $\rho'=\sum_j\xi'_j\ket{\psi'_j}\bra{\psi'_j}$ has the following form:
\begin{equation}
\begin{aligned}
\rho' & =\sum_j \frac{\xi_j}{16}\left(\ket{\psi_j}\bra{\psi_j}+\sum_{i=1}^{15} \mathbf{U}_i\ket{ \psi_j}\bra{\psi_j}\mathbf{U}_i\right) \\
& =\frac{1}{16}\left(\rho+\sum_{i=1}^{15} \mathbf{U}_i \rho \mathbf{U}_i\right).
\end{aligned}
\end{equation}
The density matrix $\rho'$ has identical diagonal elements as $\rho$, while its off-diagonal elements completely vanish.
Next, our objective is to show that
\begin{equation}
\chi(\Phi,\{\xi'_j,\ket{\psi'}_j\}) \geqslant \chi\left(\Phi,\left\{\xi_j,\ket{\psi_j}\right\}\right).
\label{Eq_14}
\end{equation}
Given that von Neumann entropy remains unchanged under unitary operations \cite{nielsen2002quantum}, one may write \{$S\left(\Phi\left(\ket{\psi'_j}\bra{\psi'_j}\right)\right)=S\left(\Phi\left(\ket{\psi_j}\bra{\psi_j}\right)\right)$. Hence, the second term of the Holevo quantity becomes
\begin{equation}
\begin{aligned}
\sum_j \xi'_j S\left(\Phi\left(\ket{\psi'_j}\bra{\psi'_j}\right)\right) & =16 \sum_j \frac{\xi_j}{16} S\left(\Phi\left(\left|\psi_j\right\rangle\left\langle\psi_j\right|\right)\right) \\
& =\sum_j \xi_j S\left(\Phi\left(\left|\psi_j\right\rangle\left\langle\psi_j\right|\right)\right).
\end{aligned}
\label{Eq_15}
\end{equation}
    
Now, we use the fact that von Neumann entropy is a concave function and find the output entropy corresponding to the $\rho'$ as 
\begin{equation}
\begin{aligned}
S\left(\Phi({\rho'})\right) & =S\left(\Phi\left(\frac{1}{16} \rho+\frac{1}{16} \sum_{i=1}^{15} \mathbf{U}_i \rho \mathbf{U}_i\right)\right) \\
& \geqslant \frac{1}{16} S\left(\Phi(\rho)\right)+\frac{1}{16} \sum_{i=1}^{15} S\left(\Phi\left(\mathbf{U}_i \rho \mathbf{U}_i\right)\right) \\
& =S\left(\Phi(\rho)\right).
\end{aligned}
\label{Eq_16}
\end{equation}
Hence, the validity of the Eq. (\ref{Eq_14}) is proved from Eqs. (\ref{Eq_15}) and (\ref{Eq_16}). We can conclude from the above proof that one can construct an ensemble with the same diagonal elements as any given ensemble of pure states, such that the off-diagonal entries of the density matrix vanish and the Holevo quantity of this new ensemble is at least as large as that of the original ensemble. 

We introduce a generic input state $\{\xi_j, \ket{\psi_j} \}$:
\begin{equation}
\begin{aligned}
\ket{\psi}_j&=a_j\ket{00}+b_j\ket{01}+c_j\ket{02}+d_j\ket{10}+e_j\ket{11}\\
&+f_j\ket{12}+g_j\ket{20}+h_j\ket{21}+k_j\ket{22},
\end{aligned}
\label{Eq_generic}
\end{equation}
where $a_j,\; b_j,\; c_j,\; d_j,\; e_j,\; f_j,\; g_j,\; h_j\; \text{and}\; k_j \in \mathbb{C}$ and satisfy the normalization condition. We can write the corresponding density matrix $\rho=\sum_j \xi_j\ket{\psi_j}\bra{\psi_j}$.

Eventually, the corresponding density matrix $\rho'$, which yields the upper bound of Holevo quantity, becomes the diagonal matrix:
\begin{equation}
    \rho'=\begin{pmatrix}
        \alpha &&&&&&&&\\
        &\beta_1&&&&&&&\\
        &&\beta_2&&&&&&\\
        &&&\beta_3&&&&&\\
        &&&&\gamma&&&&\\
         &&&&&\beta_4&&&\\
          &&&&&&\beta_5&&\\
           &&&&&&&\beta_6&\\
            &&&&&&&&\delta\\
    \end{pmatrix},
\label{Eq_diag}
\end{equation}
where we have assumed
$$
\begin{aligned}
\alpha=\sum_j \xi_j|a_j|^2,\;\beta_1=\sum_j\xi_j|b_j|^2,\;\beta_2=\sum_j \xi_j|c_j|^2,\\
\beta_3=\sum_j \xi_j|d_j|^2,\;\gamma=\sum_j \xi_j|e_j|^2,\;\beta_4=\sum_j \xi_j|f_j|^2,\\
\beta_5=\sum_j \xi_j|g_j|^2,\;\beta_6=\sum_j\xi_j|h_j|^2,\;\delta=\sum_j \xi_j|k_j|^2\;.
\end{aligned}
$$
Now, we utilize the covariance property of the channel with respect to the unitary swap operations defined in Eq. (\ref{swap}).
We start with the ensemble $\{\xi'_j,\ket{\psi'_j}\}$ defined in Eq. (\ref{Eq_diag}), and create another ensemble by replacing each state $\ket{\psi'_j}$ with the set of states $\{\ket{\psi'_j},\mathbf{V}_i\ket{\psi'_j} \} $; each one occurs with probability $\xi'_j / 6$. The new ensemble is denoted by $\{\bar{\xi}_j,\ket{\bar{\psi_j}}\}$ and the new density operator is
\begin{equation}
\begin{aligned}
\bar{\rho} & =\sum_j \frac{\xi'_j}{6}\left(\ket{\psi'_j}\bra{\psi'_j}+\sum_{i=1}^{5} \mathbf{V}_i\ket{ \psi'_j}\bra{\psi'_j}\mathbf{V}_i\right) \\
& =\frac{1}{6}\left(\rho'+\sum_{i=1}^{5} \mathbf{V}_i \rho' \mathbf{V}_i\right).
\end{aligned}
\end{equation}
We will prove that $\{\bar{\xi}_j,\ket{\bar{\psi}_j}\}$ has the Holevo quantity $\chi$, which is greater or equal to that of the ensemble $\{\xi'_j,\ket{\psi'_j}\}$.
The first term of the Holevo quantity $\chi$ takes the following form:
\begin{equation}
\begin{aligned}
S\left(\Phi({\bar{\rho}})\right) & =S\left(\Phi\left(\frac{1}{6} \rho'+\frac{1}{6} \sum_{i=1}^{5} \mathbf{V}_i \rho' \mathbf{V}_i\right)\right) \\
& \geqslant \frac{1}{6} S\left(\Phi(\rho')\right)+\frac{1}{6} \sum_{i=1}^{5} S\left(\Phi\left(\mathbf{V}_i \rho' \mathbf{V}_i\right)\right) \\
& =S\left(\Phi(\rho')\right).
\end{aligned}
\label{Eq_19}
\end{equation}
Now, we prove that the second term of $\chi$ remains invariant under swap unitaries:
\begin{equation}
\begin{aligned}
\sum_j \bar{\xi}_j S\left(\Phi\left(\ket{\bar{\psi}_j}\bra{\bar{\psi}_j}\right)\right) & =6 \sum_j \frac{\xi'_j}{6} S\left(\Phi\left(\left|\psi'_j\right\rangle\left\langle\psi'_j\right|\right)\right) \\
& =\sum_j \xi'_j S\left(\Phi\left(\left|\psi'_j\right\rangle\left\langle\psi'_j\right|\right)\right).
\end{aligned}
\label{Eq_20}
\end{equation}
The above two equations (\ref{Eq_19}) and (\ref{Eq_20}) prove that the Holevo quantity of the ensemble $\{\bar{\xi}_j,\ket{\bar{\psi}_j}\}$ yields the upper bound of that of the ensemble $\{\xi'_j,\ket{\psi'_j}\}$. Hence, one can infer that the Holevo quantity of the ensemble $\{\bar{\xi}_j,\ket{\bar{\psi}_j}\}$ is at least as large as that of the original ensemble $\{\xi_j,\ket{\psi_j}\}$.
 
 The subsequent section examines the single-shot classical capacity of two specific channel types: the single decay channel and the V-type decay channel \cite{macchiavello2020bounding}. These channels are obtained by imposing constraints on the decay parameters, and we have computed their classical capacity using some algebraic inequality and the convex property of binary Shannon entropy. Due to the complex structure of eigenvalues of the output state in the case of $\Lambda$-type decay channel and three decay rate channels, we have not been able to get the analytic expression of the single-shot classical capacity.  
\subsubsection{V-type decay channel}
\label{subsection_v}
The lowermost energy level in this damping channel only interacts with the two higher energy levels, and the transition from $\ket{2}\rightarrow\ket{1} $ is not permitted.
The Kraus operators that represent the V-type decay channel corresponding to two qutrit systems can be derived by setting $p_3$ to zero in Equation (\ref{kraus}). The resulting Kraus operators are provided below:
\begin{align}
E_{00}=&\ket{00}\bra{00}+\sqrt{1-p_1}\ket{11}\bra{11}+\sqrt{1-p_2}\ket{22}\bra{22},\nonumber\\
&+\sum_{\substack{i,j=0\\i\neq j}}^{2}\ket{ij}\bra{ij}\nonumber\\
E_{11}=&\sqrt{p_1}\ket{00}\bra{11},\, E_{22}=\sqrt{p_2}\ket{00}\bra{22}.\nonumber\\
\end{align} 
These are the same Kraus operators used for V-type transition in a three-level system in Ref. \cite{xu2022enhancing}.
If the quantum channel $\Phi_{(p_1,p_2,0)}$ acts on the generic state given by Eq. (\ref{Eq_generic}), the output density matrix becomes
\begin{equation}
\begin{aligned}
\rho''_j&=E_{00}(\ket{\psi_j}\bra{\psi_j}) E_{00}^\dagger+E_{11}(\ket{\psi_j}\bra{\psi_j}) E_{11}^\dagger\\
&+E_{22}(\ket{\psi_j}\bra{\psi_j}) E_{22}^\dagger.
\end{aligned}
\label{Eq_kraus_v}
\end{equation}
The matrix representation of the density matrix $\rho''_j$ is shown in Eq. (\ref{AP_B19}) of Appendix \ref{AP_deg} by putting $\rho=\ket{\psi_j}\bra{\psi_j}$ in the above equation. The density matrix has seven dimensional noiseless subspace span $\{\ket{00},\ket{01},\ket{10},\ket{02},\ket{20},\ket{12},\ket{21} \}$.
The two nonzero eigenvalues are 
$$
\eta^{\pm}_j=\frac{1}{2}[1\pm \sqrt{1-l_j^2} ],
$$
where $l_j^2=4(1-|a|_j^2-p_1|e|_j^2-p_2|k|_j^2)(p_1|e|_j^2+p_2|k|_j^2)$
From the expression of eigenvalues, we can see that $\eta_j$ solely depends on the absolute value of the coefficients and is independent of the phase. Therefore, we can assume the state parameters are real.

The average entropy corresponding to the output state is found as
\begin{equation}
    \sum_j \xi_j S(\Phi(\ket{\psi_j}\bra{\psi_j}))=\sum_j \xi_j H_2(\eta_j),
\end{equation}
where $H_2(x)=-x\log_2(x)-(1-x)\log_2(1-x)$ is commonly known as Shannon's binary entropy.

Finally, one can modify the ensemble $\{\Bar{\xi}_j,\ket{\Bar{\psi}_j}\}$ and obtain another ensemble $\{\Tilde{\xi}_j,\ket{\Tilde{\psi}_j}\}$ by replacing coefficients $b_j$, $c_j$, $d_j$, $f_j$, $g_j$ and $h_j$ of each $\ket{\Bar{\psi_j}}$ by $|\Tilde{b}_j|^2=|\Tilde{c}_j|^2=|\Tilde{d}_j|^2=|\Tilde{f}_j|^2= |\Tilde{g}_j|^2 =|\Tilde{h}_j|^2=(|b_j|^2+|c_j|^2+|d_j|^2+|f_j|^2+|g_j|^2+|h_j|^2)/6$. It is straightforward to prove that this ensemble $\{\Tilde{\xi}_j,\ket{\Tilde{\psi}_j}\}$ will give the same density matrix $\bar{\rho}$. It can also be checked that the Holevo quantity will be unchanged, which is clear from the expression of eigenvalues $\eta^{\pm}_j$.

The series of relations established so far demonstrate that we can find an ensemble $\{\Tilde{\xi}_j,\ket{\Tilde{\psi}_j}\}$, which enables us to determine the upper bound of the Holevo quantity of any arbitrary ensemble $\{\xi_j,\ket{\psi_j}\}$. This is because $\{\Tilde{\xi}_j,\ket{\Tilde{\psi}_j}\}$ is a subset of the original ensemble $\{\xi_j,\ket{\psi_j}\}$. Consequently, maximizing the Holevo quantity for $\{\Tilde{\xi}_j,\ket{\Tilde{\psi}_j}\}$ will also result in the maximum for the entire set $\{\xi_j,\ket{\psi_j}\}$.

\begin{figure}[ht]
\subfloat[]{\includegraphics[width=0.47\textwidth]{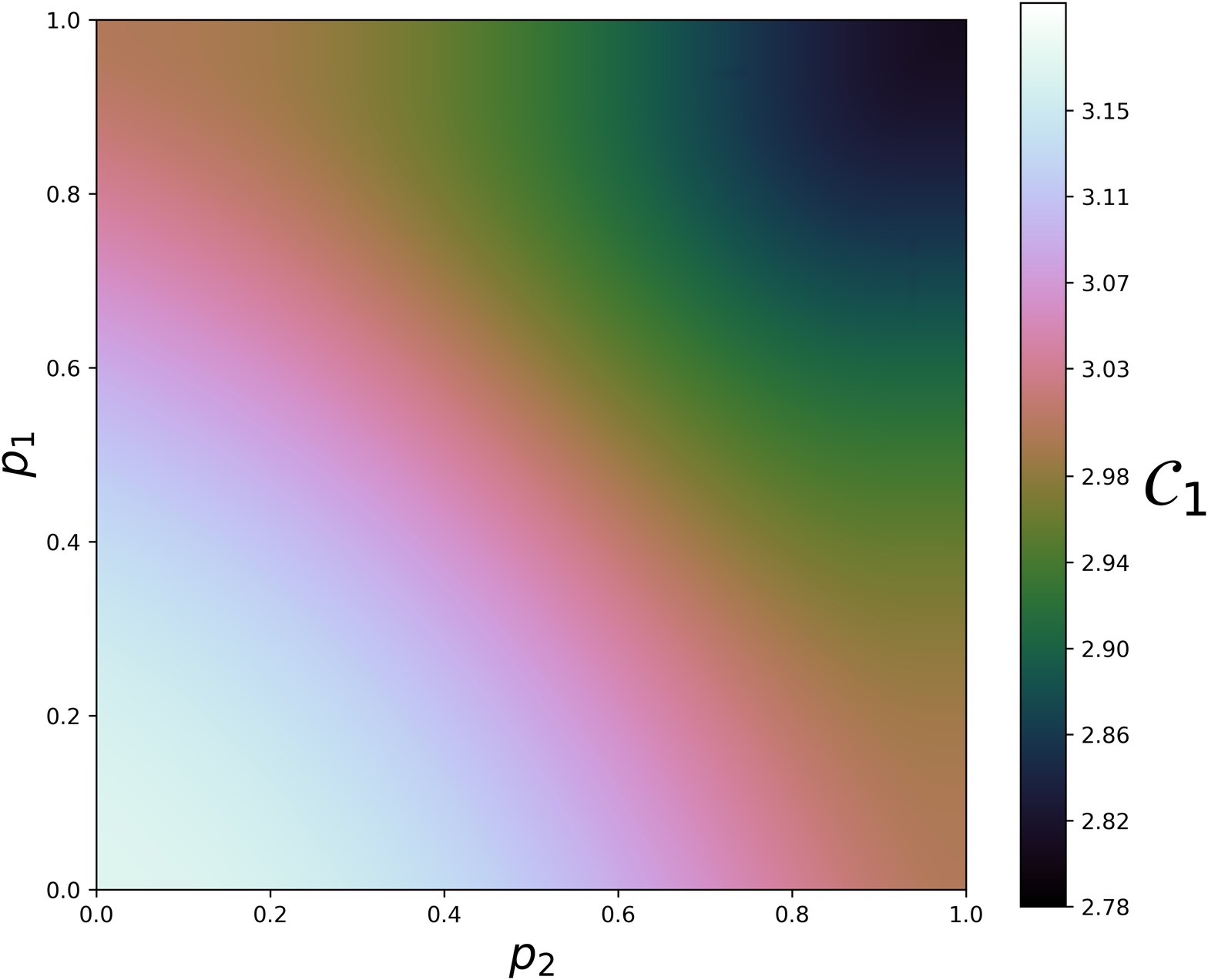}\label{fig:CC_V}}
\newline
\subfloat[]{\includegraphics[width=0.47\textwidth]{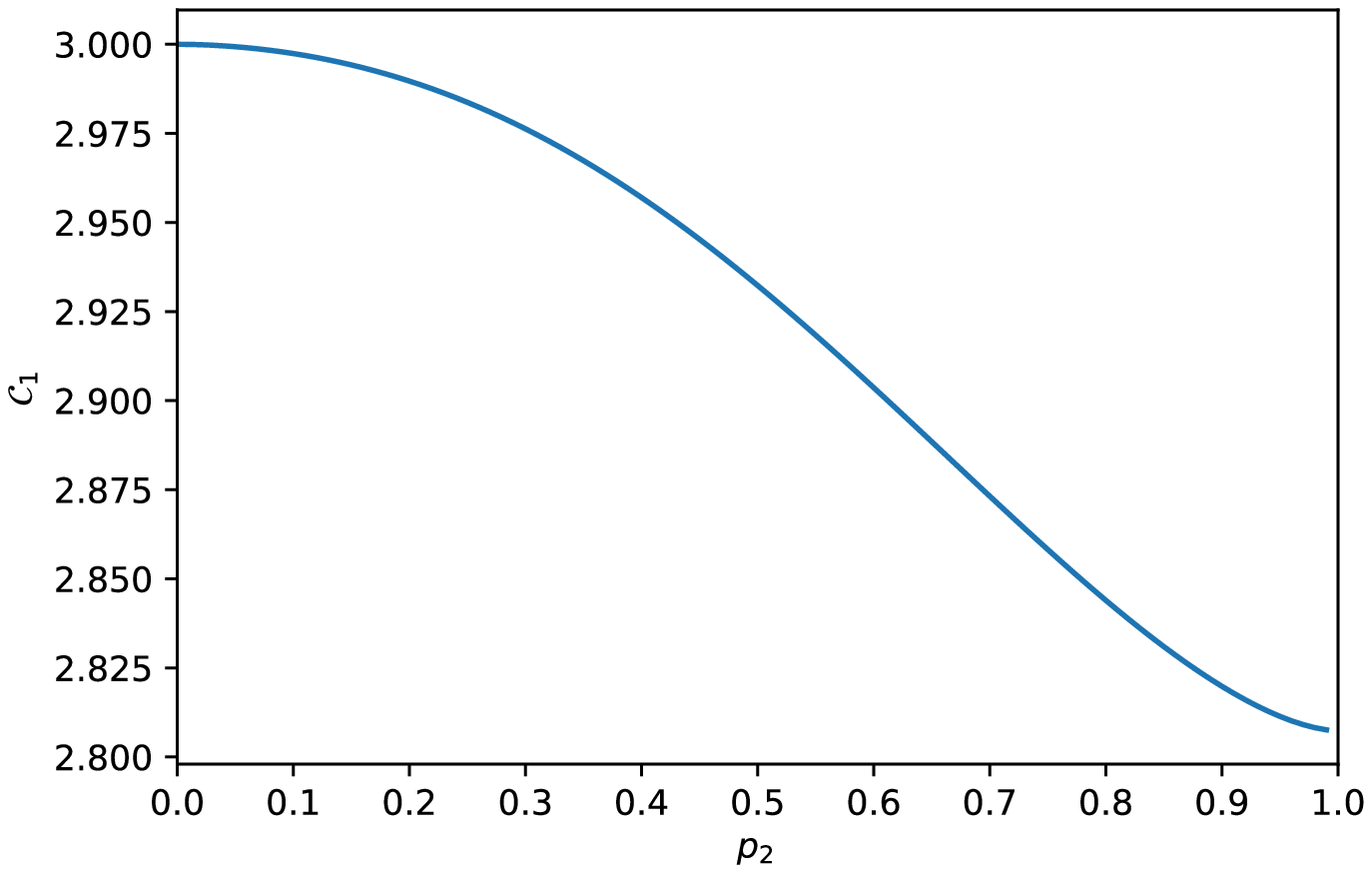}\label{fig:ccb1}}
\caption{(Color online) (a) {The upper bound of single-shot classical capacity} ($\mathcal{C}_1$) of channel $\Phi_{(p_1, p_2, 0)}$ varies according to the damping parameter $p_1$ and $p_2$. The value of $\mathcal{C}_1$ is obtained through numerical optimization. (b) The dynamics of $\mathcal{C}_1$ with respect to $p_2$ when the first excited state is completely damped {\it i.e.,} $p_1=1$.}
    \label{fig:CC}
\end{figure}

In summary, we have to investigate the classical capacity of the ensemble $\{\Tilde{\xi}_j,\ket{\Tilde{\psi}_j}\}$, where the states have the following form:
\begin{equation}
\begin{aligned}
\ket{\Tilde{\psi}}_j&=a_j\ket{00}+b_j\ket{01}\pm b_j\ket{02}\pm b_j\ket{10}+e_j\ket{11}\\
&\pm b_j\ket{12}\pm b_j\ket{20}\pm b_j\ket{21}+k_j\ket{22}.
\end{aligned}
\label{Eq_state}
\end{equation}
The corresponding density matrix $\Bar{\rho}$, which gives the maximum Holevo bound, is a diagonal matrix with the elements \{$\alpha,\beta,\beta,\beta,\gamma,\beta,\beta,\beta,\delta$\}, where
\begin{equation}
\begin{aligned} 
&\alpha=\sum_j \xi_j |a_j|^2,\; \beta=\sum_j \xi_j |b_j|^2,\;\gamma=\sum_j \xi_j |c_j|^2,\\
&\delta=\sum_j \xi_j |d_j|^2
\end{aligned}
\label{Eq_beta}
\end{equation}
and they satisfy normalization relation $|a_j|^2+6|b_j|^2+|c_j|^2+|d_j|^2=1$.
{The entropy of the output state for the channel $\Phi_{(p_1,p_2,0)}$ with the input state $\bar{\rho}$ is}
\begin{equation}
\begin{aligned}
S(\Phi(\Bar{\rho}))=&-(\alpha+p_1\gamma+p_2\delta)\log_2{(\alpha+p_1\gamma+p_2\delta)}\\
&-6\beta\log_2{\beta}-\gamma(1-p_1)\log_2{((1-p_1)\gamma)}\\
&-\delta(1-p_2)\log_2{((1-p_2)\delta)}.
\end{aligned}
\end{equation}
\begin{widetext}
Now, we utilize the following inequality to calculate the lower bound of the second term of the Holevo quantity:
\begin{equation}
\begin{aligned}
\hspace{2cm}&\sum_j \xi_j H_2 \left\{\frac{1+\sqrt{1-[(6|b_j|^2+|e_j|^2+|k_j|^2)^2-(6|b_j|^2+(1-2p_1)|e_j|^2+(1-2p_2)|k_j|^2)^2]}}{2}\right \}\\
&\geq \sum_j \xi_j H_2 \left\{\frac{1+\sqrt{1-[2p_1|e_j|^2+2p_2|k_j|^2]^2}}{2}\right \}\geq H_2 \left\{\frac{1+\sqrt{1-[2p_1\sum_j \xi_j|e_j|^2+2p_2\sum_j \xi_j|k_j|^2]^2}}{2}\right \}\\
&= H_2 \left\{\frac{1+\sqrt{1-[2p_1\gamma+2p_2\delta]^2}}{2}\right \}.\\
\end{aligned}
\end{equation}
The first inequality we obtain using the relation $X^2-Y^2\geq(X-Y)^2$, when $X \geq Y$ and the second inequality is obtained using the convexity property of the binary entropy function $H_2(\frac{1+\sqrt{1-x^2}}{2})$.
The complete expression of Holevo quantity for the V-type decay channel is obtained by maximizing it over all possible values of $\alpha,\; \beta,\; \gamma\; \text{and}\; \delta$, which is
\begin{equation}
\begin{aligned}
    \chi(\Phi, \{\Tilde{\xi}_j,\ket{\Tilde{\psi}_j}\})&= \max_{\alpha,\beta,\gamma,\delta}\Big((\alpha+p_1\gamma+p_2\delta)\log_2{(\alpha+p_1\gamma+p_2\delta)}-6\beta\log_2{\beta}\\
    &-\gamma(1-p_1)\log_2{((1-p_1)\gamma)}-\delta(1-p_2)\log_2{((1-p_2)\delta)}\\
&+H_2 \left\{\frac{1+\sqrt{1-[2p_1\gamma+2p_2\delta]^2}}{2}\right \}\Big).
\end{aligned}
\label{CC_v}
\end{equation}
\end{widetext}
Therefore, we can conclude that Eq. (\ref{CC_v}) gives the {upper bound of the} single-shot classical capacity $\mathcal{C}_1$ of the V-type decay channel. In Fig. (\ref{fig:CC_V}), we have shown the {upper bound of} $\mathcal{C}_1$ with respect to the decay rates $p_1$ and $p_2$. In the case of complete damping ($p_1=1$) of the energy level $\ket{11}$, the output density matrix becomes eight-dimensional, indicating the maximum value of the capacity $\log_2 8$ at $p_2=0$. The Fig. (\ref{fig:ccb1}) illustrates the decay of the upper bound of $\mathcal{C}_1$ as a function of $p_2$, under the condition that the energy level $\ket{11}$ is completely damped.
\subsubsection{Single decay channel}
\begin{figure}[ht]
\subfloat[]{\includegraphics[width=0.48\textwidth]{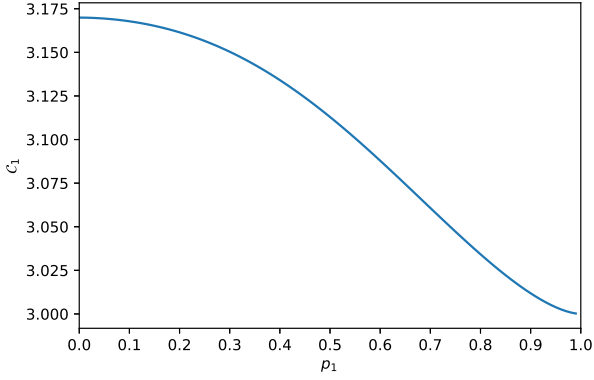}\label{fig:cca}}
\newline
\subfloat[]{\includegraphics[width=0.48\textwidth]{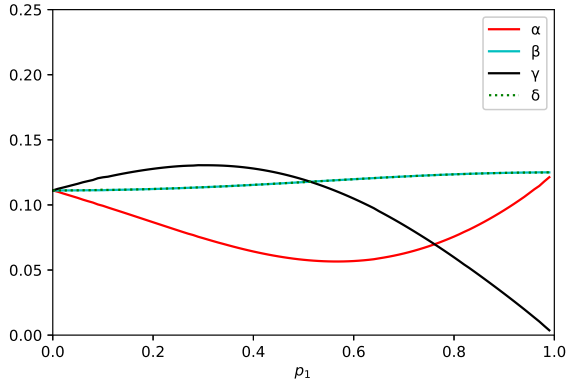}\label{fig:ccb}}
\caption{(Color online) (a) The {upper bound of single-shot classical capacity} ($\mathcal{C}_1$) of channel $\Phi_{(p_1, 0, 0)}$ varies according to the damping parameter $p_1$. The value of $\mathcal{C}_1$ is obtained through numerical optimization. (b) The populations $\alpha$, $\beta$, $\gamma$ and $\delta$ with respect to the damping parameter $p_1$ during the optimization.}
\label{Fig_cc}
\end{figure}

The upper bound of the capacity $\mathcal{C}_1$ for single decay channel \textit{i.e.}, only one of the three damping parameters $p_i$ is non-zero, can be calculated from Eq. (\ref{CC_v}) by setting one of the damping parameters $p_1$ or $p_2$ equals to zero. Since the Kraus operators for the mappings $\Phi_{(p_1,0,0)}$, $\Phi_{(0,p_2,0)}$ and $\Phi_{(0,0,p_3)}$ has the same form, the corresponding Holevo quantity will also be same. Hence, we can write the expression of Holevo quantity for the mapping $\Phi_{(p_1,0,0)}$ as

\begin{equation}
\begin{aligned}
&\chi(\Phi, \{\Tilde{\xi}_j,\ket{\tilde{\psi}_j}\})=\\
&\max_{\alpha,\beta,\gamma,\delta}\Big((\alpha+p_1\gamma)\log_2{(\alpha+p_1\gamma)}-6\beta\log_2{\beta}\\
    &-\gamma(1-p_1)\log_2{((1-p_1)\gamma)}-\delta\log_2{\delta}\\
&+H_2 \left\{\frac{1+\sqrt{1-[2p_1\gamma]^2}}{2}\right \}\Big).
\end{aligned}
\label{CC_1}
\end{equation}
The above equation is a upper bound of the Holevo quantity, which is the {upper bound of the} single-shot classical capacity of the map $\Phi_{(p_1,0,0)}$.
After performing the optimization over all possible values of $\alpha, \beta, \gamma\; \text{and}\; \delta$, one can obtain the variation of $\mathcal{C}_1$ with respect to the damping parameter $p_1$, which is depicted in Fig. (\ref{fig:cca}). Fig. (\ref{fig:ccb}) displays the state parameters  $\alpha, \beta, \gamma\; \text{and}\; \delta$ against $p_1$ during the optimization process. 

\section{Quantum capacity}\label{Sec_QC}
The quantum capacity, $\mathcal{Q}$, represents the fundamental measure of a channel's capability to transmit and convey quantum information reliably. {The asymptotic expression formally defining the quantum capacity of the channel $\Phi$ is }\cite{lloyd1997capacity,barnum1998information} 
\begin{equation}\label{QC1}
 \mathcal{Q}=\lim _{n \rightarrow \infty} \frac{\mathcal{Q}_n}{n}, \quad Q_n=\max _{\rho^{(n)}} I_c\left(\Phi^{\otimes n}, \rho^{(n)}\right),   
\end{equation}
{where the input state for n instances of channel usage is represented by $\rho^{(n)}$} and the coherent information is
\begin{equation}
I_c\left(\Phi^{\otimes n}, \rho^{(n)}\right)=S\left(\Phi^{\otimes n}\left(\rho^{(n)}\right)\right)-S\left(\tilde{\Phi}^{\otimes n}\left( \rho^{(n)}\right)\right),
\end{equation}
with $S(\rho)=-\tr[\rho\log_2\rho]$ {is the well-known expression of von Neumann entropy corresponding to the state $\rho$ and $\tilde{\Phi}$ is the complementary map of $\Phi$.} For a degradable channel,{ the coherent information shows additivity property}, and the quantum capacity $\mathcal{Q}$ reduces to its single-shot capacity $\mathcal{Q}_1$. {However, an anti-degradable channel is a zero-capacity channel because of the no-cloning principle}. It is to be noted that the optimization process outlined in Eq. (\ref{QC1}) must be conducted over the set of density matrices $\rho^{(n)}$ corresponding to n uses of the channel.

Let us consider the generic input state $\{\xi_j, \ket{\psi_j} \}$ and the corresponding density matrix $\rho=\sum_j \xi_j\ket{\psi_j}\bra{\psi_j}$ as defined earlier in Eq. (\ref{Eq_generic}).
We aim to identify the category of input states that enables us to find the quantum capacity, specifically by maximizing the coherent information. To accomplish this goal, we observe that it is possible to construct a diagonal density operator for any two-qutrit density operator $\rho$, as shown below:
\begin{equation}
\rho'=\frac{1}{16}\Big(\rho+\sum_{i=1}^{15}\mathbf{U}_i\rho\mathbf{U}_i\Big).
\end{equation}
Eventually, the density matrix $\rho'$ becomes the diagonal matrix with elements \{$\alpha,\beta_1,\beta_2,\beta_3,\gamma,\beta_4,\beta_5,\beta_6,\delta$\}.
Now, we will prove that the coherent information of $\rho'$ is greater than or equal to that associated with the state $\rho$:
\begin{equation}
\begin{aligned}
I_c(\Phi,\rho')&=I_c\Bigg(\Phi, \frac{1}{16}\Big(\rho+\sum_{i=1}^{15}\mathbf{U}_i\rho\mathbf{U}_i \Big)\Bigg)\\
&\geq \frac{1}{16}I_c(\Phi,\rho)+\frac{1}{16}\sum_{i=1}^{15}I_c\Big(\Phi, \mathbf{U}_i\rho\mathbf{U}_i\Big)\\
&=\frac{1}{16}I_c(\Phi,\rho)+\frac{1}{16}\sum_{i=1}^{15}S\Big(\Phi\Big( \mathbf{U}_i\rho\mathbf{U}_i\Big)\Big)\\
&-\frac{1}{16}\sum_{i=1}^{15}S\Big(\Tilde{\Phi}\Big( \mathbf{U}_i\rho\mathbf{U}_i\Big)\Big)=I_c(\Phi,\rho).
\end{aligned}
\end{equation}
We have utilized the property of degradable channels in the inequality above, which states that coherent information of degradable channels exhibits concave behaviour. {We use the fact that the von Neumann entropy is invariant under unitary operations and arrive at the following conclusion:} $S(\Phi( \mathbf{U}_i\rho\mathbf{U}_i))=S(\Phi(\rho))$.

Now, we can make a new state $\Bar{\rho}$:
$$
\Bar{\rho}=\frac{1}{6}\Big(\rho+\sum_{i=1}^{5}\mathbf{V}_i\rho'\mathbf{V}_i \Big),
$$
where $\mathbf{V}_i$ are defined in Eq. (\ref{swap}). Here the density matrix $\Bar{\rho}$ becomes a diagonal matrix with elements \{$\alpha,\beta,\beta,\beta,\gamma,\beta,\beta,\beta,\delta$\}, where $\beta$ is defined in Eq. (\ref{Eq_beta}).

In the same way as above, using the concavity property of degradable channel, we can show that the coherent information $I_c(\Phi,\Bar{\rho})\geq I_c(\Phi,\rho')$.
We may conclude that by optimizing the coherent information of the diagonal state $\Bar{\rho}$, we can derive the quantum capacity. Therefore, within the degradable region, the expression of quantum capacity is
\begin{equation}
  \begin{aligned}
      \mathcal{Q}\left(\Phi\right) & =\mathcal{Q}_1\left(\Phi\right) =\max _{\bar{\rho}} I_c\left(\Phi, \bar{\rho}\right) \\
& =\max _{\bar{\rho}}\left\{S\left(\Phi\left(\bar{\rho}\right)\right)-S(\Tilde{\Phi}\left(\bar{\rho}\right))\right\}.\\
  \end{aligned}  
\end{equation}
Our next task involves calculating the quantum capacity of the fully correlated MAD channel. Nevertheless, there is an obstacle that we need to overcome. First, we need to check whether the channel is degradable or non-degradable. 
We have shown in Appendix \ref{AP_deg} that our fully correlated two qutrit MAD channel $\Phi_{(p_1,p_2,p_3)}$ is non-degradable. It is not anti-degradable also.
Hence, we simplify the problem by setting one or two decay parameters in such a way that the Kraus operators required to represent the map are {less than four, and then we show the resulting maps exhibit degradability properties in some range of decay parameters.}
In the following subsection, we systematically examine the quantum capacity of a fully correlated MAD channel under various conditions, one by one.
\subsubsection{Single decay channel}
The instances of the fully correlated MAD channel that we are analyzing in this context involve situations where only one of the three damping parameters, $p_i$, has a non-zero value. The associated maps for this single decay are $\Phi_{(p_1,0,0)}$, $\Phi_{(0,p_2,0)}$ and $\Phi_{(0,0,p_3)}$ respectively.

We observed that two non-zero Kraus operators corresponding to the mapping $\Phi_{(p_1,0,0)}$ are
\begin{equation}
\begin{aligned}
E_{00}=&\ket{00}\bra{00}+\sqrt{1-p_1}\ket{11}\bra{11}+\ket{22}\bra{22}\nonumber\\
&+\sum_{\substack{i,j=0\\i\neq j}}^{2}\ket{ij}\bra{ij}\nonumber\\
E_{11}=&\sqrt{p_1}\ket{00}\bra{11}.\nonumber\\
\end{aligned} 
\end{equation}
The expression for the transformation $\Phi_{(p_1,0,0)}(\rho)$ and the corresponding complementary map $\tilde{\Phi}_{(p_1,0,0)}(\rho)$ according to the equation (\ref{Eq_commap}) are given in the Appendix \ref{AP_deg}. 
{It is noteworthy to mention that the structure of the Kraus operators for the single decay map admits the partial coherent direct sum (PCDS) structure. According to Ref.} \cite{chessa2021partially}, {a PCDS map is degradable if its diagonal blocks are also degradable. We can use that method also to find out the degradability condition for the map $\Phi_{(p_1,0,0)}$.} However, we conducted the degradability analysis using the matrix inversion method without considering the PCDS structure, as reported in Appendix \ref{AP_deg}.

From the channel degradability analysis, we have seen that the channel is degradable for $p_1\leq \frac{1}{2}$. Even though the channel is not degradable or anti-degradable for $p_1\geq \frac{1}{2}$, we can still compute its quantum capacity in this range using the monotonicity constraint of the quantum capacity function discussed in the Appendix \ref{Ap_comp}. 

Consequently, the quantum capacity in the degradable region ($0\leq p_1 \leq 1/2$) is obtained as follows:
\begin{equation}\label{Eq_48}
\begin{aligned}
\mathcal{Q}\left(\Phi\right) & =\max_{\bar{\rho}} I_c\left(\Phi, \bar{\rho}\right) \\
& =\max _{\bar{\rho}}\left\{S\left(\Phi\left(\bar{\rho}\right)\right)-S(\Tilde{\Phi}\left(\bar{\rho}\right))\right\}\\
&=\max_{\alpha,\beta,\gamma,\delta}\{-(\alpha+p_1\gamma)\log_2{(\alpha+p_1\gamma)}-6\beta\log_2{\beta}\\
&-((1-p_1)\gamma)\log_2{((1-p_1)\gamma)}-\delta\log_2{\delta}\\
&+(1-p_1\gamma)\log_2{(1-p_1\gamma)}+p_1\gamma\log_2{(p_1\gamma})\}.\\
\end{aligned}
\end{equation}
{The above equation yields the value of $\mathcal{Q}$ equal to $\log_28$ at $p_1=1/2$. This value serves as the upper bound of $\mathcal{Q}$ for the region $1/2 < p_1 \leq 1$, which is evident from the monotonic behaviour of the quantum capacity function. Again, from the Eq. (\ref{AP_B6}) in Appendix \ref{AP_deg}, one can observe that the transformation has eight-dimensional decoherence-free subspace spanning over $\ket{00}$, $\ket{01}$, $\ket{02}$, $\ket{10}$, $\ket{12}$, $\ket{20}$, $\ket{21}$ and $\ket{22}$ bases. Hence, the lower bound of the $\mathcal{Q}$ for the single decay map is $\log_28$. Since the lower bound of $\mathcal{Q}$ coincides with the upper bound, we can conclude that quantum capacity is $\log_2 8$ in the non-degradable region.}

\begin{figure}[ht]
\subfloat[]{\includegraphics[width=0.48\textwidth]{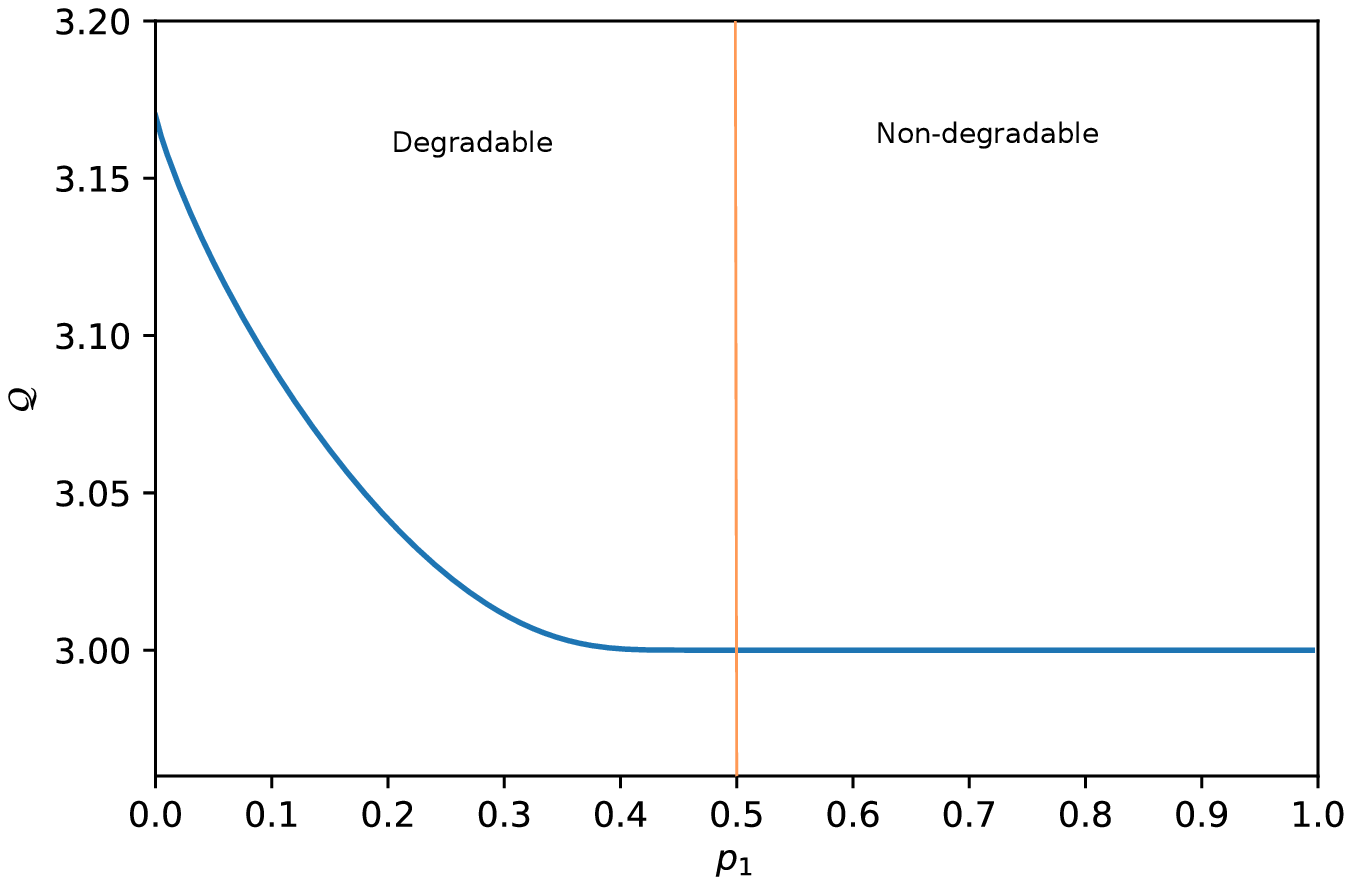}\label{fig:2a}}
\newline
\subfloat[]{\includegraphics[width=0.48\textwidth]{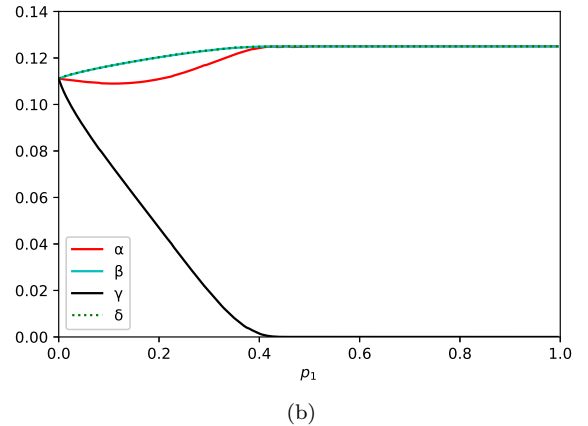}\label{fig:2b}}
\caption{(Color online) (a) The quantum capacity ($\mathcal{Q}$) of channel $\Phi_{(p_1, 0, 0)}$ varies according to the damping parameter $p_1$. 
(b) The populations $\alpha$, $\beta$, $\gamma$ and $\delta$ refer to the states which optimize the quantum capacity formula corresponding to the $\Phi_{(p_1,0,0)}$ channel with respect to the damping parameter $p_1$.}
\label{Fig_3}
\end{figure}

The results are depicted in Fig. (\ref{Fig_3}). In the case of $p_1=0$, the value of the quantum capacity becomes $\log_29$, which is obviously the maximum value of $\mathcal{Q}$ for the nine-dimensional density matrix.

In the above section, we have calculated the quantum capacity for the mapping $\Phi_{(p_1,0,0)}$. It can be readily observed that the other group of transformation $\Phi_{(0,p_2,0)}$ and $\Phi_{(0,0,p_3)}$ can be transformed into each other by simply swapping energy levels. Therefore, the quantum capacity of these three groups should be the same as each channel can be derived from the other, {\it i.e.}, $\mathcal{Q}(\Phi_{(p,0,0)})=\mathcal{Q}(\Phi_{(0,p,0)})=\mathcal{Q}(\Phi_{(0,0,p)})$ for all the values of $p$ between 0 and 1.
\subsubsection{Double decay channel}
This section focuses primarily on a model that allows for only two possible transitions. Specifically, we consider a scenario where one of the three damping parameters, $p_1$, $p_2$, or $p_3$, is equal to zero. Further, we classify the double decay channel as a V-type or $\Lambda$-type decay channel depending upon the conditions $p_3=0$ and $p_1=0$ \cite{macchiavello2020bounding}. In the following subsection, we calculate the quantum capacity of these channels since they are degradable. However, when $p_2=0$, the channel is referred to as a $\Xi$-type decay channel and can be described by four Kraus operators. Since the $\Xi$-type decay channel is not degradable, we are unable to calculate the corresponding quantum capacity of this channel.
\subsection{V-type decay channel}
If the decay parameter $p_3$ is equal to zero, the atoms undergo a V-type transition, which is commonly referred to as a V-type decay channel. For this channel, the accessible values of $p_1$ and $p_2$ lie on the surface EFH, as shown in Fig. (\ref{Region}). The Kraus operators representing the V-type decay channel are given in Sec. (\ref{subsection_v}), where we computed the classical capacity for this type of channel.
In the Apendix \ref{AP_deg} we provide the expression for both the transformation $\Phi_{(p_1,p_2,0)}(\rho)$ and its corresponding complementary map $\tilde{\Phi}_{(p_1,p_2,0)}(\rho)$ and the degradability analysis of the channel. From the degradability analysis, we have seen that the channel is degradable for $p_1\leq1/2$ and $p_2\leq1/2$, and the quantum capacity reduces to the single-shot capacity in this region. In other ranges of $p_1$ and $p_2$, the channel is non-degradable. {In order to calculate the quantum capacity $\mathcal{Q}$ in the non-degradable region, we use the results of the capacity analysis corresponding to the maps $\Phi_{(1,p_2,0)}$ and $\Phi_{(p_1,1,0)}$ along with the monotonic property of the quantum capacity function. The quantum capacity analysis of the maps $\Phi_{(1,p_2,0)}$ and the technical details of the capacity analysis in the non-degradable region of the V-type decay channel are provided in Appendix \ref{AP_deg}.}

\begin{figure}[ht]
    \centering
    \includegraphics[width=0.47\textwidth]{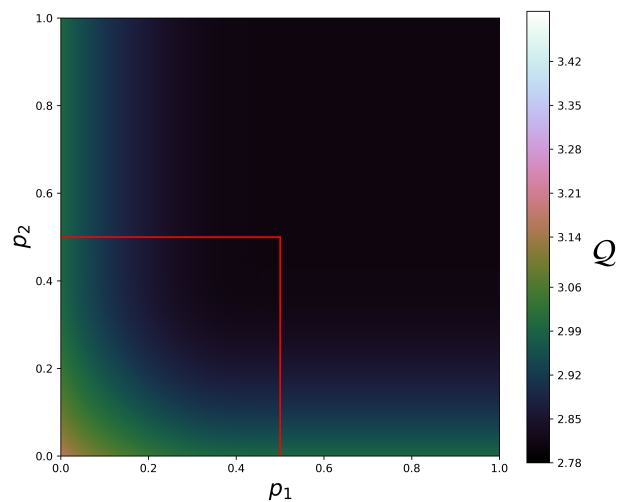}
    \caption{Contour plot of the quantum capacity ($\mathcal{Q}$) for the V-type decay channel, $\Phi_{(p_1, p_2, 0)}$ with the damping parameters $p_1$ and $p_2$. In the range $p_1$ and $p_2$ are less than or equal to 1/2 (area enclosed by the red line), the channel is degradable. If both $p_1$ and $p_2$ exceed the value 1/2, the channel is non-degradable.}
    \label{Quantum capacity of V-type channel}
\end{figure}

The following expression gives the quantum capacity in the degradable region:
\begin{equation}\label{Eq_49}
\begin{aligned}
\mathcal{Q}\left(\Phi\right) & =\max_{\alpha,\beta,\gamma,\delta}\{-(\alpha+p_1\gamma+p_2\delta)\log_2{(\alpha+p_1\gamma+p_2\delta)}\\&-6\beta\log_2{\beta}-\gamma(1-p_1)\log_2{((1-p_1)\gamma)}-\delta(1-p_2)\\
&\times\log_2{((1-p_2)\delta)}+p_1\gamma\log_2{(p_1\gamma)}+(1-p_1\gamma-p_2\delta)\\
&\times\log_2{(1-p_1\gamma p_2\delta)}+p_2\delta\log_2{(p_2\delta)} \}.\\
\end{aligned}
\end{equation}
From the above expression the values of $\mathcal{Q}(\Phi)$ can be determined on the border of the degradable region {\it i.e.}, $\Phi_{(1/2,p_2,0)}$ and $\Phi_{(p_1,1/2,0)} \forall p_1,p_2\leq1/2$ is known. Now, we calculate the quantum capacity at the edge of the parameter space, {\it i.e.}, $\mathcal{Q}(\Phi_{(1,p_2,0)})$ and $\mathcal{Q}(\Phi_{(p_1,1,0)})$. We have shown that for a specific value of $p_1$ or $p_2$, the quantum capacity has the same value at the border of the degradable region and on the edges. Thus, based on the monotonicity constraint described in Eq. (\ref{Eq_mono_v}), it can be concluded that the quantum capacity remains unchanged in the intermediate region.
The behavior of $\mathcal{Q}(\Phi_{(p_1,p_2,0)})$ with respect to $p_1$ and $p_2$ is displayed in Fig. (\ref{Quantum capacity of V-type channel}).

According to the findings presented in Ref. \cite{chessa2021quantum}, it has been demonstrated that the uncorrelated V-type decay channel exhibits anti-degradability property within the region $p_1\geq1/2$ and $p_2\geq1/2$, which implies that the quantum capacity in this particular region is effectively zero. Here, we have shown that the fully correlated V-type decay channel is not anti-degradable and has the lowest quantum capacity value $\log_27$.
\subsection{\texorpdfstring{$\Lambda$-type decay channel}{}}
\begin{figure}[ht]
    \centering
    \includegraphics[width=0.47\textwidth]{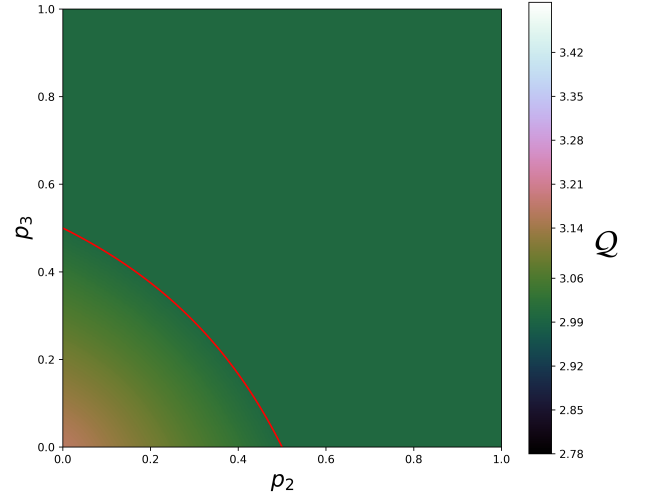}
    \caption{Contour plot of the quantum capacity ($\mathcal{Q}$) for the $\Lambda$-type decay channel, $\Phi_{(0, p_2, p_3)}$ with the damping parameters $p_2$ and $p_3$. In the range $(1-p_2)(1-p_3)\geq\frac{1}{2}$, the channel is degradable. However, in other regions, the channel is non-degradable, and the corresponding value of quantum capacity is fixed at \textit{$\log_28$}, obtained from the monotonicity principle.}
    \label{fig_4}
\end{figure}

In this type of damping channel, energy level $\ket{2}$ of the two three-level systems interacts with the lower lying levels $\ket{1}$ and $\ket{0}$ in a correlated manner. In this case, we examine the quantum capacity value for $\Phi$ that belongs to the square surface CDEH depicted in Fig. (\ref{Region}), which is defined by the condition $p_1=0$. According to Eq. (\ref{kraus}), one can write the Kraus operators that describe this channel as
$$
\begin{aligned}
E_{00}=&\ket{00}\bra{00}+\ket{11}\bra{11}+\sqrt{1-p_{23}}\ket{22}\bra{22}\nonumber\\
&+\sum_{\substack{i,j=0\\i\neq j}}^{2}\ket{ij}\bra{ij}\nonumber\\
E_{22}=&\sqrt{(1-\Theta)p_{23}}\ket{00}\bra{22}\\
E_{33}=&\sqrt{\Theta p_{23}}\ket{11}\bra{22},\nonumber
\end{aligned} 
$$
where for convenience we have taken $p_{23}=1-(1-p_2)(1-p_3)$ and $\Theta=\ln{(1-p_3)}/(\ln{(1-p_3)}+\ln{(1-p_2)})$.\\
The symmetry of the model under the exchange of $p_2$ and $p_3$ is apparent from the structure of the Kraus operators. Hence, we conclude that
\begin{equation}
    \mathcal{Q}(\Phi_{(0,p_2,p_3)})=\mathcal{Q}(\Phi_{(0,p_3,p_2)}).
\end{equation}
In Appendix \ref{AP_deg} we have shown the transformation $\Phi_{(0,p_2,p_3)}$ and complementary map $\Tilde{\Phi}_{(0,p_2,p_3)}$. We have seen that the channel is degradable for $(1-p_2)(1-p_3)\geq\frac{1}{2}$, and it is not anti-degradable in any region. The equation represents the quantum capacity of the channel in the degradable region:
\begin{equation}\label{Eq_51}
\begin{aligned}
\mathcal{Q}\left(\Phi\right) & =\max_{\alpha,\beta,\gamma,\delta}\{-(\alpha+(1-\Theta)p_{23}\delta)\log_2{(\alpha+(1-\Theta)p_{23}\delta)}\\
&-6\beta\log_2{\beta}-(\gamma+\Theta p_{23}\delta)\log_2{(\gamma+\Theta p_{23}\delta)}\\
&-((1-p_{23})\delta)\log_2{(1-p_{23})\delta)}+(\Theta p_{23}\delta)\log_2(\Theta p_{23}\delta)\\
&+((1-\Theta)p_{23}\delta)\log_2{((1-\Theta)p_{23}\delta)}+(1-p_{23}\delta)\\
&\times\log_2{(1-p_{23}\delta)} \}.\\
\end{aligned}
\end{equation}
At the border of the degradable region $(1-p_2)(1-p_3)=1/2$, (shown in the red curve in Fig. (\ref{fig_4})) the value of the quantum capacity can be determined, which is $\log_28$. {This value serves as the upper bound of $\mathcal{Q}$ in the non-degradable region.} We also observe that the output density matrix has eight-dimensional decoherence-free subspace. Hence, the lower bound of the quantum capacity is $\log_2 8$. From the composition rule and monotonicity constraint of the quantum capacity function, we conclude that $\mathcal{Q}(\Phi)=\log_28$ in the non-degradable region.
The behavior of the quantum capacity $ \mathcal{Q}(\Phi_{(0,p_2,p_3)})$ with respect to $p_2$ and $p_3$ is illustrated in Fig. (\ref{fig_4}).
 In the case of an uncorrelated $\Lambda$-type decay channel, as shown in Ref. \cite{chessa2021quantum}, the channel is anti-degradable between $p_2+p_3\geq1/2$ and $p_2+p_3\leq1$ indicates zero value of quantum capacity and beyond the range $p_2+p_3=1$ the channel is not CPTP. However, a fully correlated $\Lambda$-type decay channel is CPTP for all values of $p_2$ and $p_3$, and the lowest value of quantum capacity is $\log_28$.
\subsubsection{Three decay rate channel}
\begin{figure}[ht]
    \centering
    \includegraphics[width=0.47\textwidth]{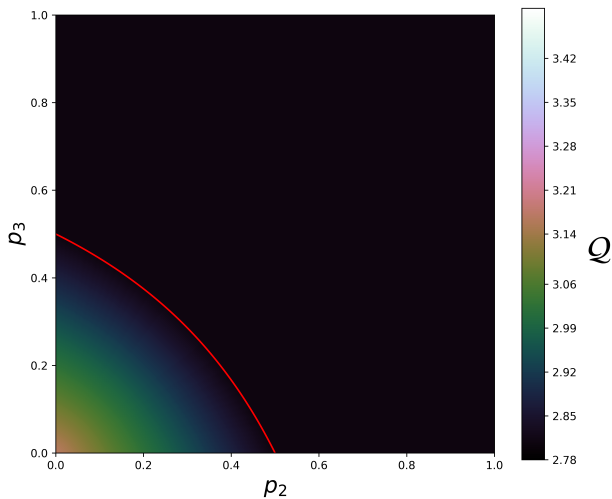}
    \caption{Contour plot of the quantum capacity ($\mathcal{Q}$) of the specific three decay rate channel, $\Phi_{(p_{23}, p_2, p_3)}$ with the damping parameters $p_2$ and $p_3$. In the range $(1-p_2)(1-p_3)\geq\frac{1}{2}$ the channel is degradable like $\Lambda$-type decay channel . However, in the other region, the channel is non-degradable, and the corresponding value of quantum capacity value is fixed at $\log_27$ obtained from the monotonicity principle.}
    \label{Fig_8}
\end{figure}

Let us consider the region $1-p_1=(1-p_2)(1-p_3)$, which is indicated by the surface BEFH  in Fig. (\ref{Region}). This special three decay rate map satisfying the above-mentioned constraint, admits the following Kraus operators:
\begin{align}
E_{00}=&\ket{00}\bra{00}+\sqrt{(1-p_{23})}\ket{11}\bra{11}\nonumber\\
&+\sqrt{(1-p_{23})}\ket{22}\bra{22}+\sum_{\substack{i,j=0\\i\neq j}}^{2}\ket{ij}\bra{ij},\nonumber\\
E_{11}=&\sqrt{p_{23}}\ket{00}\bra{11},\, E_{22}=\sqrt{p_{23}}\ket{00}\bra{22},\nonumber\\
\end{align} 
where we have denoted $p_{1}=1-(1-p_2)(1-p_3)=p_{23}$ for convenience.\\
From the structural symmetry of the Kraus operators, it is clear that under the exchange of $p_2$ and $p_3$, quantum capacity does not change. Hence, we can write
\begin{equation}
    \mathcal{Q}(\Phi_{(p_{23},p_2,p_3)})=\mathcal{Q}(\Phi_{(p_{23},p_3,p_2)}).
\end{equation}
Similar to the $\Lambda$-type decay channel this map, $\Phi_{(p_{23},p_2,p_3)}$ is also degradable in the region $(1-p_2)(1-p_3)\geq\frac{1}{2}$. 
In the degradable region, the quantum capacity for this channel is
\begin{equation}
\begin{aligned}
\mathcal{Q}\left(\Phi\right) & =\max_{\alpha,\beta,\gamma,\delta}\{-(\alpha+p_{23}\gamma+p_{23}\delta)\log_2{(\alpha+p_{23}\gamma+p_{23}\delta)}\\
&-6\beta\log_2{\beta}-\gamma(1-p_{23})\log_2{((1-p_{23})\gamma)}-(1-p_{23}\delta)\\
&\times\log_2{(1-p_{23}\delta)}+p_{23}\gamma\log_2{(p_{23}\gamma)}+p_{23}\delta\log_2{(p_{23}\delta)}\\
&+(1-p_{23}\gamma-p_{23}\delta)\log_2{(1-p_{23}\gamma- p_{23}\delta)} \}.\\
\end{aligned}
\end{equation}
In the other region, we can also calculate the quantum capacity using composition rule and monotonicity constraints like $\Lambda$-type decay channel. The output density matrix corresponding to this channel can be verified to possess a seven-dimensional decoherence-free subspace. This observation indicates that the lower bound of the quantum capacity for this channel is given by the $\log_27$. The quantum capacity for this map $\Phi_{(p_{23},p_2,p_3)}$ is displayed in Fig. (\ref{Fig_8}).
\section{Entanglement assisted capacity} \label{Sec_EAC}
{In this section, we examine} the classical and quantum capacities of the fully correlated MAD channel in the entanglement-assisted scenario. The concept of entanglement-assisted quantum capacity, $\mathcal{Q}_E$, refers to the maximum quantity of quantum information which can be transferred reliably through a given channel per each use of that channel with the assumption that the two communicating parties have access to an unlimited supply of entanglement resources beforehand. It can be expressed as \cite{bennett1999entanglement,bennett2002entanglement}
\begin{equation}
\mathcal{Q}_E=\frac{1}{2}\max _\rho I\left(\Phi, \rho\right),
\end{equation}
where the optimization process is carried out with respect to the input $\rho$ and
\begin{equation}
I\left(\Phi, \rho\right)=S(\rho)+I_c\left(\Phi, \rho\right).
\end{equation}
{The mutual information functional $I$ is equal to the coherent information functional $I_c$ with the addition of entropy $S(\rho)$ of input state.} The mutual information functional satisfies the additivity property \cite{adami1997neumann}, and because of that, no regularization is needed in the calculation of $\mathcal{Q}_E$. So we can write
\begin{equation}
I\left(\Phi, \rho\right)=S\left(\rho\right)+S\left(\Phi(\rho)\right)-S(\Tilde{\Phi}(\rho)).
\end{equation}
The covariance property of von Neumann entropy and concavity of the coherent information also apply here for mutual information.

\begin{figure}[!]
\subfloat[]{\includegraphics[width=8cm, height=6.5cm]{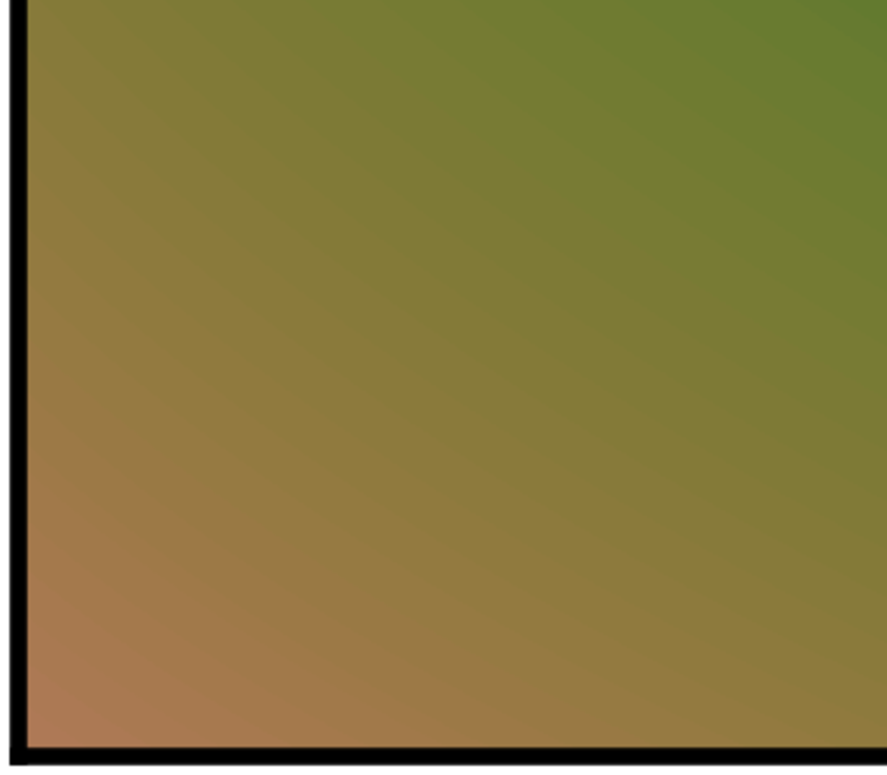}\label{fig:6a}}

\subfloat[]{\includegraphics[width=8cm, height=6.5cm]{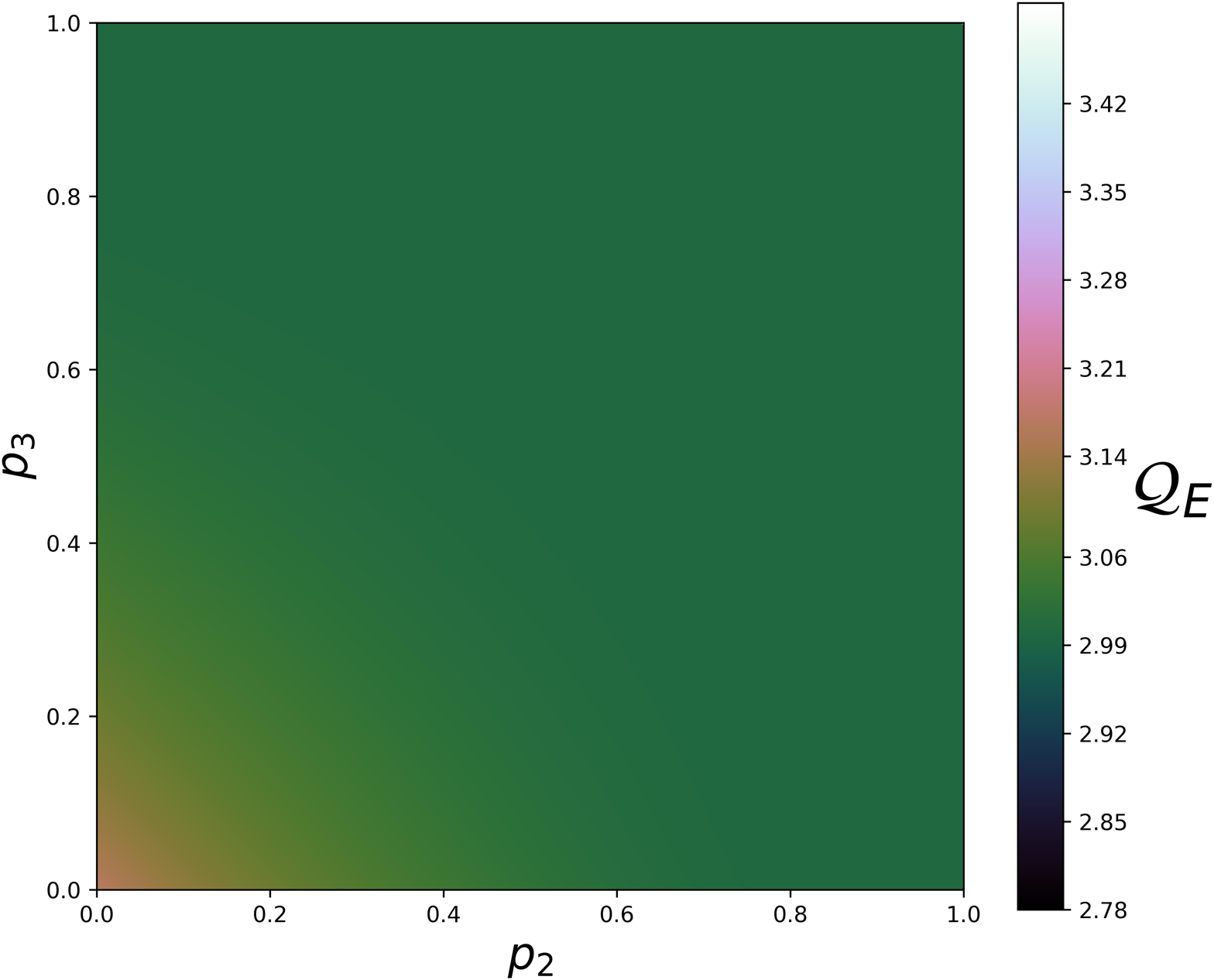}\label{fig:6b}}

\subfloat[]{\includegraphics[width=8cm, height=6.5cm]{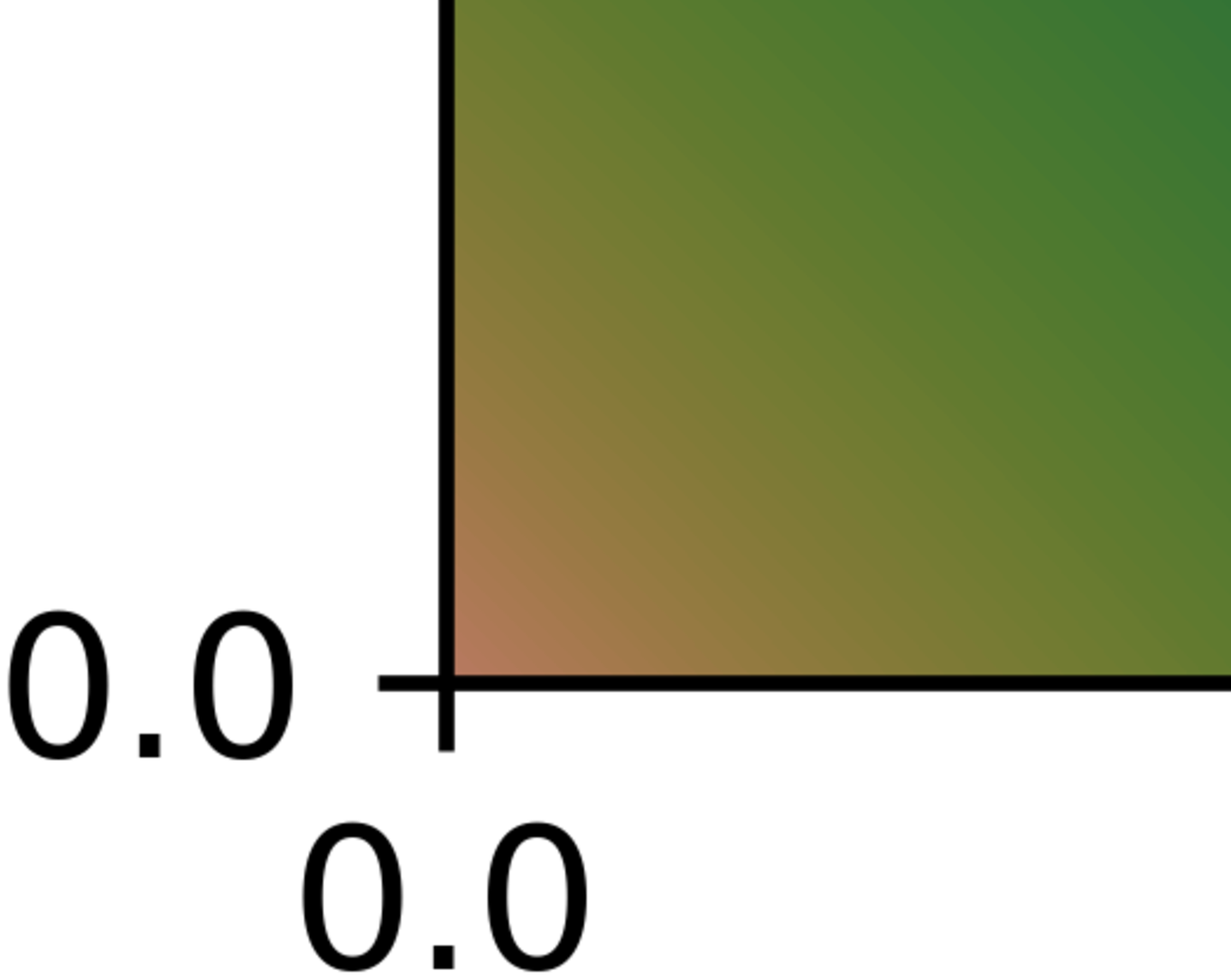}\label{fig:6c}}
\caption{(a) Contour plot of $\mathcal{Q}_E$ for the V-type decay channel, $\Phi_{(p_1,p_2,0)}$ with damping parameters $p_1$ and $p_2$. (b) Contour plot of $\mathcal{Q}_E$ for the $\Lambda$-type decay channel, $\Phi_{(0, p_2, p_3)}$ varies with damping parameters $p_2$ and $p_3$. (c) Contour plot of $\mathcal{Q}_E$ for the three decay rate channels, $\Phi_{(p, p_2, p_3)}$ varies with damping parameters $p_2$ and $p_3$.}
\label{Fig_11}
\end{figure}

{The entanglement-assisted classical capacity $\mathcal{C}_E$ refers to the optimal transmission rate of classical information with the assistance of unrestricted entanglement shared between communicating parties.} It is equal to twice the entanglement-assisted quantum capacity value, \textit{i.e.},
\begin{equation}
    \mathcal{Q}_E=\frac{1}{2}(\mathcal{C}_E).
\end{equation}
The explicit expressions of entanglement-assisted quantum capacity for single decay channel, double decay and triple decay channel are given in Appendix \ref{AP_deg}.
{In Figs.} (\ref{Fig_5}) and (\ref{Fig_10}), {we have illustrated the dynamics of $\mathcal{Q}_E$ for two different maps: first one is for the single decay map $\Phi_{(1,p_1,0)}$, while the second one corresponds to the map $\Phi_{(p,p,p)}$ where all the decay rates are equal.}
In addition, the corresponding populations of $\alpha$, $\beta$, $\gamma$ and $\delta$ during the optimization process for a single decay channel have been plotted. In Fig. (\ref{Fig_11}), the dynamics of $\mathcal{Q}_E$ for the V-type decay channel, $\Lambda$-type decay channel and special three decay rate channel have been displayed. 
\begin{figure}[ht]
\subfloat[]{\includegraphics[width=0.48\textwidth]{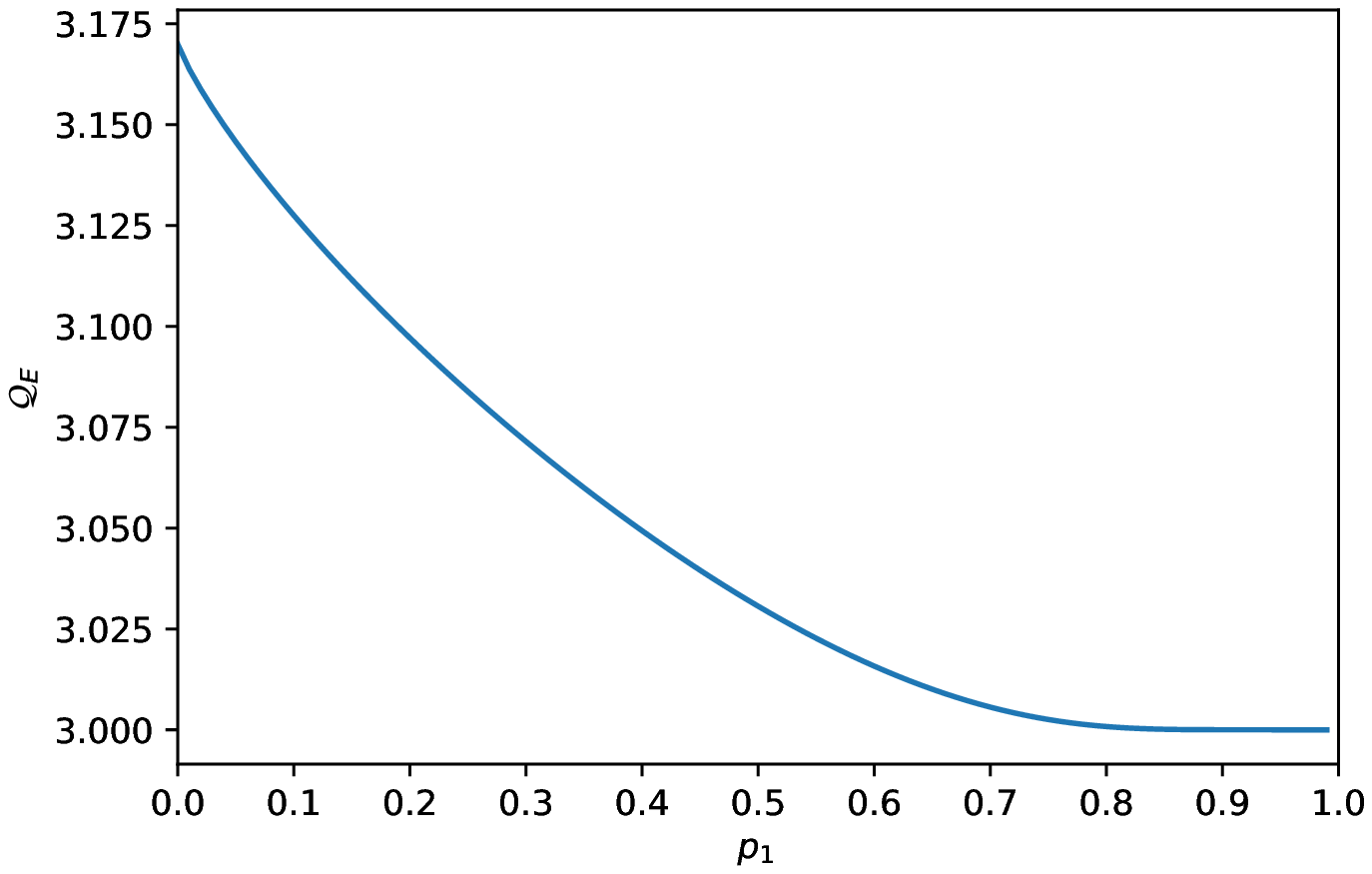}\label{fig:5a}}
\newline
\subfloat[]{\includegraphics[width=0.48\textwidth]{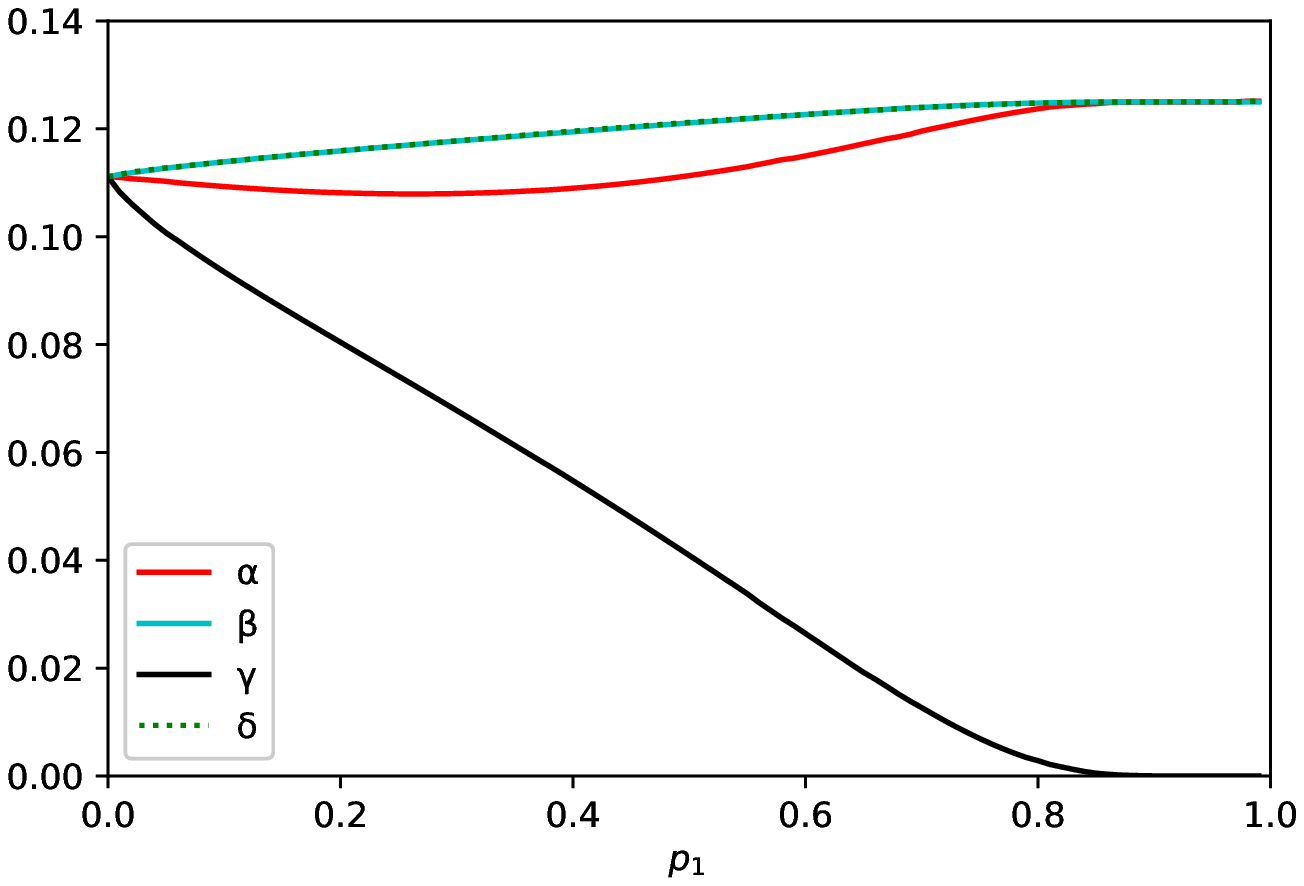}\label{fig:5b}}
\caption{(Color online) (a) The plot of $\mathcal{Q}_E$ for the single decay channel $\Phi_{(p_1, 0, 0)}$ with respect to the damping parameter $p_1$. {The solution of the optimization problem given in Eq. (\ref{QE_Single}) determines the values of $\mathcal{Q}_E$ at different $p_1$.} (b) The populations $\alpha$, $\beta$, $\gamma$ and $\delta$ refer to the states that optimize the $\mathcal{C}_E$ formula for the $\Phi_{(p_1,0,0)}$ channel with respect to the damping parameter $p_1$.}
\label{Fig_5}
\end{figure}
\begin{figure}[ht]
\subfloat[]{\includegraphics[width=0.48\textwidth]{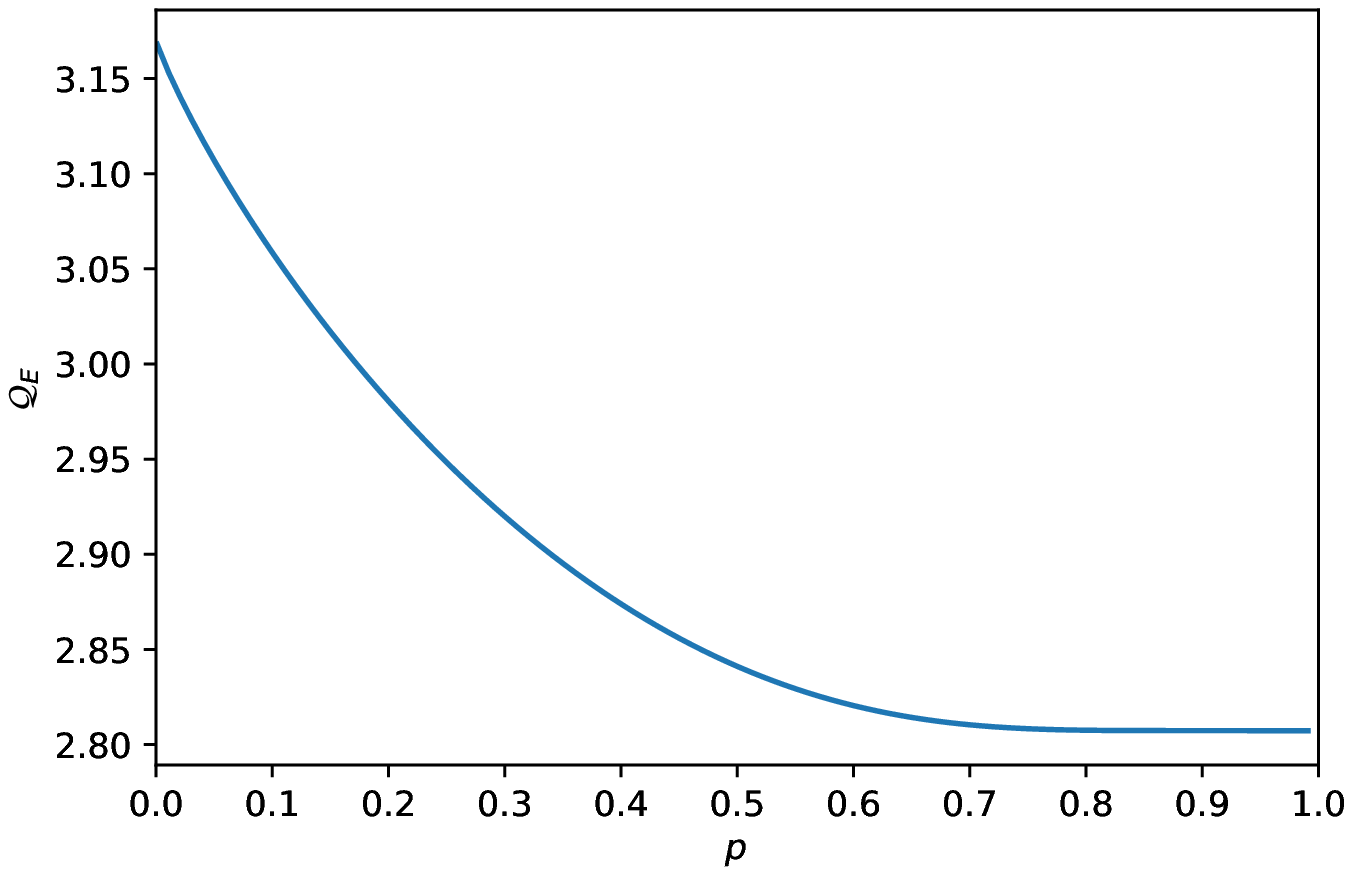}\label{fig:10a}}
\newline
\subfloat[]{\includegraphics[width=0.48\textwidth]{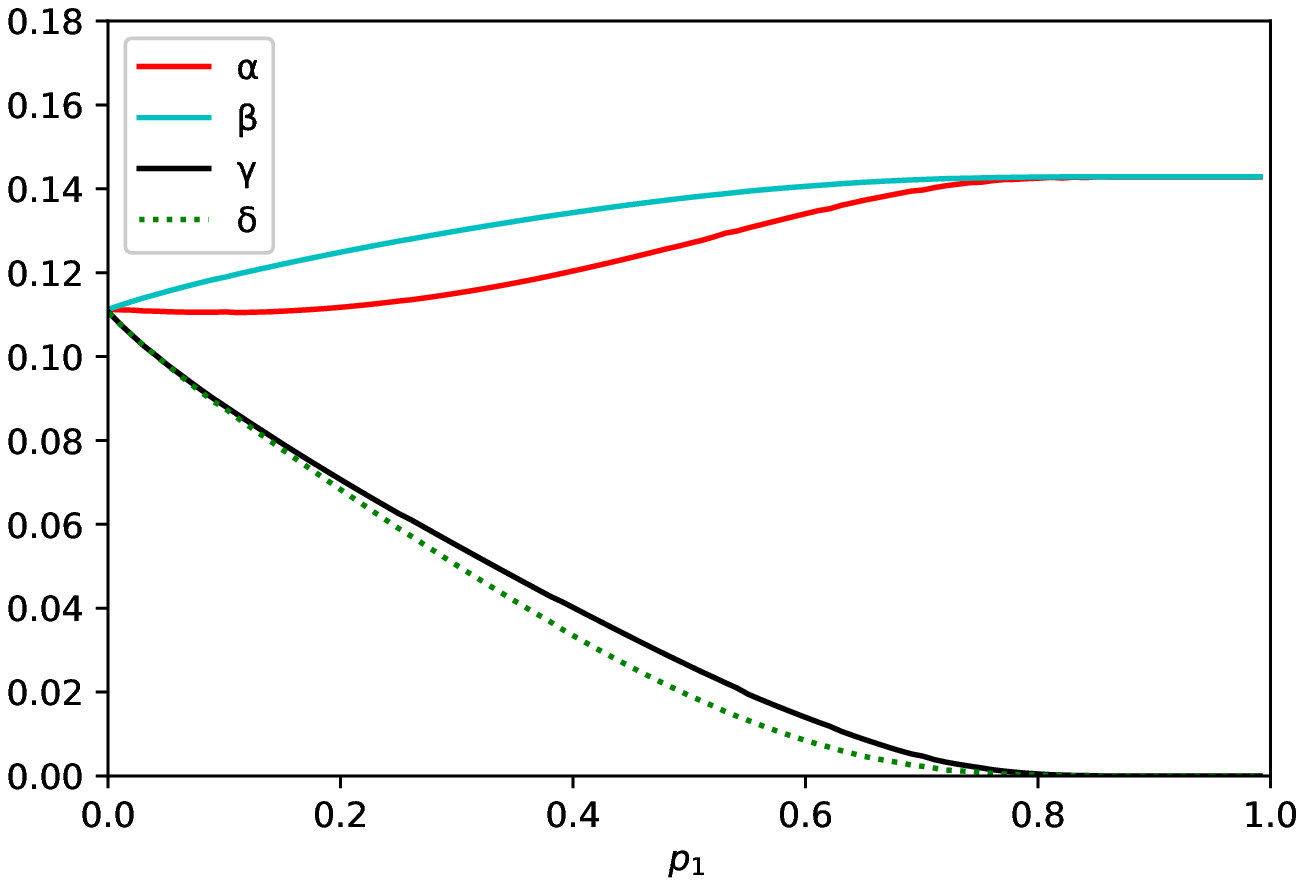}\label{fig:10b}}
\caption{(Color online) (a) The plot of $\mathcal{Q}_E$ for the channel $\Phi_{(p,p,p)}$ with respect to the damping parameter $p$. {The solution of the optimization problem given in Eq. (\ref{AP_B28}) in Appendix \ref{AP_deg}, which determines the values of $\mathcal{Q}_E$ at different $p_1$.} (b) The populations $\alpha$, $\beta$, $\gamma$ and $\delta$ refer to the states that optimize the $\mathcal{C}_E$ formula corresponding to the map $\Phi_{(p,p,p)}$.}
\label{Fig_10}
\end{figure}
\section{Conclusion}\label{Sec_conclusion}
In quantum information theory, the qubit ADC model is a well-known example of quantum noise. It has been shown that correlated ADC channel has higher information transmission capacity and can protect quantum correlations efficiently. In our work, we have investigated the information capacity for a multi-dimensional version of the correlated ADC model with a special focus on dimension d=3. We have explicitly calculated the upper bound of the single-shot classical and quantum capacities of various maps associated with the fully correlated MAD channels on the qutrit space. This computation has expanded the set of models whose capacity is known. 
In Ref. \cite{chessa2021quantum}, {it has been shown that V-type and $\Lambda$-type memoryless qutrit MAD channel exhibit anti-degradability properties in some specific regions, which leads to zero quantum capacity in that region. On the other hand, the fully correlated V-type and $\Lambda$-type qutrit channels do not exhibit anti-degradability and have positive quantum capacity over the entire range of parameters.} 
We have observed that the $\Lambda$-type decay channel exhibits a higher quantum capacity compared to the V-type decay channel.
The insights this research provides can be useful in designing and optimizing quantum communication systems to operate in noisy environments. The findings of this study provide a basis for further exploration of the information capacity for MAD channels with arbitrary degrees of memory.
\begin{acknowledgments}
R.S. acknowledge the financial support provided by IISER Kolkata. We also acknowledge Abhinash Kumar Roy for the fruitful discussion. The authors would like to express their gratitude to the anonymous reviewer for providing several suggestions that have enhanced the quality of the manuscript.  
\end{acknowledgments}
\appendix
\section{}\label{Ap_comp}
\subsection{Composition rule}
The study presented in Ref. \cite{chessa2021quantum} established that under composition rules, MAD channels are closed, a useful property for examining their information capacities. Here, we examine whether fully correlated MAD channels exhibit similar behaviour. We notice that if $\Phi_{(p_1',p_2',p_3')}$ and $\Phi_{(p_1'',p_2'',p_3'')}$ are two maps such that they fulfil the CPTP conditions, then we have
\begin{equation}\label{Eq_comp}
 \Phi_{(p_1',p_2',p_3')}\circ \Phi_{(p_1'',p_2'',p_3'')}=\Phi_{(p_1,p_2,p_3)}.
\end{equation}
The significance of the above equation in addressing our current problem lies in the channel data-processing inequalities \cite{keyl2002fundamentals,khatri2020information}. If a CPTP map $\Phi_{(p_1,p_2,p_3)}$ is obtained by combining two CPTP maps $\Phi_{(p'_1,p'_2,p'_3)}$ and $\Phi_{(p''_1,p''_2,p''_3)}$ then the channel data processing inequality indicates that any information capacity function $\mathcal{F}$ such as classical capacity $\mathcal{C}$, quantum capacity $\mathcal{Q}$, entanglement-assisted quantum capacity $\mathcal{Q}_E$, etc., must satisfy the relation described below \cite{wolf2007quantum}
\begin{equation}\label{Eq_inequality}
\mathcal{F}(\Phi_{(p_1,p_2,p_3)}) \leq \min \left\{\mathcal{F}\left(\Phi_{(p'_1,p'_2,p'_3)}\right), \mathcal{F}\left(\Phi_{(p''_1,p''_2,p''_3)}\right)\right\}.
\end{equation}
The new rate vector $\left(p_1, p_2, p_3\right)$ for V-type decay channel from Eq. (\ref{Eq_comp}) is
\begin{equation}
\begin{aligned}
p_1 &=p_1^{\prime}+p_1^{\prime \prime}-p_1^{\prime} p_1^{\prime \prime}, \\
p_2 &=p_2^{\prime}+p_2^{\prime \prime}-p_2^{\prime} p_2^{\prime \prime}, \\
\end{aligned}
\end{equation}
and the new rate vector of components for $\Lambda$-type decay channel
\begin{equation}
\begin{aligned}
p_2 &=p_2^{\prime}+p_2^{\prime \prime}-p_2^{\prime} p_2^{\prime \prime}, \\
p_3 &=p_3^{\prime}+p_3^{\prime \prime}-p_3^{\prime} p_3^{\prime \prime},\\
\end{aligned}
\end{equation}
which also satisfies CPTP conditions. 
The aforementioned inequality (\ref{Eq_inequality}) can be utilized to predict the monotonic behaviour of the capacity $\mathcal{F}(\Phi_{(p_1,p_2,p_3)})$  with respect to the decay rates $\{p_1, p_2, p_3 \}$  by applying it to Eq. (\ref{Eq_comp}).
The lower and upper bounds obtained from this inequality can prove to be valuable in expanding the capacity formula to domains where the map is non-degradable. In specific cases, for a single-decay fully correlated MAD channels defined by the single non-zero decay parameter (for example, $p_1$), we obtain
\begin{equation}
\Phi_{\left(p_1^{\prime}, 0,0\right)} \circ \Phi_{\left(p_1^{\prime\prime}, 0,0\right)}=\Phi_{\left(p_1^{\prime\prime}, 0,0\right)} \circ \Phi_{\left(p_1^{\prime}, 0,0\right)}=\Phi_{\left(p_1, 0,0\right)}.
\end{equation}
In this way we can infer that the cpacity functional $\mathcal{F}(\Phi_{(p_1,p_2,p_3)})$ is a non-increasing function with respect to decay parameter $p_1$:
\begin{equation}\label{Eq_monosingle}
\mathcal{F}\left(\Phi_{\left(p_1, 0,0\right)}\right) \geq \mathcal{F}\left(\Phi_{\left(p^{\prime}, 0,0\right)}\right), \quad \forall p_1 \leq p^{\prime}.
\end{equation}
Similarly, for other single decay maps $\Phi_{(0,p_2,0)}$ and $\Phi_{(0,0,p_3)}$, we can deduce the above inequality and prove their non-increasing behaviour. 
We can write the mapping $\Phi_{(p_1,p_2,p_3)}$ in terms of the composition form:
\begin{equation}
\Phi_{(p_1,p_2,p_3)}=\Phi_{(0,0,\Bar{p}_3)}\circ\Phi_{(0,\bar{p}_2,0)}\circ\Phi_{(p_1,0,0)}.
\end{equation}

where, $\Bar{p}_2=p_1+(1-\Theta)p_{123}$ and $\Bar{p}_3=\frac{\Theta p_{123}}{1-p_1-(1-\Theta)p_{123}}$.
Alternatively, we can write the given composition:
\begin{equation}   
\Phi_{(p_1,p_2,p_3)}=\Phi_{(0,\bar{p}_2,0)}\circ\Phi_{(0,0,\Bar{p}_3)}\circ\Phi_{(p_1,0,0)}.
\end{equation}
where, $\Bar{p}_2=\frac{p_1+(1-\Theta)p_{123}}{1-\Theta p_{123}}$ and $\Bar{p}_3=\Theta p_{123}$. To make the notation simpler in the above equation, we have written $\Theta(\Gamma)\equiv\Theta$ and $p_{123}=(1-p_1)-(1-p_2)(1-p_3)$.
For V-type decay channel setting $p_3=0$, we obtain the relation 
\begin{equation}
\Phi_{(p_1,p_2,0)}=\Phi_{(0,p_2,0)}\circ\Phi_{(p_1,0,0)}=\Phi_{(p_1,0,0)}\circ\Phi_{(0,p_2,0)}.
\end{equation}
The Eq. (\ref{Eq_inequality}) leads to the following condition:
\begin{equation}
  \mathcal{F}(\Phi_{(p_1,p_2,0)}) \leq \min \left\{\mathcal{F}\left(\Phi_{(0,p_2,0)}\right), \mathcal{F}\left(\Phi_{(p_1,0,0)}\right)\right\}. \label{Eq_mono_v} 
\end{equation}
For $\Lambda$-type decay channel, $\Phi_{(0,p_2,p_3)}$ can be written as the following composition:
\begin{equation}
\Phi_{(0,p_2,p_3)}=\Phi_{(0,0,\Tilde{p}_3)}\circ\Phi_{(0,\Tilde{p}_2,0)}=\Phi_{(0,\Tilde{p_2},0)}\circ\Phi_{(0,0,\Tilde{p}_3)}.
\end{equation}
and we obtain the inequality:
\begin{equation}
  \mathcal{F}(\Phi_{(0,p_2,p_3)}) \leq \min \left\{\mathcal{F}\left(\Phi_{(0,\Tilde{p}_2,0)}\right), \mathcal{F}\left(\Phi_{(0,0,\Tilde{p}_3)}\right)\right\},  
\end{equation}
where, $\Tilde{p_2}=(1-\Theta)p_{23}$ and $\Tilde{p_3}=\frac{\Theta p_{23}}{1-(1-\Theta)p_{23}}$ and $p_{23}=1-(1-p_2)(1-p_3)$.

\section{}\label{AP_deg}
\begin{widetext}
\subsection{\texorpdfstring{Degradibility of $\Phi_{(p_1,p_2,p_3)}$ map}{}} 
First, we write the matrix form of the Kraus operator for the fully correlated MAD channel, which will be used in different cases with different conditions,
    \begin{equation}
\begin{split}
  E_{00}=
  {\begin{bmatrix}
   \textbf{I}_{4\times 4} &&&\\
   &\sqrt{1-p_1}&&\\
   &&\textbf{I}_{3\times 3}&\\
   &&&\sqrt{(1-p_2)(1-p_3)}\\
    \end{bmatrix}}_{9\times 9},
  E_{11}=
{\left[\begin{array}{ccc|ccc} 
	 &&&\sqrt{p_1}&\dots&0\\
    &\textbf{O}_{4\times 5}&&\vdots&\ddots&\vdots\\
    &&& 0 &\dots&0\\
    \hline
   &\textbf{O}_{4\times 4}&&& \textbf{O}_{4\times 4}
\end{array}\right] }_{9\times 9},
\\
E_{22}=
{\begin{bmatrix}
  0 & \dots & 0 &\sqrt{p_1+(1-\Theta(\Gamma))p_{123}}\\
  0 & \dots & 0 & 0\\
  \vdots & \ddots &\vdots &\vdots\\
  0 & \dots & 0 & 0\\
    \end{bmatrix}}_{9\times 9},
E_{33}=
{\left[\begin{array}{ccc|ccc} 
	 &&&0&\dots&0\\
    &\textbf{O}_{4\times 5}&&\vdots&\ddots&\vdots\\
    &&& 0 &\dots&\sqrt{\Theta(\Gamma)p_{123}}\\
    \hline
   &\textbf{O}_{4\times 4}&&& \textbf{O}_{4\times 4}
\end{array}\right] }_{9\times 9},
   \end{split}
\end{equation}
where $\Theta(\Gamma)=\Gamma_{3}/(\Gamma_3+\Gamma_2-\Gamma_1)$,  $p_{123}=e^{-\Gamma_1 t}-e^{-(\Gamma_2+\Gamma_3)t}=(1-p_1)-(1-p_2)(1-p_3) $ and  \textbf{O} and \textbf{I} are respectively null matrix and identity matrix respectively.{ The map is completely positive when $(1-p_1)\geq(1-p_2)(1-p_3)$. If the map $\Phi_{(p_1,p_2,p_3)}$ acts on the input state $\rho$ given in Eq. (\ref{Input}), the output state is given by} \\
{$\Phi_{(p_1,p_2,p_3)}(\rho)=$}
\begin{equation}\label{M_0}
\resizebox{\textwidth}{!}{$
\left( 
\begin{array}{ccccccccc}
\Tilde{\rho}_{00}& \rho_{01} & \rho_{02} & \rho_{03} & \sqrt{\Tilde{p_1}} \rho_{04} & \rho_{05} & \rho_{06} & \rho_{07} &  \sqrt{\Tilde{p_2}\Tilde{p_3}}\rho_{08} \\
 \rho_{10} & \rho_{11} & \rho_{12} & \rho_{13} & \sqrt{\Tilde{p_1}} \rho_{14} & \rho_{15} & \rho_{16} & \rho_{17} &  \sqrt{\Tilde{p_2}\Tilde{p_3}}\rho_{18}\\
 \rho_{20} & \rho_{21} & \rho_{22} & \rho_{23} & \sqrt{\Tilde{p_1}} \rho_{24} & \rho_{25} & \rho_{26} & \rho_{27} &  \sqrt{\Tilde{p_2}\Tilde{p_3}}\rho_{28} \\
 \rho_{30} & \rho_{31} & \rho_{32} & \rho_{33} & \sqrt{\Tilde{p_1}} \rho_{34} & \rho_{35} & \rho_{36} & \rho_{37} &  \sqrt{\Tilde{p_2}\Tilde{p_3}}\rho_{38} \\
 \sqrt{\Tilde{p_1}} \rho_{40} & \sqrt{\Tilde{p_1}} \rho_{41} & \sqrt{\Tilde{p_1}} \rho_{42} & \sqrt{\Tilde{p_1}} \rho_{43} & \Tilde{\rho}_{44} & \sqrt{\Tilde{p_1}} \rho_{45} & \sqrt{\Tilde{p_1}} \rho_{46} & \sqrt{\Tilde{p_1}} \rho_{47} & \sqrt{\Tilde{p_1}}\sqrt{\Tilde{p_2}\Tilde{p_3}}\rho_{48} \\
 \rho_{50} & \rho_{51} & \rho_{52} & \rho_{53} & \sqrt{\Tilde{p_1}} \rho_{54} & \rho_{55} & \rho_{56} & \rho_{57} &  \sqrt{\Tilde{p_2}\Tilde{p_3}}\rho_{58} \\
 \rho_{60} & \rho_{61} & \rho_{62} & \rho_{63} & \sqrt{\Tilde{p_1}} \rho_{64} & \rho_{65} & \rho_{66} & \rho_{67} &  \sqrt{\Tilde{p_2}\Tilde{p_3}}\rho_{68} \\
 \rho_{70} & \rho_{71} & \rho_{72} & \rho_{73} & \sqrt{\Tilde{p_1}} \rho_{74} & \rho_{75} & \rho_{76} & \rho_{77} &  \sqrt{\Tilde{p_2}\Tilde{p_3}} \rho_{78}\\
  \sqrt{\Tilde{p_2}\Tilde{p_3}}\rho_{80} &  \sqrt{\Tilde{p_2}\Tilde{p_3}}\rho_{81} &  \sqrt{\Tilde{p_2}\Tilde{p_3}}\rho_{82} &  \sqrt{\Tilde{p_2}\Tilde{p_3}}\rho_{83} & \sqrt{\Tilde{p_1}}  \sqrt{\Tilde{p_2}\Tilde{p_3}}\rho_{84} &  \sqrt{\Tilde{p_2}\Tilde{p_3}}\rho_{85} &  \sqrt{\Tilde{p_2}\Tilde{p_3}}\rho_{86} & \sqrt{\Tilde{p_2}\Tilde{p_3}} \rho_{87} & \Tilde{p_2}\Tilde{p_3}\rho_{88} \\
\end{array}
\right)
$}
\end{equation}
{where
$\Tilde{p_1}=1-p_1$, $\Tilde{p_2}=1-p_2$, $\Tilde{p_3}=1-p_3$
$\Tilde{\rho}_{00} =\rho_{00}+p_1\rho_{44}+(p_1+(1-\Theta(\Gamma))(1-p_1-(1-p_2)(1-p_3)))\rho_{88}$\\
$\Tilde{\rho}_{44}=(1-p_1)\rho_{44}+\Theta(\Gamma)(1-p_1-(1-p_2)(1-p_3))\rho_{88}$.
The complementary map $\Tilde{\Phi}_{(p_1,p_2,p_3)}$ corresponding to the above map is}\\
{$\Tilde{\Phi}_{(p_1,p_2,p_3)}(\rho)=$}
\begin{equation}\label{CM_0}
\resizebox{\textwidth}{!}{$
    \left(
\begin{array}{cccccccccccccccc}
 1-{p_1}\rho_{44}-p_{23}\rho_{88} & 0 & 0 & 0 & 0 & \sqrt{{p_1}} \rho_{04} & 0 & 0 & 0 & 0 & \sqrt{p_1+(1-\Theta(\Gamma)p_{123} }\rho_{08} & 0 & 0 & 0 & 0 & \sqrt{1-{p_1}}\sqrt{\Theta(\Gamma)p_{123}} \rho_{48}  \\
 0 & 0 & 0 & 0 & 0 & 0 & 0 & 0 & 0 & 0 & 0 & 0 & 0 & 0 & 0 & 0 \\
 0 & 0 & 0 & 0 & 0 & 0 & 0 & 0 & 0 & 0 & 0 & 0 & 0 & 0 & 0 & 0 \\
 0 & 0 & 0 & 0 & 0 & 0 & 0 & 0 & 0 & 0 & 0 & 0 & 0 & 0 & 0 & 0 \\
 0 & 0 & 0 & 0 & 0 & 0 & 0 & 0 & 0 & 0 & 0 & 0 & 0 & 0 & 0 & 0 \\
 \sqrt{{p_1}} \rho_{40} & 0 & 0 & 0 & 0 & {p_1} \rho_{44} & 0 & 0 & 0 & 0 & \sqrt{{p_1}}\sqrt{p_1+(1-\Theta(\Gamma)p_{123} } \rho_{48}  & 0 & 0 & 0 & 0 & 0 \\
 0 & 0 & 0 & 0 & 0 & 0 & 0 & 0 & 0 & 0 & 0 & 0 & 0 & 0 & 0 & 0 \\
 0 & 0 & 0 & 0 & 0 & 0 & 0 & 0 & 0 & 0 & 0 & 0 & 0 & 0 & 0 & 0 \\
 0 & 0 & 0 & 0 & 0 & 0 & 0 & 0 & 0 & 0 & 0 & 0 & 0 & 0 & 0 & 0 \\
 0 & 0 & 0 & 0 & 0 & 0 & 0 & 0 & 0 & 0 & 0 & 0 & 0 & 0 & 0 & 0 \\
 \sqrt{p_1+(1-\Theta(\Gamma)p_{123} }\rho_{80}  & 0 & 0 & 0 & 0 & \sqrt{{p_1}}\sqrt{p_1+(1-\Theta(\Gamma)p_{123} } \rho_{84}  & 0 & 0 & 0 & 0 & {p_1+(1-\Theta(\Gamma)p_{123} }\rho_{88}  & 0 & 0 & 0 & 0 & 0 \\
 0 & 0 & 0 & 0 & 0 & 0 & 0 & 0 & 0 & 0 & 0 & 0 & 0 & 0 & 0 & 0 \\
 0 & 0 & 0 & 0 & 0 & 0 & 0 & 0 & 0 & 0 & 0 & 0 & 0 & 0 & 0 & 0 \\
 0 & 0 & 0 & 0 & 0 & 0 & 0 & 0 & 0 & 0 & 0 & 0 & 0 & 0 & 0 & 0 \\
 0 & 0 & 0 & 0 & 0 & 0 & 0 & 0 & 0 & 0 & 0 & 0 & 0 & 0 & 0 & 0 \\
 \sqrt{1-{p_1}}\sqrt{p_1+(1-\Theta(\Gamma)p_{123} } \rho_{84}  & 0 & 0 & 0 & 0 & 0 & 0 & 0 & 0 & 0 & 0 & 0 & 0 & 0 & 0 &  1-{p_1}\rho_{44}-p_{23}\rho_{88} \\
\end{array}
\right)
$}
\end{equation}
{where, $p_{23}=1-(1-p_2)(1-p_3)$ and $p_{123}=(1-p_1)-(1-p_2)(1-p_3)$.}
\end{widetext}
{First, let's begin by investigating whether the map is degradable or anti-degradable. In this context, we follow the method outlined in Sec. \ref{Sec_IIIA}. 
The channel is said to be degradable when the mapping $\Phi_D=\Tilde{\Phi}\circ\Phi^{-1}$ is a CPTP map. The map $\Phi_D$ is said to be CPTP if the Choi matrix associated with $\Phi_D$ is positive.
Here, we have checked the complete positivity of the mapping $\Phi_D$ by writing the states $\rho$, $\Phi(\rho)$ and $\Tilde{\Phi}(\rho)$ in Liouville-Fock space finding the transformation matrix $\mathcal{M}_{\Phi}$ between $\rho$ and $\Phi(\rho)$, and the transformation matrix $\mathcal{M}_{\Tilde{\Phi}}$ between $\rho$ and $\Tilde{\Phi}(\rho)$ and finally checking the positivity of the following matrix :}
\begin{equation}\label{Eq_D}  \mathcal{M}_{\Phi_D}=\mathcal{M}_{\Tilde{\Phi}} \mathcal{M}^{-1}_{\Phi}.
\end{equation}
{Similarly, the channel is said to be anti-degradable when the matrix:}
\begin{equation}\label{Eq_AD}
\mathcal{M}_{\Phi_{AD}}=\mathcal{M}_{\Phi} \mathcal{{M}}^{-1}_{\Tilde{\Phi}}
\end{equation}
{is a positive matrix.}
{First we calculate the supermaps $\mathcal{M}_{\Phi}$ and $\mathcal{M}_{\tilde{\Phi}}$ from Eqs.} (\ref{M_0}) and (\ref{CM_0}). { Then,  following Eq. }(\ref{Eq_D}),{ we observed that $\mathcal{M}_{\Phi_D}$ is positive when $(1-p_1)<(1-p_2)(1-p_3)$. However, as previously shown in the main text, the map $\Phi_{(p_1, p_2, p_3)}$ is CPTP when it fulfils the condition $(1-p_1) \geq (1-p_2)(1-p_3)$. {Hence the map $\Phi_{D}$ is not CPTP. Now, we have to prove that any other degrading CPTP maps do not exist. In this way, we can ensure the non-degradability of the map $\Phi_{(p_1, p_2, p_3)}$. To prove this, we use Theorem 3 given in Ref.} \cite{bradler2015pitfalls}, {which is as follows:}
\begin{theorem}
{let $\Phi:$ $\mathcal{O}(\mathcal{H}_S)\rightarrow\mathcal{O}(\mathcal{H}_{S'})$ be a quantum channel with the complementary channel $\Tilde{\Phi}:$ $\mathcal{O}(\mathcal{H}_S)\rightarrow\mathcal{O}(\mathcal{H}_{\mathcal{E}'})$ and the corresponding super-operator $\mathcal{M}_{\Phi}$ be a full rank matrix: $rank[\mathcal{M}_{\Phi}]=\min[d^2_S,d^2_{S'}]$. Then, if a degrading map $\Phi_D:$ $\mathcal{O}(\mathcal{H}_{S'})\rightarrow\mathcal{O}(\mathcal{H}_{\mathcal{E}'})$ exists, it is unique iff $d_{S'}\leq d_{S}$ }.
\end{theorem}
{We observe that the dimension of the input state $\rho$ and output state $\Phi_{(p_1,p_2,p_3)}(\rho)$ are same \textit{i.e.,} $d_{S}=d_{S'}$ and the super-operator $\mathcal{M}_{\Phi}$ is a triangular matrix, which is a full rank matrix except in two cases: when $p_1=1$ or when $(1-p_2)(1-p_3)=0$. Hence, in the cases $p_1\neq 1$ and $(1-p_2)(1-p_3)\neq 0$, the map $\mathcal{M}_{\Phi}$ is a full rank matrix and since the $\Phi_D$ is not CPTP the map $\Phi$ is not degradable. 

Now, in the cases when $p_1=1$ and $(1-p_2)(1-p_3)=0$, the map $\mathcal{M}_{\Phi}$ is not full rank matrix and, hence, not invertible.}  
{If we carefully notice $\Phi_{(p_1,p_2,p_3)}$ and $\Tilde{\Phi}_{(p_1,p_2,p_3)}$ given in Eqs.} (\ref{M_0}) and (\ref{CM_0}) {after imposing constraints $p_1=1$ and $(1-p_2)(1-p_3)=0$, we find many elements that are present in $\Phi$ but absent in $\Tilde{\Phi}$. For instance, when $p_1=1$ we find the elements $\rho_{04}, \rho_{48}$ of the input state $\rho$ are present in $\Tilde{\Phi}$ but absent in $\Phi$. Similarly, when $(1-p_2)(1-p_3)=0$, we find the components $\rho_{08}$, $\rho_{48}$ of the input state $\rho$ are present in $\Tilde{\Phi}$ but absent in $\Phi$. Therefore, there is no linear map that we can apply to $\Phi(\rho)$ to obtain $\Tilde{\Phi}(\rho)$.
Hence, the channel $\Phi_{(p_1,p_2,p_3)}$ is not degradable.

Now, we can confirm the anti-degradability of the channel by establishing that $\ker \Tilde{\Phi} \not\subset \ker \Phi$ }\cite{cubitt2008structure}. { we find many elements---e.g., the elements $\ket{00}\bra{01}$, $\ket{00}\bra{02}$, etc.---that are present in $\Phi_{(p_1,p_2,p_3)}$ but absent in $\Tilde{\Phi}{(p_1,p_2,p_3)}$. Hence, the kernel of $\Tilde{\Phi}{(p_1,p_2,p_3)}$ cannot be a subset of $\Phi_{(p_1,p_2,p_3)}$. However, we can directly conclude that the channel is not anti-degradable by examining the matrix $\Phi_{(p_1,p_2,p_3)}$ since it possesses a seven-dimensional decoherence-free subspace. Hence, the lower bound of quantum capacity is $\log_27$. But, for an anti-degradable channel, the quantum capacity should be zero.  Hence, the channel $\Phi_{(p_1,p_2,p_3)}$ is not anti-degradable.}

\subsubsection{Single decay channel}
The instances of the fully correlated qutrit MAD channel that we are analyzing in this context involve situations where only one of the three damping parameters, $p_i$ has a non-zero value. The associated maps for this single decay are $\Phi_{(p_1,0,0)}$, $\Phi_{(0,p_2,0)}$ and $\Phi_{(0,0,p_3)}$ respectively.
 Note that the mapping $\Phi_{(0,p_2,0)}$ and $\Phi_{(0,0,p_3)}$ can be obtained from the mapping $\Phi_{(p_1,0,0)}$ by swapping the energy levels $\ket{00}\leftrightarrow\ket{11}$ and $\ket{00}\leftrightarrow\ket{22}$. 
Consequently, the quantum capacity of the three single decay channels is the same. The transformation $\Phi_{(p_1,0,0)}(\rho)$ can be obtained according to the equation $\rho_t=\sum_n E_n \rho {E_n}^\dagger$, as:
\begin{widetext}
\begin{equation}\label{AP_B6}
\resizebox{\textwidth}{!}{$
\Phi=
\left(
\begin{array}{ccccccccc}
 p_1 \rho_{44}+\rho_{00} & \rho_{01} & \rho_{02} & \rho_{03} & \sqrt{1-p_1} \rho_{04} & \rho_{05} & \rho_{06} & \rho_{07} & \rho_{08} \\
 \rho_{10} & \rho_{11} & \rho_{12} & \rho_{13} & \sqrt{1-p_1} \rho_{14} & \rho_{15} & \rho_{16} & \rho_{17} & \rho_{18} \\
 \rho_{20} & \rho_{21} & \rho_{22} & \rho_{23} & \sqrt{1-p_1} \rho_{24} & \rho_{25} & \rho_{26} & \rho_{27} & \rho_{28} \\
 \rho_{30} & \rho_{31} & \rho_{32} & \rho_{33} & \sqrt{1-p_1} \rho_{34} & \rho_{35} & \rho_{36} & \rho_{37} & \rho_{38} \\
 \sqrt{1-p_1} \rho_{40} & \sqrt{1-p_1} \rho_{41} & \sqrt{1-p_1} \rho_{42} & \sqrt{1-p_1} \rho_{43} & \left(1-p_1\right) \rho_{44} & \sqrt{1-p_1} \rho_{45} & \sqrt{1-p_1} \rho_{46} & \sqrt{1-p_1} \rho_{47} & \sqrt{1-p_1} \rho_{48} \\
 \rho_{50} & \rho_{51} & \rho_{52} & \rho_{53} & \sqrt{1-p_1} \rho_{54} & \rho_{55} & \rho_{56} & \rho_{57} & \rho_{58} \\
 \rho_{60} & \rho_{61} & \rho_{62} & \rho_{63} & \sqrt{1-p_1} \rho_{64} & \rho_{65} & \rho_{66} & \rho_{67} & \rho_{68} \\
 \rho_{70} & \rho_{71} & \rho_{72} & \rho_{73} & \sqrt{1-p_1} \rho_{74} & \rho_{75} & \rho_{76} & \rho_{77} & \rho_{78} \\
 \rho_{80} & \rho_{81} & \rho_{82} & \rho_{83} & \sqrt{1-p_1} \rho_{84} & \rho_{85} & \rho_{86} & \rho_{87} & \rho_{88} \\
\end{array}
\right).
$}
\end{equation}
\end{widetext}
The corresponding complementary channel $\Tilde{\Phi}_{(p_1,0,0)}(\rho)$ calculated according to the Eq. (\ref{Eq_commap}) in the main text:
\begin{equation}
\Tilde{\Phi}=
   \left(
\begin{array}{cccc}
 1-p_1 \rho_{44} & 0 & 0 & \sqrt{p_1} \rho_{04} \\
 0 & 0 & 0 & 0 \\
 0 & 0 & 0 & 0 \\
 \sqrt{p_1} \rho_{40} & 0 & 0 &  p_1\rho_{44} \\
\end{array}
\right).
\end{equation}
For single decay channel we have found that $\mathcal{M}_{\Phi_D}$ is positive when, $p_1\leq\frac{1}{2}$. Therefore, the channel is degradable for $p_1\leq\frac{1}{2}$. {This degrading map is unique since $\mathcal{M}_{\Phi}$ for the single decay map is a full rank matrix and the dimension of $\rho$ and $\Phi_{(p_1,0,0)}(\rho)$ are same: $d_S=d_{S'}$ }\cite{bradler2015pitfalls}. 
{ From Eq.} (\ref{AP_B6}) {we observe that the channel has eight-dimensional noiseless subspace. Hence, we can say that the single decay channel is not anti-degradable.}
In the degradable region, the expression of quantum capacity is given in Eq. (\ref{Eq_48}).
However, we can calculate the quantum capacity for $p_1\geq1/2$ using the monotonic behaviour of the $\mathcal{Q}$ and the lower bound discussed in the following section.
From Eq. (\ref{AP_B6}) we can see that the transformation is noiseless over the subspace $\ket{00}$, $\ket{01}$, $\ket{02}$, $\ket{10}$, $\ket{12}$, $\ket{20}$, $\ket{21}$ and $\ket{22}$. Therefore, the lower bound of quantum capacity and classical capacity is 
\begin{equation}
    \mathcal{Q}(\Phi),\; \mathcal{C}(\Phi)\geq \log_28=3.
\end{equation}
Now, in the degradable region, we already know the quantum capacity value. At $p=1/2$ the quantum capacity $\mathcal{Q}(\Phi)=3$ and the lower bound of  $\mathcal{Q}(\Phi)$ is 3. It is also shown in Eq. (A6) that the $\mathcal{Q}(\Phi)$ is a nonincreasing function. Therefore, beyond $p=1/2$ the value of the quantum capacity becomes constant, which is $\log_28=3$.
\subsubsection{\texorpdfstring{$\Phi_{(1,p_2,0)}$ channel}{}}
In this subsection, we calculate the quantum capacity for the map $\Phi_{(1,p_2,0)}$ when the first excited state is completely damped and transition $\ket{22}\leftrightarrow\ket{00}$ is possible. The value of this quantum capacity will be useful to calculate that of a V-type decay channel in the non-degradable region, which we will show in the next subsection. The Kraus operator describing this channel is,
\begin{equation}
 \begin{aligned}
E_{00}=&\ket{00}\bra{00}+\bra{11}+\sqrt{1-p_2}\ket{22}\bra{22},\nonumber\\
&+\sum_{\substack{i,j=0\\i\neq j}}^{2}\ket{ij}\bra{ij}\nonumber\\
E_{11}=&\ket{00}\bra{11},\, E_{22}=\sqrt{p_2}\ket{00}\bra{22}.\nonumber\\
\end{aligned}    
\end{equation}
The transformation of $\rho$ under this channel can be written as:
\begin{widetext}

\begin{equation}
\resizebox{\textwidth}{!}{$
\Phi=
    \left(
\begin{array}{ccccccccc}
 p_2 \rho_{88}+\rho_{00}+\rho_{44} & \rho_{01} & \rho_{02} & \rho_{03} & 0 & \rho_{05} & \rho_{06} & \rho_{07} & \sqrt{1-p_2} \rho_{08} \\
 \rho_{10} & \rho_{11} & \rho_{12} & \rho_{13} & 0 & \rho_{15} & \rho_{16} & \rho_{17} & \sqrt{1-p_2} \rho_{18} \\
 \rho_{20} & \rho_{21} & \rho_{22} & \rho_{23} & 0 & \rho_{25} & \rho_{26} & \rho_{27} & \sqrt{1-p_2} \rho_{28} \\
 \rho_{30} & \rho_{31} & \rho_{32} & \rho_{33} & 0 & \rho_{35} & \rho_{36} & \rho_{37} & \sqrt{1-p_2} \rho_{38} \\
 0 & 0 & 0 & 0 & 0 & 0 & 0 & 0 & 0 \\
 \rho_{50} & \rho_{51} & \rho_{52} & \rho_{53} & 0 & \rho_{55} & \rho_{56} & \rho_{57} & \sqrt{1-p_2} \rho_{58} \\
 \rho_{60} & \rho_{61} & \rho_{62} & \rho_{63} & 0 & \rho_{65} & \rho_{66} & \rho_{67} & \sqrt{1-p_2} \rho_{68} \\
 \rho_{70} & \rho_{71} & \rho_{72} & \rho_{73} & 0 & \rho_{75} & \rho_{76} & \rho_{77} & \sqrt{1-p_2} \rho_{78} \\
 \sqrt{1-p_2} \rho_{80} & \sqrt{1-p_2} \rho_{81} & \sqrt{1-p_2} \rho_{82} & \sqrt{1-p_2} \rho_{83} & 0 & \sqrt{1-p_2} \rho_{85} & \sqrt{1-p_2} \rho_{86} & \sqrt{1-p_2} \rho_{87} & \left(1-p_2\right) \rho_{88} \\
\end{array}
\right).
$}
\label{Eq_6}
\end{equation}
The corresponding complementary map is given by;
\begin{equation}
\Tilde{\Phi}=
    \left(
\begin{array}{ccccccccc}
 1-\rho_{44}-p_2\rho_{88} & 0 & 0 & 0 & \rho_{04} & 0 & 0 & 0 & \sqrt{p_2} \rho_{08} \\
 0 & 0 & 0 & 0 & 0 & 0 & 0 & 0 & 0 \\
 0 & 0 & 0 & 0 & 0 & 0 & 0 & 0 & 0 \\
 0 & 0 & 0 & 0 & 0 & 0 & 0 & 0 & 0 \\
 \rho_{40} & 0 & 0 & 0 & \rho_{44} & 0 & 0 & 0 & \sqrt{p_2} \rho_{48} \\
 0 & 0 & 0 & 0 & 0 & 0 & 0 & 0 & 0 \\
 0 & 0 & 0 & 0 & 0 & 0 & 0 & 0 & 0 \\
 0 & 0 & 0 & 0 & 0 & 0 & 0 & 0 & 0 \\
 \sqrt{p_2} \rho_{80} & 0 & 0 & 0 & \sqrt{p_2} \rho_{84} & 0 & 0 & 0 & p_2 \rho_{88} \\
\end{array}
\right).
\end{equation}
\end{widetext}
{The super-operator $\mathcal{M}_{\Phi}$ corresponding to the map $\Phi_{(1,p_2,0)}$ is not a full rank matrix (not invertible). We observe that components of input state $\rho$ like $\rho_{04}$, $\rho_{48}$ that are belong to $\Phi_{(1,p_2,0)}$ are not present in  $\Tilde{\Phi}_{(1,p_2,0)}$. Hence, no linear map exists that can transform $\Phi_{(1,p_2,0)}$ into $\Tilde{\Phi}_{(1,p_2,0)}$. Therefore, we can conclude that the channel is not degradable. 
On the other hand, we find many elements---e.g., the elements $\ket{00}\bra{01}$, $\ket{00}\bra{02}$, etc.---that are present in $\Phi_{(1,p_2,0)}$ but absent in $\Tilde{\Phi}{(1,p_2,0)}$. Hence, the kernel of $\Tilde{\Phi}{(1,p_2,0)}$ cannot be a subset of $\Phi_{(1,p_2,0)}$. Moreover, the channel $\Tilde{\Phi}_{(1,p_2,0)}$ has seven-dimensional noiseless subspace, which suggests that the channel has positive quantum capacity. Therefore, the channel is not anti-degradable.}

However, we are still able to calculate the quantum capacity by simulating the output state $\Phi_{(1,p_2,0)}(\rho)$ by $\Phi_{(1,p_2,0)}(\Omega)=\phi_{p_2}(\Omega)$ where $\Omega$ is the density matrix span over eight-dimensional space: $\ket{00}$, $\ket{01}$, $\ket{02}$, $\ket{10}$,$\ket{12}$, $\ket{20}$, $\ket{21}$ and $\ket{22}$. Specifically, we can write 
\begin{equation}\label{Eq_8} \mathcal{Q}_{\Phi_{1,p_2,0}}=\mathcal{Q}^{(1)}_{\Phi_{(1,p_2,0)}}=\mathcal{Q}_{\phi_{p_2}}.
\end{equation}
Now, our goal is to show that the quantum capacity of the map $\phi_{p_2}$ is equal to that of the $\Phi_{(1,p_2,0)}$ {\it i.e.}, we have to prove Eq. (\ref{Eq_8}).
It is obvious that $\mathcal{Q}_{\phi}$ is the natural lower bound of the $\mathcal{Q}_{\Phi}$. The coherent information $\max_{\rho} I_c(\Phi_{1,p_2,0},\rho) \geq \max_{\Omega} I_c(\Phi_{1,p_2,0},\Omega)$.
We can write the mathematical expression, 
\begin{equation}
    \mathcal{Q}_{\Phi}\geq \mathcal{Q}_{\phi}.
\end{equation}

Now, we will prove $\mathcal{Q}_{\phi}$ is the upper bound of  $\mathcal{Q}_{\Phi}$ as following way. The map $\phi_{p_2}$ acts on eight dimension Hilbert space span over $\ket{00}$, $\ket{01}$, $\ket{02}$, $\ket{10}$,$\ket{12}$, $\ket{20}$, $\ket{21}$ and $\ket{22}$. For the generic density matrix $\Omega$, the output state is
\begin{widetext}
    \begin{equation}
\resizebox{\textwidth}{!}{$
        \phi_{p_2}=
        \left(
\begin{array}{cccccccc}
 \Omega_{00}+\Omega_{44}+p_2 \Omega_{88} & \Omega_{01} & \Omega_{02} & \Omega_{03} & \Omega_{05} & \Omega_{06} & \Omega_{07} & \sqrt{1-p_2} \Omega_{08} \\
 \Omega_{10} & \Omega_{11} & \Omega_{12} & \Omega_{13} & \Omega_{15} & \Omega_{16} & \Omega_{17} & \sqrt{1-p_2} \Omega_{18} \\
 \Omega_{20} & \Omega_{21} & \Omega_{22} & \Omega_{23} & \Omega_{25} & \Omega_{26} & \Omega_{27} & \sqrt{1-p_2} \Omega_{28} \\
 \Omega_{30} & \Omega_{31} & \Omega_{32} & \Omega_{33} & \Omega_{35} & \Omega_{36} & \Omega_{37} & \sqrt{1-p_2} \Omega_{38} \\
 \Omega_{50} & \Omega_{51} & \Omega_{52} & \Omega_{53} & \Omega_{55} & \Omega_{56} & \Omega_{57} & \sqrt{1-p_2} \Omega_{58} \\
 \Omega_{60} & \Omega_{61} & \Omega_{62} & \Omega_{63} & \Omega_{65} & \Omega_{66} & \Omega_{67} & \sqrt{1-p_2} \Omega_{68} \\
 \Omega_{70} & \Omega_{71} & \Omega_{72} & \Omega_{73} & \Omega_{75} & \Omega_{76} & \Omega_{77} & \sqrt{1-p_2} \Omega_{78} \\
 \sqrt{1-p_2} \Omega_{80} & \sqrt{1-p_2} \Omega_{81} & \sqrt{1-p_2} \Omega_{82} & \sqrt{1-p_2} \Omega_{83} & \sqrt{1-p_2} \Omega_{85} & \sqrt{1-p_2} \Omega_{86} & \sqrt{1-p_2} \Omega_{87} & \left(1-p_2\right) \Omega_{88} \\
\end{array}
\right)
$}  
\label{Eq_10}
\end{equation}
and the complementary map has the following form:
\begin{equation}
\Tilde{\phi}=
  \left(
\begin{array}{cccccccc}
 1-p_2\Omega_{88} & 0 & 0 & 0 & 0 & 0 & 0 & \sqrt{p_2} \Omega_{08} \\
 0 & 0 & 0 & 0  & 0 & 0 & 0 & 0 \\
 0 & 0 & 0 & 0 & 0 & 0 & 0 & 0 \\
 0 & 0 & 0 & 0  & 0 & 0 & 0 & 0 \\
 0 & 0 & 0 & 0  & 0 & 0 & 0 & 0 \\
 0 & 0 & 0 & 0  & 0 & 0 & 0 & 0 \\
 0 & 0 & 0 & 0  & 0 & 0 & 0 & 0 \\
 \sqrt{p_2} \Omega_{80} & 0 & 0 & 0 & 0 & 0 & 0 & p_2 \Omega_{88} \\
\end{array}
\right), 
\end{equation}
where for $i,j=0,1,2,3,5,6,7,8$ we set $\Omega_{i,j}=\bra{i}\Omega\Ket{j}$. The basis $\ket{i}$ and $\ket{j}$ are $\ket{0}\equiv\ket{00}$, $\ket{1}\equiv\ket{01}$, $\ket{2}\equiv\ket{02}$, $\ket{3}\equiv\ket{10}$, $\ket{4}\equiv\ket{11}$, $\ket{5}\equiv\ket{12}$, $\ket{6}\equiv\ket{20}$,$\ket{7}\equiv\ket{21}$ and $\ket{8}\equiv\ket{22}$ respectively.
\end{widetext}
Explicitly, we can write
\begin{equation}\label{Eq_12}
    \Phi_{(1,p_2,0)}=\phi_{p_2}\circ\epsilon,
\end{equation}
where $\epsilon$ is a CPTP map: $\mathcal{H}(S)\rightarrow\mathcal{H}(S')$ that transforms the state of the system $S$ to $S'$ by completely eliminating level $\ket{11}$ and transferring its population to $\ket{00}$ level, {\it i.e.},
\begin{equation}
    \epsilon(\rho)=
    \left(
\begin{array}{cccccccc}
 \rho_{00}+\rho_{44} & \rho_{01} & \rho_{02} & \rho_{03} & \rho_{05} & \rho_{06} & \rho_{07} & \rho_{08} \\
 \rho_{10} & \rho_{11} & \rho_{12} & \rho_{13} & \rho_{15} & \rho_{16} & \rho_{17} & \rho_{18} \\
 \rho_{20} & \rho_{21} & \rho_{22} & \rho_{23} & \rho_{25} & \rho_{26} & \rho_{27} & \rho_{28} \\
 \rho_{30} & \rho_{31} & \rho_{32} & \rho_{33} & \rho_{35} & \rho_{36} & \rho_{37} & \rho_{38} \\
 \rho_{50} & \rho_{51} & \rho_{52} & \rho_{53} & \rho_{55} & \rho_{56} & \rho_{57} & \rho_{58} \\
 \rho_{60} & \rho_{61} & \rho_{62} & \rho_{63} & \rho_{65} & \rho_{66} & \rho_{67} & \rho_{68} \\
 \rho_{70} & \rho_{71} & \rho_{72} & \rho_{73} & \rho_{75} & \rho_{76} & \rho_{77} & \rho_{78} \\
 \rho_{80} & \rho_{81} & \rho_{82} & \rho_{83} & \rho_{85} & \rho_{86} & \rho_{87} & \rho_{88} \\
\end{array}
\right)
\end{equation}
where, $\rho_{ij}=\bra{i}\rho\Ket{j}$.
From Eq. (\ref{Eq_12}) using the composition rule as described in the Appendix \ref{Ap_comp}, we can say $\mathcal{Q}_{\phi}$ is the upper bound of $\mathcal{Q}_{\Phi}$ {\it i.e.},
\begin{equation}
   \mathcal{Q}_{\Phi}\leq \mathcal{Q}_{\phi}.  
\end{equation}
Finally, we can calculate  $\mathcal{Q}_{\Phi}$ by finding out the degradability of $\mathcal{Q}_{\phi}$ and using the monotonic behaviour of the $\mathcal{Q}_{\phi}$ in the non-degradable region.
First, we check the degradability of the mapping $\phi_{p_2}$ given in Eq. (\ref{Eq_10}). By checking the positivity of the super-operator $\mathcal{M}_{\phi_D}$, we can figure out that for $p_2\leq 1/2$ the channel is degradable. In this regime, the computed quantum capacity is 
\begin{equation}
\begin{aligned}
   \mathcal{Q}_{\phi}=&\max_{\Omega_d}\{S(\phi_{p_2}(\Omega_D))-S(\Tilde{\phi}_{p_2}(\Omega_D))\}\\
   =&-(\alpha+p_2\delta)\log_2(\alpha+p_2\delta)-6\beta\log_2\beta-\delta(1-p_2)\\
   &\times\log_2(\delta(1-p_2)) +(1-p_2\delta)\log_2((1-p_2\delta))\\
   &+p_2\delta\log_2(p_2\delta).
   \end{aligned}
\end{equation}
where the maximization is performed over the density matrix $\Omega_d=\alpha\ket{00}\bra{00}+\beta\ket{01}\bra{01}+\beta\ket{02}\bra{02}+\beta\ket{10}\bra{10}+\beta\ket{12}\bra{12}+\beta\ket{20}\bra{20}+\beta\ket{21}\bra{21}+\beta\ket{22}\bra{22}$ which spans over eight-dimensional subspace because of the complete elimination of the level $\ket{11}$.
{However, the map has a seven-dimensional noiseless subspace. Therefore, the lower bound of the quantum capacity is $\log_27$. Hence, we can also conclude that the channel is not anti-degradable.}
By finding out the $\mathcal{Q}_{\phi}$ at $p_2=1/2$ and using its monotonic property as described in the Appendix, we find the quantum capacity beyond $p\geq1/2$, which is displayed in Fig. (\ref{fig_1}).
\begin{figure}
    \centering
    \includegraphics[height=5.2cm, width=8cm]{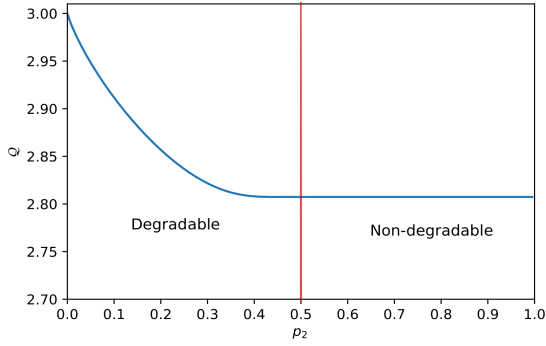}
    \caption{Quantum capacity of the map $\Phi_{(1,p_2,0)}$ varies with decay rate $p_2$.}
    \label{fig_1}
\end{figure}
Since by swapping energy level we will get the map $\Phi_{(p_1,1,0)}$ from $\Phi_{(1,p_2,0)}$. The above analysis also applies to the mapping $\Phi_{(p_1,1,0)}$.
\subsubsection{V-type decay channel}
The lowermost energy level in this damping channel only interacts with the two higher energy levels, and the transition from $\ket{22}\rightarrow\ket{11} $ is not permitted.
The transformation $\Phi_{(p_1,p_2,0)}$ takes the followimg form:
\begin{widetext}
\begin{equation}\label{AP_B19}
\resizebox{\textwidth}{!}{$
\Phi=
    \left(
\begin{array}{ccccccccc}
 p_1 \rho_{44}+p_2 \rho_{88}+\rho_{00} & \rho_{01} & \rho_{02} & \rho_{03} & \sqrt{1-p_1} \rho_{04} & \rho_{05} & \rho_{06} & \rho_{07} & \sqrt{1-p_2} \rho_{08} \\
 \rho_{10} & \rho_{11} & \rho_{12} & \rho_{13} & \sqrt{1-p_1} \rho_{14} & \rho_{15} & \rho_{16} & \rho_{17} & \sqrt{1-p_2} \rho_{18} \\
 \rho_{20} & \rho_{21} & \rho_{22} & \rho_{23} & \sqrt{1-p_1} \rho_{24} & \rho_{25} & \rho_{26} & \rho_{27} & \sqrt{1-p_2} \rho_{28} \\
 \rho_{30} & \rho_{31} & \rho_{32} & \rho_{33} & \sqrt{1-p_1} \rho_{34} & \rho_{35} & \rho_{36} & \rho_{37} & \sqrt{1-p_2} \rho_{38} \\
 \sqrt{1-p_1} \rho_{40} & \sqrt{1-p_1} \rho_{41} & \sqrt{1-p_1} \rho_{42} & \sqrt{1-p_1} \rho_{43} & \left(1-p_1\right) \rho_{44} & \sqrt{1-p_1} \rho_{45} & \sqrt{1-p_1} \rho_{46} & \sqrt{1-p_1} \rho_{47} & \sqrt{1-p_1} \sqrt{1-p_2} \rho_{48} \\
 \rho_{50} & \rho_{51} & \rho_{52} & \rho_{53} & \sqrt{1-p_1} \rho_{54} & \rho_{55} & \rho_{56} & \rho_{57} & \sqrt{1-p_2} \rho_{58} \\
 \rho_{60} & \rho_{61} & \rho_{62} & \rho_{63} & \sqrt{1-p_1} \rho_{64} & \rho_{65} & \rho_{66} & \rho_{67} & \sqrt{1-p_2} \rho_{68} \\
 \rho_{70} & \rho_{71} & \rho_{72} & \rho_{73} & \sqrt{1-p_1} \rho_{74} & \rho_{75} & \rho_{76} & \rho_{77} & \sqrt{1-p_2} \rho_{78} \\
 \sqrt{1-p_2} \rho_{80} & \sqrt{1-p_2} \rho_{81} & \sqrt{1-p_2} \rho_{82} & \sqrt{1-p_2} \rho_{83} & \sqrt{1-p_1} \sqrt{1-p_2} \rho_{84} & \sqrt{1-p_2} \rho_{85} & \sqrt{1-p_2} \rho_{86} & \sqrt{1-p_2} \rho_{87} & \left(1-p_2\right) \rho_{88} \\
\end{array}
\right).
$}
\end{equation}
The corresponding complementary channel $\Tilde{\Phi}_{(p_1,p_2,0)}(\rho)$ calculated according to the Eq. (\ref{Eq_commap}) in the main text:
\begin{equation}
\Tilde{\Phi}=
    \left(
\begin{array}{ccccccccc}
 1-p_1 \rho_{44}-p_2 \rho_{88} & 0 & 0 & 0 & \sqrt{p_1} \rho_{04} & 0 & 0 & 0 & \sqrt{p_2} \rho_{08} \\
 0 & 0 & 0 & 0 & 0 & 0 & 0 & 0 & 0 \\
 0 & 0 & 0 & 0 & 0 & 0 & 0 & 0 & 0 \\
 0 & 0 & 0 & 0 & 0 & 0 & 0 & 0 & 0 \\
 \sqrt{p_1} \rho_{40} & 0 & 0 & 0 & p_1 \rho_{44} & 0 & 0 & 0 & \sqrt{p_1} \sqrt{p_2} \rho_{48} \\
 0 & 0 & 0 & 0 & 0 & 0 & 0 & 0 & 0 \\
 0 & 0 & 0 & 0 & 0 & 0 & 0 & 0 & 0 \\
 0 & 0 & 0 & 0 & 0 & 0 & 0 & 0 & 0 \\
 \sqrt{p_2} \rho_{80} & 0 & 0 & 0 & \sqrt{p_1} \sqrt{p_2} \rho_{84} & 0 & 0 & 0 & p_2 \rho_{88} \\
\end{array}
\right).
\end{equation}
\end{widetext}
Note that the map $\Phi_{(p_2,p_1,0)}$ can be obtained from the map $\Phi_{(p_1,p_2,0)}$ by swapping the levels $\ket{11}\leftrightarrow\ket{22}$. Hence, the quantum capacity of these two maps is the same.
Now, we calculate the range of the values of $p_1$ and $p_2$ for which the channel is degradable. We obtain $\mathcal{M}_{\Phi_D}$ as given in  Eq. (\ref{Eq_D}), a $81\times 81$ matrix which is positive when $(1-2p_1)\geq0$ and $(1-2p_1)\geq0$. Hence, $\Phi_D$ is CPTP when $p_1\leq1/2$ and $p_2\leq1/2$ and the map is degrading in this range. 
{The super-operator $\mathcal{M}_{\Phi}$ corresponding to V-type decay channel is a full rank matrix except $p_1=1$. Now, when $p_1\neq 1$, we can say the degrading map is unique according to Theorem 3 given in Ref.} \cite{bradler2015pitfalls}. { In the case of $p_1=1$, the channel becomes $\Phi_{(1,p_2,0)}$, which is neither degradable nor anti-degradable, as discussed in the earlier section. We find the elements $\rho_{01}, \rho_{02}$ etc., of the input state $\rho$ are present in $\Tilde{\Phi}_{(p_1,p_2,0)}$ but absent in $\Phi_{(p_1,p_2,0)}$. Therefore, there is no linear map that we can apply to $\Tilde{\Phi}(\rho)$ to obtain $\Phi(\rho)$. Moreover, the V-type decay channel has a seven-dimensional noiseless subspace, indicating that the lower bound of quantum capacity is $\log_27$. Hence, we can conclude that the V-type decay channel is not anti-degradable.}
The expression of quantum capacity in the degradable regime is given in Eq. (\ref{Eq_49}).
 Again, we use the composition rule as given in the Appendix \ref{Ap_comp} section to calculate the quantum capacity in the non-degradable region. 

It is important to note that the capacities are also known at the edges of the parameter space. This is because when one of the rates is zero, the system reduces to the single-decay fully correlated MAD channel and the expression of quantum capacity for that given in the main text. If one of the levels is completely damped, then the channel reduces to $\Phi_{(1,p_2,0)}$ or $\Phi_{(p_1,1,0)}$ for which the analysis of quantum capacity is done in the previous section. 
We also know the quantum capacity value on the border of the degradable region {\it i.e.}, $\mathcal{Q}_{\Phi_{(1/2,p_2,0)}}$ and $\mathcal{Q}_{\Phi_{(p_1,1/2,0)}}$ is known for all values of $p_1,p_2\leq1/2$. We find that the quantum capacity on the edge and the border of the degradable region are the same. Accordingly, we can say
\begin{equation}
\begin{aligned}
    \mathcal{Q}_{\Phi_{(1/2,p_2,0)}}=\mathcal{Q}_{\Phi_{(1,p_2,0)}}\; \forall p_2\leq 1/2\\
    \mathcal{Q}_{\Phi_{(p_1,1/2,0)}}=\mathcal{Q}_{\Phi_{(p_1,1,0)}}\; \forall p_1\leq 1/2.
    \end{aligned}
\end{equation}
{The value of the $\mathcal{Q}_{\Phi}$ at $p_1=1/2$ and $p_2=1/2$ is equal to $\log_27$. Hence,  $\mathcal{Q}_{\Phi_{(1/2,p_2,0)}}= \mathcal{Q}_{\Phi_{(p_1,1/2,0)}}=\log_27\;\forall p_1\geq1/2\; \text{and}\; p_2\geq1/2$, which is also the upper bound of $\mathcal{Q}_{\Phi}$ in the region $p_1>1/2$ and $p_2>1/2$. On the other hand, the seven-dimensional noiseless subspace of the transformation indicates that the lower bound of $\mathcal{Q}_{\Phi}$ is equal to $\log_27$. Since the lower bound and upper bound of quantum capacity are the same, the value of $\mathcal{Q}_{\Phi}$ in the region $p_1>1/2$ and $p_2>1/2$ is $\log_27$.}
Hence, using the monotonicity constraint, we can say the value of quantum capacity in the entire non-degradable region, which is shown in Fig. (\ref{Quantum capacity of V-type channel}).

\subsubsection{ \texorpdfstring{$\Lambda$-type decay channel}{}}
In damping channels of this type, energy level $\ket{2}$ interacts with the lower lying levels $\ket{1}$ and $\ket{0}$.
\begin{widetext}
The transformation $\Phi_{(0,p_2,p_3)}(\rho)$ can be written as
\begin{equation}
\resizebox{\textwidth}{!}{$
\Phi=
\left(
\begin{array}{ccccccccc}
 (1-\Theta) p_{23} \rho_{88}+\rho_{00} & \rho_{01} & \rho_{02} & \rho_{03} & \rho_{04} & \rho_{05} & \rho_{06} & \rho_{07} & \sqrt{1-p_{23}} \rho_{08} \\
 \rho_{10} & \rho_{11} & \rho_{12} & \rho_{13} & \rho_{14} & \rho_{15} & \rho_{16} & \rho_{17} & \sqrt{1-p_{23}} \rho_{18} \\
 \rho_{20} & \rho_{21} & \rho_{22} & \rho_{23} & \rho_{24} & \rho_{25} & \rho_{26} & \rho_{27} & \sqrt{1-p_{23}} \rho_{28} \\
 \rho_{30} & \rho_{31} & \rho_{32} & \rho_{33} & \rho_{34} & \rho_{35} & \rho_{36} & \rho_{37} & \sqrt{1-p_{23}} \rho_{38} \\
 \rho_{40} & \rho_{41} & \rho_{42} & \rho_{43} & \Theta p_{23} \rho_{88}+\rho_{44} & \rho_{45} & \rho_{46} & \rho_{47} & \sqrt{1-p_{23}} \rho_{48} \\
 \rho_{50} & \rho_{51} & \rho_{52} & \rho_{53} & \rho_{54} & \rho_{55} & \rho_{56} & \rho_{57} & \sqrt{1-p_{23}} \rho_{58} \\
 \rho_{60} & \rho_{61} & \rho_{62} & \rho_{63} & \rho_{64} & \rho_{65} & \rho_{66} & \rho_{67} & \sqrt{1-p_{23}} \rho_{68} \\
 \rho_{70} & \rho_{71} & \rho_{72} & \rho_{73} & \rho_{74} & \rho_{75} & \rho_{76} & \rho_{77} & \sqrt{1-p_{23}} \rho_{78} \\
 \sqrt{1-p_{23}} \rho_{80} & \sqrt{1-p_{23}} \rho_{81} & \sqrt{1-p_{23}} \rho_{82} & \sqrt{1-p_{23}} \rho_{83} & \sqrt{1-p_{23}} \rho_{84} & \sqrt{1-p_{23}} \rho_{85} & \sqrt{1-p_{23}} \rho_{86} & \sqrt{1-p_{23}} \rho_{87} & 1-p_{23} \rho_{88} \\
\end{array}
\right).
$}
\label{Eq_lambda}
\end{equation}
The corresponding complementary map $\Tilde{\Phi}_{(0,p_2,p_3)}(\rho)$ is
\begin{equation}
\Tilde{\Phi}=
    \left(
\begin{array}{ccccccccc}
 1-(1-p_{23}) \rho_{88} & 0 & 0 & 0 & \rho_{08} \sqrt{(1-\Theta) p_{23}} & 0 & 0 & 0 & \rho_{48} \sqrt{\Theta p_{23}} \\
 0 & 0 & 0 & 0 & 0 & 0 & 0 & 0 & 0 \\
 0 & 0 & 0 & 0 & 0 & 0 & 0 & 0 & 0 \\
 0 & 0 & 0 & 0 & 0 & 0 & 0 & 0 & 0 \\
 \rho_{80} \sqrt{(1-\Theta) p_{23}} & 0 & 0 & 0 & (1-\Theta) p_{23} \rho_{88} & 0 & 0 & 0 & 0 \\
 0 & 0 & 0 & 0 & 0 & 0 & 0 & 0 & 0 \\
 0 & 0 & 0 & 0 & 0 & 0 & 0 & 0 & 0 \\
 0 & 0 & 0 & 0 & 0 & 0 & 0 & 0 & 0 \\
 \rho_{84} \sqrt{\Theta p_{23}} & 0 & 0 & 0 & 0 & 0 & 0 & 0 & \theta p_{23} \rho_{88} \\
\end{array}
\right).
\end{equation}
\end{widetext}
Now, we find the values of $p_2$ and $p_3$ for which the channel is degradable. {We observe the matrix $\mathcal{M}_{\Phi}$ is a full rank matrix and $d_S=d_{S'}$.} We obtain $\mathcal{M}_{\Phi_D}$ as given in Eq. (\ref{Eq_D}), a $81\times 81$ matrix which is positive when $(1-p_2)(1-p_3)\geq\frac{1}{2}$. {Therefore, in this range where $(1-p_2)(1-p_3)\geq\frac{1}{2}$, the channel $\Phi_{(0,p_2,p_3)}$ is degradable, and the degrading map is unique.} \cite{bradler2015pitfalls}.
{The presence of an eight-dimensional noiseless subspace in $\Phi_{(0,p_2,p_3)}$ indicates that the channel has positive quantum capacity for all possible values of $p_2$ and $p_3$. This observation indirectly suggests that the $\Lambda$-type decay channel is not anti-degradable.}
The expression of the quantum capacity in the degradable region is given in Eq. (\ref{Eq_51}).
When $p_1$ or $p_2$ is zero, the channel becomes a single decay channel for which the quantum capacity is known. As we stated, the transformation $\Phi_{(0,p_2,p_3)}$ has eight-dimensional noiseless subspace span over $\ket{00}$, $\ket{01}$, $\ket{02}$, $\ket{10}$, $\ket{11}$, $\ket{12}$, $\ket{20}$ and $\ket{21}$. Hence, the lower bound of the quantum capacity value is $\log_28=3$, which is the same at the boundary of the degradable region $(1-p_2)(1-p_3)=1/2$. Hence, from the monotonicity constraint, we can obtain the value of quantum capacity in the non-degradable region, which is equal to $\mathcal{Q}_{(0,p_2,p_3)}$ at $(1-p_2)(1-p_3)=1/2$. The plot of Quantum capacity with $p_2$ and $p_3$ is shown in Fig. (\ref{fig_4}). 
\subsection{Entanglement Assisted Capacity}
Here, we have given the expression of entanglement-assisted quantum capacity $\mathcal{Q}_E$ for different possible maps.
\subsubsection{Single decay channel}
In the case of a single decay fully correlated MAD channel, the expression of entanglement-assisted quantum capacity is as follows:
\begin{equation}\label{QE_Single}
\begin{aligned}
\mathcal{Q}_E\left(\Phi\right) & =\max _{\bar{\rho}} I\left(\Phi, \bar{\rho}\right) \\
& =\max _{\bar{\rho}}\left\{S\left(\Phi(\bar{\rho})\right)+I_c\left(\Phi, \bar{\rho}\right)\right\}\\
&=\max_{\alpha,\beta,\gamma,\delta}\{-\alpha\log_2{\alpha}-\gamma\log_2{\gamma}-(\alpha+p_1\gamma)\\
&\times\log_2{(\alpha+p_1\gamma)}-12\beta\log_2{\beta}-((1-p_1)\gamma)\\
&\times\log_2{((1-p_1)\gamma)}-2\delta\log_2{\delta}+(1-p_1\gamma)\\
&\times\log_2{(1-p_1\gamma)}+p_1\gamma\log_2{(p_1\gamma)}\}.\\
\end{aligned}
\end{equation}
To obtain the value of $\mathcal{Q}_E$for a given $p_1$ numerically optimization is performed over all possible values of $\alpha$, $\beta$, $\gamma$ and $\delta$. 
The corresponding dynamics of $\mathcal{Q}_E$ with the decay rate $p_1$ are shown in Fig. (\ref{Fig_5}).
The above result is also true for other single decay mappings, namely, $\Phi_{(0,p_2,0)}$ and $\Phi_{(0,0,p_3)}$  because of the symmetry in the transformation.
\subsubsection{V-type decay channel}
Similarly, for the V-type decay channel, we also calculate the entanglement-assisted quantum capacity, which is 
\begin{equation}
\begin{aligned}
\mathcal{Q}\left(\Phi\right) & =\max_{\alpha,\beta,\gamma,\delta}\{-\alpha \log_2\alpha -\gamma \log_2 \gamma -\delta\log_2\delta-\\&(\alpha+p_1\gamma+p_2\delta)\log_2{(\alpha+p_1\gamma+p_2\delta)}\\&-12\beta\log_2{\beta}-\gamma(1-p_1)\log_2{((1-p_1)\gamma)}-\delta(1-p_2)\\
&\times\log_2{((1-p_2)\delta)}+p_1\gamma\log_2{(p_1\gamma)}+(1-p_1\gamma-p_2\delta)\\
&\times\log_2{(1-p_1\gamma p_2\delta)}+p_2\delta\log_2{(p_2\delta)} \}.\\
\end{aligned}
\end{equation}
After performing the numerical optimization over $\alpha$, $\beta$, $\gamma$ and $\delta$, the obtained value of $\mathcal{Q}_E$ is plotted in the contour plot Fig. (\ref{Fig_11}) with respect to $p_1$ and $p_2$. 
\subsubsection{\texorpdfstring{$\Lambda$-type decay channel}{}}
For $\Lambda$-type decay channel the equation of $\mathcal{Q}_E$ is given below:
\begin{equation}
\begin{aligned}
\mathcal{Q}\left(\Phi\right) & =\max_{\alpha,\beta,\gamma,\delta}\{-\alpha \log_2\alpha-12\beta\log_2{\beta} -\gamma \log_2 \gamma -\delta\log_2\delta\\
&-(\alpha+(1-\Theta)p_{23}\delta)\log_2{(\alpha+(1-\Theta)p_{23}\delta)}\\
&-(\gamma+\Theta p_{23}\delta)\log_2{(\gamma+\Theta p_{23}\delta)}+(\Theta p_{23}\delta)\log_2(\Theta p_{23}\delta)\\
&-((1-p_{23})\delta)\log_2{(1-p_{23})\delta)}+(1-p_{23}\delta)\log_2{(1-p_{23}\delta)}\\
&+((1-\Theta)p_{23}\delta)\log_2{((1-\Theta)p_{23}\delta)} \}.\\
\end{aligned}
\end{equation}
The corresponding contour plot of $\mathcal{Q}_E$ is displayed in Fig. (\ref{Fig_11}).
\subsubsection{Three decay rate channel}
Now, we consider a transformation $\Phi_{(p,p,p)}$ in which all the decay rates are the same. For this mapping, we could not calculate the quantum capacity as the channel is neither degradable nor anti-degradable. However, the entanglement-assisted quantum capacity can be calculated as follows:
\begin{equation}\label{AP_B28}
\begin{aligned}
\mathcal{Q}\left(\Phi\right) & =\max_{\alpha,\beta,\gamma,\delta}\{-\alpha \log_2\alpha -\gamma \log_2 \gamma -\delta\log_2\delta\\
&-(\alpha+p\gamma+p\delta)\log_2{(\alpha+p\gamma+p\delta)}-12\beta\log_2{\beta}\\
&-((1-p)\gamma+ (1-p)\delta)\log_2{((1-p)\gamma+ (1-p)\delta)}\\
&-((1-p)^2\delta)\log_2{((1-p)^2\delta)}+(p\gamma)\log_2{(p\gamma)}\\
&+2(1-p\gamma+(p^2-2p)\delta)\log_2{(1-p\gamma)+(p^2-2p)\delta)}\\
&+(p\delta)\log_2(p\delta)\}.\\
\end{aligned}
\end{equation}
We have plotted the $\Phi_{(p,p,p)}$ against decay parameter $p$ in Fig. (\ref{Fig_10}).
\nocite{*}

\bibliography{apssamp}

\providecommand{\noopsort}[1]{}\providecommand{\singleletter}[1]{#1}%
\begin{thebibliography}{57}%
\makeatletter
\providecommand \@ifxundefined [1]{%
 \@ifx{#1\undefined}
}%
\providecommand \@ifnum [1]{%
 \ifnum #1\expandafter \@firstoftwo
 \else \expandafter \@secondoftwo
 \fi
}%
\providecommand \@ifx [1]{%
 \ifx #1\expandafter \@firstoftwo
 \else \expandafter \@secondoftwo
 \fi
}%
\providecommand \natexlab [1]{#1}%
\providecommand \enquote  [1]{``#1''}%
\providecommand \bibnamefont  [1]{#1}%
\providecommand \bibfnamefont [1]{#1}%
\providecommand \citenamefont [1]{#1}%
\providecommand \href@noop [0]{\@secondoftwo}%
\providecommand \href [0]{\begingroup \@sanitize@url \@href}%
\providecommand \@href[1]{\@@startlink{#1}\@@href}%
\providecommand \@@href[1]{\endgroup#1\@@endlink}%
\providecommand \@sanitize@url [0]{\catcode `\\12\catcode `\$12\catcode
  `\&12\catcode `\#12\catcode `\^12\catcode `\_12\catcode `\%12\relax}%
\providecommand \@@startlink[1]{}%
\providecommand \@@endlink[0]{}%
\providecommand \url  [0]{\begingroup\@sanitize@url \@url }%
\providecommand \@url [1]{\endgroup\@href {#1}{\urlprefix }}%
\providecommand \urlprefix  [0]{URL }%
\providecommand \Eprint [0]{\href }%
\providecommand \doibase [0]{https://doi.org/}%
\providecommand \selectlanguage [0]{\@gobble}%
\providecommand \bibinfo  [0]{\@secondoftwo}%
\providecommand \bibfield  [0]{\@secondoftwo}%
\providecommand \translation [1]{[#1]}%
\providecommand \BibitemOpen [0]{}%
\providecommand \bibitemStop [0]{}%
\providecommand \bibitemNoStop [0]{.\EOS\space}%
\providecommand \EOS [0]{\spacefactor3000\relax}%
\providecommand \BibitemShut  [1]{\csname bibitem#1\endcsname}%
\let\auto@bib@innerbib\@empty
\bibitem [{\citenamefont {Nielsen}\ and\ \citenamefont
  {Chuang}(2002)}]{nielsen2002quantum}%
  \BibitemOpen
  \bibfield  {author} {\bibinfo {author} {\bibfnamefont {M.~A.}\ \bibnamefont
  {Nielsen}}\ and\ \bibinfo {author} {\bibfnamefont {I.}~\bibnamefont
  {Chuang}},\ }\href@noop {} {\bibinfo {title} {Quantum computation and quantum
  information}} (\bibinfo {year} {2002})\BibitemShut {NoStop}%
\bibitem [{\citenamefont {Holevo}(2019)}]{holevo2019quantum}%
  \BibitemOpen
  \bibfield  {author} {\bibinfo {author} {\bibfnamefont {A.~S.}\ \bibnamefont
  {Holevo}},\ }\bibfield  {title} {\bibinfo {title} {Quantum systems, channels,
  information},\ }in\ \href@noop {} {\emph {\bibinfo {booktitle} {Quantum
  Systems, Channels, Information}}}\ (\bibinfo  {publisher} {de Gruyter},\
  \bibinfo {year} {2019})\BibitemShut {NoStop}%
\bibitem [{\citenamefont {Shannon}(2001)}]{shannon2001mathematical}%
  \BibitemOpen
  \bibfield  {author} {\bibinfo {author} {\bibfnamefont {C.~E.}\ \bibnamefont
  {Shannon}},\ }\bibfield  {title} {\bibinfo {title} {A mathematical theory of
  communication},\ }\href@noop {} {\bibfield  {journal} {\bibinfo  {journal}
  {ACM SIGMOBILE mobile computing and communications review}\ }\textbf
  {\bibinfo {volume} {5}},\ \bibinfo {pages} {3} (\bibinfo {year}
  {2001})}\BibitemShut {NoStop}%
\bibitem [{\citenamefont {Watrous}(2018)}]{watrous2018theory}%
  \BibitemOpen
  \bibfield  {author} {\bibinfo {author} {\bibfnamefont {J.}~\bibnamefont
  {Watrous}},\ }\href@noop {} {\emph {\bibinfo {title} {The theory of quantum
  information}}}\ (\bibinfo  {publisher} {Cambridge university press},\
  \bibinfo {year} {2018})\BibitemShut {NoStop}%
\bibitem [{\citenamefont {Wilde}(2013)}]{wilde2013quantum}%
  \BibitemOpen
  \bibfield  {author} {\bibinfo {author} {\bibfnamefont {M.~M.}\ \bibnamefont
  {Wilde}},\ }\href@noop {} {\emph {\bibinfo {title} {Quantum information
  theory}}}\ (\bibinfo  {publisher} {Cambridge University Press},\ \bibinfo
  {year} {2013})\BibitemShut {NoStop}%
\bibitem [{\citenamefont {Gyongyosi}\ \emph {et~al.}(2018)\citenamefont
  {Gyongyosi}, \citenamefont {Imre},\ and\ \citenamefont
  {Nguyen}}]{gyongyosi2018survey}%
  \BibitemOpen
  \bibfield  {author} {\bibinfo {author} {\bibfnamefont {L.}~\bibnamefont
  {Gyongyosi}}, \bibinfo {author} {\bibfnamefont {S.}~\bibnamefont {Imre}},\
  and\ \bibinfo {author} {\bibfnamefont {H.~V.}\ \bibnamefont {Nguyen}},\
  }\bibfield  {title} {\bibinfo {title} {A survey on quantum channel
  capacities},\ }\href@noop {} {\bibfield  {journal} {\bibinfo  {journal} {IEEE
  Communications Surveys \& Tutorials}\ }\textbf {\bibinfo {volume} {20}},\
  \bibinfo {pages} {1149} (\bibinfo {year} {2018})}\BibitemShut {NoStop}%
\bibitem [{\citenamefont {Bennett}\ \emph {et~al.}(1999)\citenamefont
  {Bennett}, \citenamefont {Shor}, \citenamefont {Smolin},\ and\ \citenamefont
  {Thapliyal}}]{bennett1999entanglement}%
  \BibitemOpen
  \bibfield  {author} {\bibinfo {author} {\bibfnamefont {C.~H.}\ \bibnamefont
  {Bennett}}, \bibinfo {author} {\bibfnamefont {P.~W.}\ \bibnamefont {Shor}},
  \bibinfo {author} {\bibfnamefont {J.~A.}\ \bibnamefont {Smolin}},\ and\
  \bibinfo {author} {\bibfnamefont {A.~V.}\ \bibnamefont {Thapliyal}},\
  }\bibfield  {title} {\bibinfo {title} {Entanglement-assisted classical
  capacity of noisy quantum channels},\ }\href@noop {} {\bibfield  {journal}
  {\bibinfo  {journal} {Physical Review Letters}\ }\textbf {\bibinfo {volume}
  {83}},\ \bibinfo {pages} {3081} (\bibinfo {year} {1999})}\BibitemShut
  {NoStop}%
\bibitem [{\citenamefont {Poshtvan}\ and\ \citenamefont
  {Karimipour}(2022)}]{poshtvan2022capacities}%
  \BibitemOpen
  \bibfield  {author} {\bibinfo {author} {\bibfnamefont {A.}~\bibnamefont
  {Poshtvan}}\ and\ \bibinfo {author} {\bibfnamefont {V.}~\bibnamefont
  {Karimipour}},\ }\bibfield  {title} {\bibinfo {title} {Capacities of the
  covariant pauli channel},\ }\href@noop {} {\bibfield  {journal} {\bibinfo
  {journal} {Physical Review A}\ }\textbf {\bibinfo {volume} {106}},\ \bibinfo
  {pages} {062408} (\bibinfo {year} {2022})}\BibitemShut {NoStop}%
\bibitem [{\citenamefont {Buscemi}\ and\ \citenamefont
  {Datta}(2010)}]{buscemi2010quantum}%
  \BibitemOpen
  \bibfield  {author} {\bibinfo {author} {\bibfnamefont {F.}~\bibnamefont
  {Buscemi}}\ and\ \bibinfo {author} {\bibfnamefont {N.}~\bibnamefont
  {Datta}},\ }\bibfield  {title} {\bibinfo {title} {The quantum capacity of
  channels with arbitrarily correlated noise},\ }\href@noop {} {\bibfield
  {journal} {\bibinfo  {journal} {IEEE Transactions on Information theory}\
  }\textbf {\bibinfo {volume} {56}},\ \bibinfo {pages} {1447} (\bibinfo {year}
  {2010})}\BibitemShut {NoStop}%
\bibitem [{\citenamefont {Fanizza}\ \emph {et~al.}(2021)\citenamefont
  {Fanizza}, \citenamefont {Kianvash},\ and\ \citenamefont
  {Giovannetti}}]{fanizza2021estimating}%
  \BibitemOpen
  \bibfield  {author} {\bibinfo {author} {\bibfnamefont {M.}~\bibnamefont
  {Fanizza}}, \bibinfo {author} {\bibfnamefont {F.}~\bibnamefont {Kianvash}},\
  and\ \bibinfo {author} {\bibfnamefont {V.}~\bibnamefont {Giovannetti}},\
  }\bibfield  {title} {\bibinfo {title} {Estimating quantum and private
  capacities of gaussian channels via degradable extensions},\ }\href@noop {}
  {\bibfield  {journal} {\bibinfo  {journal} {Physical Review Letters}\
  }\textbf {\bibinfo {volume} {127}},\ \bibinfo {pages} {210501} (\bibinfo
  {year} {2021})}\BibitemShut {NoStop}%
\bibitem [{\citenamefont {King}(2003)}]{king2003capacity}%
  \BibitemOpen
  \bibfield  {author} {\bibinfo {author} {\bibfnamefont {C.}~\bibnamefont
  {King}},\ }\bibfield  {title} {\bibinfo {title} {The capacity of the quantum
  depolarizing channel},\ }\href@noop {} {\bibfield  {journal} {\bibinfo
  {journal} {IEEE Transactions on Information Theory}\ }\textbf {\bibinfo
  {volume} {49}},\ \bibinfo {pages} {221} (\bibinfo {year} {2003})}\BibitemShut
  {NoStop}%
\bibitem [{\citenamefont {Singh}\ and\ \citenamefont
  {Datta}(2022)}]{singh2022detecting}%
  \BibitemOpen
  \bibfield  {author} {\bibinfo {author} {\bibfnamefont {S.}~\bibnamefont
  {Singh}}\ and\ \bibinfo {author} {\bibfnamefont {N.}~\bibnamefont {Datta}},\
  }\bibfield  {title} {\bibinfo {title} {Detecting positive quantum capacities
  of quantum channels},\ }\href@noop {} {\bibfield  {journal} {\bibinfo
  {journal} {npj Quantum Information}\ }\textbf {\bibinfo {volume} {8}},\
  \bibinfo {pages} {50} (\bibinfo {year} {2022})}\BibitemShut {NoStop}%
\bibitem [{\citenamefont {Macchiavello}\ and\ \citenamefont
  {Sacchi}(2016{\natexlab{a}})}]{macchiavello2016detecting}%
  \BibitemOpen
  \bibfield  {author} {\bibinfo {author} {\bibfnamefont {C.}~\bibnamefont
  {Macchiavello}}\ and\ \bibinfo {author} {\bibfnamefont {M.~F.}\ \bibnamefont
  {Sacchi}},\ }\bibfield  {title} {\bibinfo {title} {Detecting lower bounds to
  quantum channel capacities},\ }\href@noop {} {\bibfield  {journal} {\bibinfo
  {journal} {Physical Review Letters}\ }\textbf {\bibinfo {volume} {116}},\
  \bibinfo {pages} {140501} (\bibinfo {year} {2016}{\natexlab{a}})}\BibitemShut
  {NoStop}%
\bibitem [{\citenamefont {Macchiavello}\ and\ \citenamefont
  {Sacchi}(2016{\natexlab{b}})}]{macchiavello2016witnessing}%
  \BibitemOpen
  \bibfield  {author} {\bibinfo {author} {\bibfnamefont {C.}~\bibnamefont
  {Macchiavello}}\ and\ \bibinfo {author} {\bibfnamefont {M.~F.}\ \bibnamefont
  {Sacchi}},\ }\bibfield  {title} {\bibinfo {title} {Witnessing quantum
  capacities of correlated channels},\ }\href@noop {} {\bibfield  {journal}
  {\bibinfo  {journal} {Physical Review A}\ }\textbf {\bibinfo {volume} {94}},\
  \bibinfo {pages} {052333} (\bibinfo {year} {2016}{\natexlab{b}})}\BibitemShut
  {NoStop}%
\bibitem [{\citenamefont {Cuevas}\ \emph {et~al.}(2017)\citenamefont {Cuevas},
  \citenamefont {Proietti}, \citenamefont {Ciampini}, \citenamefont {Duranti},
  \citenamefont {Mataloni}, \citenamefont {Sacchi},\ and\ \citenamefont
  {Macchiavello}}]{cuevas2017experimental}%
  \BibitemOpen
  \bibfield  {author} {\bibinfo {author} {\bibfnamefont {{\'A}.}~\bibnamefont
  {Cuevas}}, \bibinfo {author} {\bibfnamefont {M.}~\bibnamefont {Proietti}},
  \bibinfo {author} {\bibfnamefont {M.~A.}\ \bibnamefont {Ciampini}}, \bibinfo
  {author} {\bibfnamefont {S.}~\bibnamefont {Duranti}}, \bibinfo {author}
  {\bibfnamefont {P.}~\bibnamefont {Mataloni}}, \bibinfo {author}
  {\bibfnamefont {M.~F.}\ \bibnamefont {Sacchi}},\ and\ \bibinfo {author}
  {\bibfnamefont {C.}~\bibnamefont {Macchiavello}},\ }\bibfield  {title}
  {\bibinfo {title} {Experimental detection of quantum channel capacities},\
  }\href@noop {} {\bibfield  {journal} {\bibinfo  {journal} {Physical Review
  Letters}\ }\textbf {\bibinfo {volume} {119}},\ \bibinfo {pages} {100502}
  (\bibinfo {year} {2017})}\BibitemShut {NoStop}%
\bibitem [{\citenamefont {Banaszek}\ \emph {et~al.}(2004)\citenamefont
  {Banaszek}, \citenamefont {Dragan}, \citenamefont {Wasilewski},\ and\
  \citenamefont {Radzewicz}}]{banaszek2004experimental}%
  \BibitemOpen
  \bibfield  {author} {\bibinfo {author} {\bibfnamefont {K.}~\bibnamefont
  {Banaszek}}, \bibinfo {author} {\bibfnamefont {A.}~\bibnamefont {Dragan}},
  \bibinfo {author} {\bibfnamefont {W.}~\bibnamefont {Wasilewski}},\ and\
  \bibinfo {author} {\bibfnamefont {C.}~\bibnamefont {Radzewicz}},\ }\bibfield
  {title} {\bibinfo {title} {Experimental demonstration of
  entanglement-enhanced classical communication over a quantum channel with
  correlated noise},\ }\href@noop {} {\bibfield  {journal} {\bibinfo  {journal}
  {Physical Review Letters}\ }\textbf {\bibinfo {volume} {92}},\ \bibinfo
  {pages} {257901} (\bibinfo {year} {2004})}\BibitemShut {NoStop}%
\bibitem [{\citenamefont {Paladino}\ \emph {et~al.}(2002)\citenamefont
  {Paladino}, \citenamefont {Faoro}, \citenamefont {Falci},\ and\ \citenamefont
  {Fazio}}]{paladino2002decoherence}%
  \BibitemOpen
  \bibfield  {author} {\bibinfo {author} {\bibfnamefont {E.}~\bibnamefont
  {Paladino}}, \bibinfo {author} {\bibfnamefont {L.}~\bibnamefont {Faoro}},
  \bibinfo {author} {\bibfnamefont {G.}~\bibnamefont {Falci}},\ and\ \bibinfo
  {author} {\bibfnamefont {R.}~\bibnamefont {Fazio}},\ }\bibfield  {title}
  {\bibinfo {title} {Decoherence and 1/f noise in josephson qubits},\
  }\href@noop {} {\bibfield  {journal} {\bibinfo  {journal} {Physical Review
  Letters}\ }\textbf {\bibinfo {volume} {88}},\ \bibinfo {pages} {228304}
  (\bibinfo {year} {2002})}\BibitemShut {NoStop}%
\bibitem [{\citenamefont {Makhlin}\ \emph {et~al.}(2001)\citenamefont
  {Makhlin}, \citenamefont {Sch{\"o}n},\ and\ \citenamefont
  {Shnirman}}]{makhlin2001quantum}%
  \BibitemOpen
  \bibfield  {author} {\bibinfo {author} {\bibfnamefont {Y.}~\bibnamefont
  {Makhlin}}, \bibinfo {author} {\bibfnamefont {G.}~\bibnamefont {Sch{\"o}n}},\
  and\ \bibinfo {author} {\bibfnamefont {A.}~\bibnamefont {Shnirman}},\
  }\bibfield  {title} {\bibinfo {title} {Quantum-state engineering with
  josephson-junction devices},\ }\href@noop {} {\bibfield  {journal} {\bibinfo
  {journal} {Reviews of modern physics}\ }\textbf {\bibinfo {volume} {73}},\
  \bibinfo {pages} {357} (\bibinfo {year} {2001})}\BibitemShut {NoStop}%
\bibitem [{\citenamefont {Giovannetti}(2005)}]{giovannetti2005dynamical}%
  \BibitemOpen
  \bibfield  {author} {\bibinfo {author} {\bibfnamefont {V.}~\bibnamefont
  {Giovannetti}},\ }\bibfield  {title} {\bibinfo {title} {A dynamical model for
  quantum memory channels},\ }\href@noop {} {\bibfield  {journal} {\bibinfo
  {journal} {Journal of Physics A: Mathematical and General}\ }\textbf
  {\bibinfo {volume} {38}},\ \bibinfo {pages} {10989} (\bibinfo {year}
  {2005})}\BibitemShut {NoStop}%
\bibitem [{\citenamefont {Cimini}\ \emph {et~al.}(2020)\citenamefont {Cimini},
  \citenamefont {Gianani}, \citenamefont {Sacchi}, \citenamefont
  {Macchiavello},\ and\ \citenamefont {Barbieri}}]{cimini2020experimental}%
  \BibitemOpen
  \bibfield  {author} {\bibinfo {author} {\bibfnamefont {V.}~\bibnamefont
  {Cimini}}, \bibinfo {author} {\bibfnamefont {I.}~\bibnamefont {Gianani}},
  \bibinfo {author} {\bibfnamefont {M.~F.}\ \bibnamefont {Sacchi}}, \bibinfo
  {author} {\bibfnamefont {C.}~\bibnamefont {Macchiavello}},\ and\ \bibinfo
  {author} {\bibfnamefont {M.}~\bibnamefont {Barbieri}},\ }\bibfield  {title}
  {\bibinfo {title} {Experimental witnessing of the quantum channel capacity in
  the presence of correlated noise},\ }\href@noop {} {\bibfield  {journal}
  {\bibinfo  {journal} {Physical Review A}\ }\textbf {\bibinfo {volume}
  {102}},\ \bibinfo {pages} {052404} (\bibinfo {year} {2020})}\BibitemShut
  {NoStop}%
\bibitem [{\citenamefont {Caruso}\ \emph {et~al.}(2014)\citenamefont {Caruso},
  \citenamefont {Giovannetti}, \citenamefont {Lupo},\ and\ \citenamefont
  {Mancini}}]{caruso2014quantum}%
  \BibitemOpen
  \bibfield  {author} {\bibinfo {author} {\bibfnamefont {F.}~\bibnamefont
  {Caruso}}, \bibinfo {author} {\bibfnamefont {V.}~\bibnamefont {Giovannetti}},
  \bibinfo {author} {\bibfnamefont {C.}~\bibnamefont {Lupo}},\ and\ \bibinfo
  {author} {\bibfnamefont {S.}~\bibnamefont {Mancini}},\ }\bibfield  {title}
  {\bibinfo {title} {Quantum channels and memory effects},\ }\href@noop {}
  {\bibfield  {journal} {\bibinfo  {journal} {Reviews of Modern Physics}\
  }\textbf {\bibinfo {volume} {86}},\ \bibinfo {pages} {1203} (\bibinfo {year}
  {2014})}\BibitemShut {NoStop}%
\bibitem [{\citenamefont {Macchiavello}\ and\ \citenamefont
  {Palma}(2002)}]{macchiavello2002entanglement}%
  \BibitemOpen
  \bibfield  {author} {\bibinfo {author} {\bibfnamefont {C.}~\bibnamefont
  {Macchiavello}}\ and\ \bibinfo {author} {\bibfnamefont {G.~M.}\ \bibnamefont
  {Palma}},\ }\bibfield  {title} {\bibinfo {title} {Entanglement-enhanced
  information transmission over a quantum channel with correlated noise},\
  }\href@noop {} {\bibfield  {journal} {\bibinfo  {journal} {Physical Review
  A}\ }\textbf {\bibinfo {volume} {65}},\ \bibinfo {pages} {050301} (\bibinfo
  {year} {2002})}\BibitemShut {NoStop}%
\bibitem [{\citenamefont {Bowen}\ and\ \citenamefont
  {Mancini}(2004)}]{bowen2004quantum}%
  \BibitemOpen
  \bibfield  {author} {\bibinfo {author} {\bibfnamefont {G.}~\bibnamefont
  {Bowen}}\ and\ \bibinfo {author} {\bibfnamefont {S.}~\bibnamefont
  {Mancini}},\ }\bibfield  {title} {\bibinfo {title} {Quantum channels with a
  finite memory},\ }\href@noop {} {\bibfield  {journal} {\bibinfo  {journal}
  {Physical Review A}\ }\textbf {\bibinfo {volume} {69}},\ \bibinfo {pages}
  {012306} (\bibinfo {year} {2004})}\BibitemShut {NoStop}%
\bibitem [{\citenamefont {Kretschmann}\ and\ \citenamefont
  {Werner}(2005)}]{kretschmann2005quantum}%
  \BibitemOpen
  \bibfield  {author} {\bibinfo {author} {\bibfnamefont {D.}~\bibnamefont
  {Kretschmann}}\ and\ \bibinfo {author} {\bibfnamefont {R.~F.}\ \bibnamefont
  {Werner}},\ }\bibfield  {title} {\bibinfo {title} {Quantum channels with
  memory},\ }\href@noop {} {\bibfield  {journal} {\bibinfo  {journal} {Physical
  Review A}\ }\textbf {\bibinfo {volume} {72}},\ \bibinfo {pages} {062323}
  (\bibinfo {year} {2005})}\BibitemShut {NoStop}%
\bibitem [{\citenamefont {Plenio}\ and\ \citenamefont
  {Virmani}(2007)}]{plenio2007spin}%
  \BibitemOpen
  \bibfield  {author} {\bibinfo {author} {\bibfnamefont {M.}~\bibnamefont
  {Plenio}}\ and\ \bibinfo {author} {\bibfnamefont {S.}~\bibnamefont
  {Virmani}},\ }\bibfield  {title} {\bibinfo {title} {Spin chains and channels
  with memory},\ }\href@noop {} {\bibfield  {journal} {\bibinfo  {journal}
  {Physical Review Letters}\ }\textbf {\bibinfo {volume} {99}},\ \bibinfo
  {pages} {120504} (\bibinfo {year} {2007})}\BibitemShut {NoStop}%
\bibitem [{\citenamefont {Caruso}\ \emph {et~al.}(2010)\citenamefont {Caruso},
  \citenamefont {Giovannetti},\ and\ \citenamefont
  {Palma}}]{caruso2010teleportation}%
  \BibitemOpen
  \bibfield  {author} {\bibinfo {author} {\bibfnamefont {F.}~\bibnamefont
  {Caruso}}, \bibinfo {author} {\bibfnamefont {V.}~\bibnamefont
  {Giovannetti}},\ and\ \bibinfo {author} {\bibfnamefont {G.~M.}\ \bibnamefont
  {Palma}},\ }\bibfield  {title} {\bibinfo {title} {Teleportation-induced
  correlated quantum channels},\ }\href@noop {} {\bibfield  {journal} {\bibinfo
   {journal} {Physical review letters}\ }\textbf {\bibinfo {volume} {104}},\
  \bibinfo {pages} {020503} (\bibinfo {year} {2010})}\BibitemShut {NoStop}%
\bibitem [{\citenamefont {Lupo}\ \emph {et~al.}(2010)\citenamefont {Lupo},
  \citenamefont {Giovannetti},\ and\ \citenamefont
  {Mancini}}]{lupo2010capacities}%
  \BibitemOpen
  \bibfield  {author} {\bibinfo {author} {\bibfnamefont {C.}~\bibnamefont
  {Lupo}}, \bibinfo {author} {\bibfnamefont {V.}~\bibnamefont {Giovannetti}},\
  and\ \bibinfo {author} {\bibfnamefont {S.}~\bibnamefont {Mancini}},\
  }\bibfield  {title} {\bibinfo {title} {Capacities of lossy bosonic memory
  channels},\ }\href@noop {} {\bibfield  {journal} {\bibinfo  {journal}
  {Physical Review Letters}\ }\textbf {\bibinfo {volume} {104}},\ \bibinfo
  {pages} {030501} (\bibinfo {year} {2010})}\BibitemShut {NoStop}%
\bibitem [{\citenamefont {Sk}\ and\ \citenamefont
  {Panigrahi}(2022)}]{sk2022protecting}%
  \BibitemOpen
  \bibfield  {author} {\bibinfo {author} {\bibfnamefont {R.}~\bibnamefont
  {Sk}}\ and\ \bibinfo {author} {\bibfnamefont {P.~K.}\ \bibnamefont
  {Panigrahi}},\ }\bibfield  {title} {\bibinfo {title} {Protecting quantum
  coherence and entanglement in a correlated environment},\ }\href@noop {}
  {\bibfield  {journal} {\bibinfo  {journal} {Physica A: Statistical Mechanics
  and its Applications}\ }\textbf {\bibinfo {volume} {596}},\ \bibinfo {pages}
  {127129} (\bibinfo {year} {2022})}\BibitemShut {NoStop}%
\bibitem [{\citenamefont {Hu}\ and\ \citenamefont
  {Wang}(2020)}]{hu2020protecting}%
  \BibitemOpen
  \bibfield  {author} {\bibinfo {author} {\bibfnamefont {M.-L.}\ \bibnamefont
  {Hu}}\ and\ \bibinfo {author} {\bibfnamefont {H.-F.}\ \bibnamefont {Wang}},\
  }\bibfield  {title} {\bibinfo {title} {Protecting quantum fisher information
  in correlated quantum channels},\ }\href@noop {} {\bibfield  {journal}
  {\bibinfo  {journal} {Annalen der Physik}\ }\textbf {\bibinfo {volume}
  {532}},\ \bibinfo {pages} {1900378} (\bibinfo {year} {2020})}\BibitemShut
  {NoStop}%
\bibitem [{\citenamefont {Xu}\ \emph {et~al.}(2019)\citenamefont {Xu},
  \citenamefont {Zhang},\ and\ \citenamefont {Liu}}]{xu2019quantum}%
  \BibitemOpen
  \bibfield  {author} {\bibinfo {author} {\bibfnamefont {K.}~\bibnamefont
  {Xu}}, \bibinfo {author} {\bibfnamefont {G.-F.}\ \bibnamefont {Zhang}},\ and\
  \bibinfo {author} {\bibfnamefont {W.-M.}\ \bibnamefont {Liu}},\ }\bibfield
  {title} {\bibinfo {title} {Quantum dynamical speedup in correlated noisy
  channels},\ }\href@noop {} {\bibfield  {journal} {\bibinfo  {journal}
  {Physical Review A}\ }\textbf {\bibinfo {volume} {100}},\ \bibinfo {pages}
  {052305} (\bibinfo {year} {2019})}\BibitemShut {NoStop}%
\bibitem [{\citenamefont {D'Arrigo}\ \emph {et~al.}(2007)\citenamefont
  {D'Arrigo}, \citenamefont {Benenti},\ and\ \citenamefont
  {Falci}}]{d2007quantum}%
  \BibitemOpen
  \bibfield  {author} {\bibinfo {author} {\bibfnamefont {A.}~\bibnamefont
  {D'Arrigo}}, \bibinfo {author} {\bibfnamefont {G.}~\bibnamefont {Benenti}},\
  and\ \bibinfo {author} {\bibfnamefont {G.}~\bibnamefont {Falci}},\ }\bibfield
   {title} {\bibinfo {title} {Quantum capacity of dephasing channels with
  memory},\ }\href@noop {} {\bibfield  {journal} {\bibinfo  {journal} {New
  Journal of Physics}\ }\textbf {\bibinfo {volume} {9}},\ \bibinfo {pages}
  {310} (\bibinfo {year} {2007})}\BibitemShut {NoStop}%
\bibitem [{\citenamefont {Giovannetti}\ and\ \citenamefont
  {Fazio}(2005)}]{giovannetti2005information}%
  \BibitemOpen
  \bibfield  {author} {\bibinfo {author} {\bibfnamefont {V.}~\bibnamefont
  {Giovannetti}}\ and\ \bibinfo {author} {\bibfnamefont {R.}~\bibnamefont
  {Fazio}},\ }\bibfield  {title} {\bibinfo {title} {Information-capacity
  description of spin-chain correlations},\ }\href@noop {} {\bibfield
  {journal} {\bibinfo  {journal} {Physical Review A}\ }\textbf {\bibinfo
  {volume} {71}},\ \bibinfo {pages} {032314} (\bibinfo {year}
  {2005})}\BibitemShut {NoStop}%
\bibitem [{\citenamefont {Jahangir}\ \emph {et~al.}(2015)\citenamefont
  {Jahangir}, \citenamefont {Arshed},\ and\ \citenamefont
  {Toor}}]{jahangir2015quantum}%
  \BibitemOpen
  \bibfield  {author} {\bibinfo {author} {\bibfnamefont {R.}~\bibnamefont
  {Jahangir}}, \bibinfo {author} {\bibfnamefont {N.}~\bibnamefont {Arshed}},\
  and\ \bibinfo {author} {\bibfnamefont {A.}~\bibnamefont {Toor}},\ }\bibfield
  {title} {\bibinfo {title} {Quantum capacity of an amplitude-damping channel
  with memory},\ }\href@noop {} {\bibfield  {journal} {\bibinfo  {journal}
  {Quantum Information Processing}\ }\textbf {\bibinfo {volume} {14}},\
  \bibinfo {pages} {765} (\bibinfo {year} {2015})}\BibitemShut {NoStop}%
\bibitem [{\citenamefont {Khatri}\ \emph {et~al.}(2020)\citenamefont {Khatri},
  \citenamefont {Sharma},\ and\ \citenamefont {Wilde}}]{khatri2020information}%
  \BibitemOpen
  \bibfield  {author} {\bibinfo {author} {\bibfnamefont {S.}~\bibnamefont
  {Khatri}}, \bibinfo {author} {\bibfnamefont {K.}~\bibnamefont {Sharma}},\
  and\ \bibinfo {author} {\bibfnamefont {M.~M.}\ \bibnamefont {Wilde}},\
  }\bibfield  {title} {\bibinfo {title} {Information-theoretic aspects of the
  generalized amplitude-damping channel},\ }\href@noop {} {\bibfield  {journal}
  {\bibinfo  {journal} {Physical Review A}\ }\textbf {\bibinfo {volume}
  {102}},\ \bibinfo {pages} {012401} (\bibinfo {year} {2020})}\BibitemShut
  {NoStop}%
\bibitem [{\citenamefont {D'Arrigo}\ \emph {et~al.}(2013)\citenamefont
  {D'Arrigo}, \citenamefont {Benenti}, \citenamefont {Falci},\ and\
  \citenamefont {Macchiavello}}]{d2013classical}%
  \BibitemOpen
  \bibfield  {author} {\bibinfo {author} {\bibfnamefont {A.}~\bibnamefont
  {D'Arrigo}}, \bibinfo {author} {\bibfnamefont {G.}~\bibnamefont {Benenti}},
  \bibinfo {author} {\bibfnamefont {G.}~\bibnamefont {Falci}},\ and\ \bibinfo
  {author} {\bibfnamefont {C.}~\bibnamefont {Macchiavello}},\ }\bibfield
  {title} {\bibinfo {title} {Classical and quantum capacities of a fully
  correlated amplitude damping channel},\ }\href@noop {} {\bibfield  {journal}
  {\bibinfo  {journal} {Physical Review A}\ }\textbf {\bibinfo {volume} {88}},\
  \bibinfo {pages} {042337} (\bibinfo {year} {2013})}\BibitemShut {NoStop}%
\bibitem [{\citenamefont {Chessa}\ and\ \citenamefont
  {Giovannetti}(2021{\natexlab{a}})}]{chessa2021quantum}%
  \BibitemOpen
  \bibfield  {author} {\bibinfo {author} {\bibfnamefont {S.}~\bibnamefont
  {Chessa}}\ and\ \bibinfo {author} {\bibfnamefont {V.}~\bibnamefont
  {Giovannetti}},\ }\bibfield  {title} {\bibinfo {title} {Quantum capacity
  analysis of multi-level amplitude damping channels},\ }\href@noop {}
  {\bibfield  {journal} {\bibinfo  {journal} {Communications Physics}\ }\textbf
  {\bibinfo {volume} {4}},\ \bibinfo {pages} {22} (\bibinfo {year}
  {2021}{\natexlab{a}})}\BibitemShut {NoStop}%
\bibitem [{\citenamefont {Chessa}\ and\ \citenamefont
  {Giovannetti}(2021{\natexlab{b}})}]{chessa2021partially}%
  \BibitemOpen
  \bibfield  {author} {\bibinfo {author} {\bibfnamefont {S.}~\bibnamefont
  {Chessa}}\ and\ \bibinfo {author} {\bibfnamefont {V.}~\bibnamefont
  {Giovannetti}},\ }\bibfield  {title} {\bibinfo {title} {Partially coherent
  direct sum channels},\ }\href@noop {} {\bibfield  {journal} {\bibinfo
  {journal} {Quantum}\ }\textbf {\bibinfo {volume} {5}},\ \bibinfo {pages}
  {504} (\bibinfo {year} {2021}{\natexlab{b}})}\BibitemShut {NoStop}%
\bibitem [{\citenamefont {Chessa}\ and\ \citenamefont
  {Giovannetti}(2023)}]{chessa2023resonant}%
  \BibitemOpen
  \bibfield  {author} {\bibinfo {author} {\bibfnamefont {S.}~\bibnamefont
  {Chessa}}\ and\ \bibinfo {author} {\bibfnamefont {V.}~\bibnamefont
  {Giovannetti}},\ }\bibfield  {title} {\bibinfo {title} {Resonant multilevel
  amplitude damping channels},\ }\href@noop {} {\bibfield  {journal} {\bibinfo
  {journal} {Quantum}\ }\textbf {\bibinfo {volume} {7}},\ \bibinfo {pages}
  {902} (\bibinfo {year} {2023})}\BibitemShut {NoStop}%
\bibitem [{\citenamefont {Yeo}\ and\ \citenamefont
  {Skeen}(2003)}]{yeo2003time}%
  \BibitemOpen
  \bibfield  {author} {\bibinfo {author} {\bibfnamefont {Y.}~\bibnamefont
  {Yeo}}\ and\ \bibinfo {author} {\bibfnamefont {A.}~\bibnamefont {Skeen}},\
  }\bibfield  {title} {\bibinfo {title} {Time-correlated quantum
  amplitude-damping channel},\ }\href@noop {} {\bibfield  {journal} {\bibinfo
  {journal} {Physical Review A}\ }\textbf {\bibinfo {volume} {67}},\ \bibinfo
  {pages} {064301} (\bibinfo {year} {2003})}\BibitemShut {NoStop}%
\bibitem [{\citenamefont {Briegel}\ and\ \citenamefont
  {Englert}(1993)}]{briegel1993quantum}%
  \BibitemOpen
  \bibfield  {author} {\bibinfo {author} {\bibfnamefont {H.-J.}\ \bibnamefont
  {Briegel}}\ and\ \bibinfo {author} {\bibfnamefont {B.-G.}\ \bibnamefont
  {Englert}},\ }\bibfield  {title} {\bibinfo {title} {Quantum optical master
  equations: The use of damping bases},\ }\href@noop {} {\bibfield  {journal}
  {\bibinfo  {journal} {Physical Review A}\ }\textbf {\bibinfo {volume} {47}},\
  \bibinfo {pages} {3311} (\bibinfo {year} {1993})}\BibitemShut {NoStop}%
\bibitem [{\citenamefont {Choi}(1975)}]{choi1975completely}%
  \BibitemOpen
  \bibfield  {author} {\bibinfo {author} {\bibfnamefont {M.-D.}\ \bibnamefont
  {Choi}},\ }\bibfield  {title} {\bibinfo {title} {Completely positive linear
  maps on complex matrices},\ }\href@noop {} {\bibfield  {journal} {\bibinfo
  {journal} {Linear algebra and its applications}\ }\textbf {\bibinfo {volume}
  {10}},\ \bibinfo {pages} {285} (\bibinfo {year} {1975})}\BibitemShut
  {NoStop}%
\bibitem [{\citenamefont {Stinespring}(1955)}]{stinespring1955positive}%
  \BibitemOpen
  \bibfield  {author} {\bibinfo {author} {\bibfnamefont {W.~F.}\ \bibnamefont
  {Stinespring}},\ }\bibfield  {title} {\bibinfo {title} {Positive functions on
  c*-algebras},\ }\href@noop {} {\bibfield  {journal} {\bibinfo  {journal}
  {Proceedings of the American Mathematical Society}\ }\textbf {\bibinfo
  {volume} {6}},\ \bibinfo {pages} {211} (\bibinfo {year} {1955})}\BibitemShut
  {NoStop}%
\bibitem [{\citenamefont {Devetak}\ and\ \citenamefont
  {Shor}(2005)}]{devetak2005capacity}%
  \BibitemOpen
  \bibfield  {author} {\bibinfo {author} {\bibfnamefont {I.}~\bibnamefont
  {Devetak}}\ and\ \bibinfo {author} {\bibfnamefont {P.~W.}\ \bibnamefont
  {Shor}},\ }\bibfield  {title} {\bibinfo {title} {The capacity of a quantum
  channel for simultaneous transmission of classical and quantum information},\
  }\href@noop {} {\bibfield  {journal} {\bibinfo  {journal} {Communications in
  Mathematical Physics}\ }\textbf {\bibinfo {volume} {256}},\ \bibinfo {pages}
  {287} (\bibinfo {year} {2005})}\BibitemShut {NoStop}%
\bibitem [{\citenamefont {Smith}\ and\ \citenamefont
  {Smolin}(2007)}]{smith2007degenerate}%
  \BibitemOpen
  \bibfield  {author} {\bibinfo {author} {\bibfnamefont {G.}~\bibnamefont
  {Smith}}\ and\ \bibinfo {author} {\bibfnamefont {J.~A.}\ \bibnamefont
  {Smolin}},\ }\bibfield  {title} {\bibinfo {title} {Degenerate quantum codes
  for pauli channels},\ }\href@noop {} {\bibfield  {journal} {\bibinfo
  {journal} {Physical Review Letters}\ }\textbf {\bibinfo {volume} {98}},\
  \bibinfo {pages} {030501} (\bibinfo {year} {2007})}\BibitemShut {NoStop}%
\bibitem [{\citenamefont {Hastings}(2009)}]{hastings2009superadditivity}%
  \BibitemOpen
  \bibfield  {author} {\bibinfo {author} {\bibfnamefont {M.~B.}\ \bibnamefont
  {Hastings}},\ }\bibfield  {title} {\bibinfo {title} {Superadditivity of
  communication capacity using entangled inputs},\ }\href@noop {} {\bibfield
  {journal} {\bibinfo  {journal} {Nature Physics}\ }\textbf {\bibinfo {volume}
  {5}},\ \bibinfo {pages} {255} (\bibinfo {year} {2009})}\BibitemShut {NoStop}%
\bibitem [{\citenamefont {Schumacher}\ and\ \citenamefont
  {Westmoreland}(1997)}]{schumacher1997sending}%
  \BibitemOpen
  \bibfield  {author} {\bibinfo {author} {\bibfnamefont {B.}~\bibnamefont
  {Schumacher}}\ and\ \bibinfo {author} {\bibfnamefont {M.~D.}\ \bibnamefont
  {Westmoreland}},\ }\bibfield  {title} {\bibinfo {title} {Sending classical
  information via noisy quantum channels},\ }\href@noop {} {\bibfield
  {journal} {\bibinfo  {journal} {Physical Review A}\ }\textbf {\bibinfo
  {volume} {56}},\ \bibinfo {pages} {131} (\bibinfo {year} {1997})}\BibitemShut
  {NoStop}%
\bibitem [{\citenamefont {Dorlas}\ and\ \citenamefont
  {Morgan}(2008)}]{dorlas2008calculating}%
  \BibitemOpen
  \bibfield  {author} {\bibinfo {author} {\bibfnamefont {T.}~\bibnamefont
  {Dorlas}}\ and\ \bibinfo {author} {\bibfnamefont {C.}~\bibnamefont
  {Morgan}},\ }\bibfield  {title} {\bibinfo {title} {Calculating a maximizer
  for quantum mutual information},\ }\href@noop {} {\bibfield  {journal}
  {\bibinfo  {journal} {International Journal of Quantum Information}\ }\textbf
  {\bibinfo {volume} {6}},\ \bibinfo {pages} {745} (\bibinfo {year}
  {2008})}\BibitemShut {NoStop}%
\bibitem [{\citenamefont {Macchiavello}\ \emph {et~al.}(2020)\citenamefont
  {Macchiavello}, \citenamefont {Sacchi},\ and\ \citenamefont
  {Sacchi}}]{macchiavello2020bounding}%
  \BibitemOpen
  \bibfield  {author} {\bibinfo {author} {\bibfnamefont {C.}~\bibnamefont
  {Macchiavello}}, \bibinfo {author} {\bibfnamefont {M.~F.}\ \bibnamefont
  {Sacchi}},\ and\ \bibinfo {author} {\bibfnamefont {T.}~\bibnamefont
  {Sacchi}},\ }\bibfield  {title} {\bibinfo {title} {Bounding the classical
  capacity of multilevel damping quantum channels},\ }\href@noop {} {\bibfield
  {journal} {\bibinfo  {journal} {Advanced Quantum Technologies}\ }\textbf
  {\bibinfo {volume} {3}},\ \bibinfo {pages} {2000013} (\bibinfo {year}
  {2020})}\BibitemShut {NoStop}%
\bibitem [{\citenamefont {Xu}\ \emph {et~al.}(2022)\citenamefont {Xu},
  \citenamefont {Zhou}, \citenamefont {Li}, \citenamefont {Jiang},\ and\
  \citenamefont {Ian}}]{xu2022enhancing}%
  \BibitemOpen
  \bibfield  {author} {\bibinfo {author} {\bibfnamefont {R.}~\bibnamefont
  {Xu}}, \bibinfo {author} {\bibfnamefont {R.-G.}\ \bibnamefont {Zhou}},
  \bibinfo {author} {\bibfnamefont {Y.}~\bibnamefont {Li}}, \bibinfo {author}
  {\bibfnamefont {S.}~\bibnamefont {Jiang}},\ and\ \bibinfo {author}
  {\bibfnamefont {H.}~\bibnamefont {Ian}},\ }\bibfield  {title} {\bibinfo
  {title} {Enhancing robustness of noisy qutrit teleportation with markovian
  memory},\ }\href@noop {} {\bibfield  {journal} {\bibinfo  {journal} {EPJ
  Quantum Technology}\ }\textbf {\bibinfo {volume} {9}},\ \bibinfo {pages} {1}
  (\bibinfo {year} {2022})}\BibitemShut {NoStop}%
\bibitem [{\citenamefont {Lloyd}(1997)}]{lloyd1997capacity}%
  \BibitemOpen
  \bibfield  {author} {\bibinfo {author} {\bibfnamefont {S.}~\bibnamefont
  {Lloyd}},\ }\bibfield  {title} {\bibinfo {title} {Capacity of the noisy
  quantum channel},\ }\href@noop {} {\bibfield  {journal} {\bibinfo  {journal}
  {Physical Review A}\ }\textbf {\bibinfo {volume} {55}},\ \bibinfo {pages}
  {1613} (\bibinfo {year} {1997})}\BibitemShut {NoStop}%
\bibitem [{\citenamefont {Barnum}\ \emph {et~al.}(1998)\citenamefont {Barnum},
  \citenamefont {Nielsen},\ and\ \citenamefont
  {Schumacher}}]{barnum1998information}%
  \BibitemOpen
  \bibfield  {author} {\bibinfo {author} {\bibfnamefont {H.}~\bibnamefont
  {Barnum}}, \bibinfo {author} {\bibfnamefont {M.~A.}\ \bibnamefont
  {Nielsen}},\ and\ \bibinfo {author} {\bibfnamefont {B.}~\bibnamefont
  {Schumacher}},\ }\bibfield  {title} {\bibinfo {title} {Information
  transmission through a noisy quantum channel},\ }\href@noop {} {\bibfield
  {journal} {\bibinfo  {journal} {Physical Review A}\ }\textbf {\bibinfo
  {volume} {57}},\ \bibinfo {pages} {4153} (\bibinfo {year}
  {1998})}\BibitemShut {NoStop}%
\bibitem [{\citenamefont {Bennett}\ \emph {et~al.}(2002)\citenamefont
  {Bennett}, \citenamefont {Shor}, \citenamefont {Smolin},\ and\ \citenamefont
  {Thapliyal}}]{bennett2002entanglement}%
  \BibitemOpen
  \bibfield  {author} {\bibinfo {author} {\bibfnamefont {C.~H.}\ \bibnamefont
  {Bennett}}, \bibinfo {author} {\bibfnamefont {P.~W.}\ \bibnamefont {Shor}},
  \bibinfo {author} {\bibfnamefont {J.~A.}\ \bibnamefont {Smolin}},\ and\
  \bibinfo {author} {\bibfnamefont {A.~V.}\ \bibnamefont {Thapliyal}},\
  }\bibfield  {title} {\bibinfo {title} {Entanglement-assisted capacity of a
  quantum channel and the reverse shannon theorem},\ }\href@noop {} {\bibfield
  {journal} {\bibinfo  {journal} {IEEE transactions on Information Theory}\
  }\textbf {\bibinfo {volume} {48}},\ \bibinfo {pages} {2637} (\bibinfo {year}
  {2002})}\BibitemShut {NoStop}%
\bibitem [{\citenamefont {Adami}\ and\ \citenamefont
  {Cerf}(1997)}]{adami1997neumann}%
  \BibitemOpen
  \bibfield  {author} {\bibinfo {author} {\bibfnamefont {C.}~\bibnamefont
  {Adami}}\ and\ \bibinfo {author} {\bibfnamefont {N.~J.}\ \bibnamefont
  {Cerf}},\ }\bibfield  {title} {\bibinfo {title} {von neumann capacity of
  noisy quantum channels},\ }\href@noop {} {\bibfield  {journal} {\bibinfo
  {journal} {Physical Review A}\ }\textbf {\bibinfo {volume} {56}},\ \bibinfo
  {pages} {3470} (\bibinfo {year} {1997})}\BibitemShut {NoStop}%
\bibitem [{\citenamefont {Keyl}(2002)}]{keyl2002fundamentals}%
  \BibitemOpen
  \bibfield  {author} {\bibinfo {author} {\bibfnamefont {M.}~\bibnamefont
  {Keyl}},\ }\bibfield  {title} {\bibinfo {title} {Fundamentals of quantum
  information theory},\ }\href@noop {} {\bibfield  {journal} {\bibinfo
  {journal} {Physics reports}\ }\textbf {\bibinfo {volume} {369}},\ \bibinfo
  {pages} {431} (\bibinfo {year} {2002})}\BibitemShut {NoStop}%
\bibitem [{\citenamefont {Wolf}\ and\ \citenamefont
  {Perez-Garcia}(2007)}]{wolf2007quantum}%
  \BibitemOpen
  \bibfield  {author} {\bibinfo {author} {\bibfnamefont {M.~M.}\ \bibnamefont
  {Wolf}}\ and\ \bibinfo {author} {\bibfnamefont {D.}~\bibnamefont
  {Perez-Garcia}},\ }\bibfield  {title} {\bibinfo {title} {Quantum capacities
  of channels with small environment},\ }\href@noop {} {\bibfield  {journal}
  {\bibinfo  {journal} {Physical Review A}\ }\textbf {\bibinfo {volume} {75}},\
  \bibinfo {pages} {012303} (\bibinfo {year} {2007})}\BibitemShut {NoStop}%
\bibitem [{\citenamefont {Bradler}(2015)}]{bradler2015pitfalls}%
  \BibitemOpen
  \bibfield  {author} {\bibinfo {author} {\bibfnamefont {K.}~\bibnamefont
  {Bradler}},\ }\bibfield  {title} {\bibinfo {title} {The pitfalls of deciding
  whether a quantum channel is (conjugate) degradable and how to avoid them},\
  }\href@noop {} {\bibfield  {journal} {\bibinfo  {journal} {Open Systems \&
  Information Dynamics}\ }\textbf {\bibinfo {volume} {22}},\ \bibinfo {pages}
  {1550026} (\bibinfo {year} {2015})}\BibitemShut {NoStop}%
\bibitem [{\citenamefont {Cubitt}\ \emph {et~al.}(2008)\citenamefont {Cubitt},
  \citenamefont {Ruskai},\ and\ \citenamefont {Smith}}]{cubitt2008structure}%
  \BibitemOpen
  \bibfield  {author} {\bibinfo {author} {\bibfnamefont {T.~S.}\ \bibnamefont
  {Cubitt}}, \bibinfo {author} {\bibfnamefont {M.~B.}\ \bibnamefont {Ruskai}},\
  and\ \bibinfo {author} {\bibfnamefont {G.}~\bibnamefont {Smith}},\ }\bibfield
   {title} {\bibinfo {title} {The structure of degradable quantum channels},\
  }\href@noop {} {\bibfield  {journal} {\bibinfo  {journal} {Journal of
  Mathematical Physics}\ }\textbf {\bibinfo {volume} {49}},\ \bibinfo {pages}
  {102104} (\bibinfo {year} {2008})}\BibitemShut {NoStop}%
\end{thebibliography}%

\end{document}